\documentclass[reprint, nofootinbib, amsmath,amssymb, aps,floatfix]{revtex4-2}
\usepackage{gensymb}
\usepackage{textcomp}
\usepackage{lipsum}
\usepackage{graphicx}
\usepackage{dcolumn}
\usepackage{bm}
\usepackage{siunitx}
\DeclareSIUnit\gauss{G}
\DeclareSIUnit\erg{erg}

\usepackage{cancel}
\usepackage{tabularx}
\usepackage{amssymb}
\usepackage{amsmath}
\usepackage{relsize}
\usepackage{listings}
\usepackage{commath}
\usepackage{enumitem}
\usepackage{xfrac}
\usepackage{float}
\usepackage{rotating}
\usepackage{booktabs}
\usepackage{makecell}
\usepackage{mathtools}
\usepackage{caption}
\usepackage{subcaption}
\usepackage{multirow}
\usepackage[table,xcdraw]{xcolor}
\usepackage[version=4]{mhchem}
\usepackage[colorlinks,bookmarks=false,citecolor=red,linkcolor=blue,urlcolor=blue]{hyperref}

\begin{document}
\title{Game susceptibility, Correlation and Payoff capacity as a measure of Cooperative behaviour in the thermodynamic limit of some Social dilemmas}
\author{Colin Benjamin}
\email{colin.nano@gmail.com}
\author{Rajdeep Tah}
\affiliation{School of Physical Sciences, National Institute of Science Education and Research, HBNI, Jatni-752050, India}

\begin{abstract}
 Analytically, finding the origins of cooperative behavior in infinite-player games is an exciting topic of current interest. In this paper, we compare three analytical methods, i.e., Nash equilibrium mapping (NEM), Darwinian selection (DS) and Aggregate selection (AS), with a numerical Agent based method (ABM) via the game susceptibility, correlation, and payoff capacity as indicators of cooperative behaviour. While the analytical NEM model shows excellent agreement with the numerical ABM, the other analytical models like AS and DS show notable divergence with ABM in the thermodynamic limit for the indicators in question. Previously, cooperative behavior was studied by considering game magnetization and individual players' average payoff as indicators. This paper shows that game susceptibility, correlation, and payoff capacity can aid in understanding cooperative behavior in social dilemmas in the thermodynamic limit. The results obtained via NEM and ABM are in good agreement for all three indicators in question, for both Hawk-Dove and the Public goods games. After comparing the results obtained for all five indicators, we see that individual players' average payoff and payoff capacity serve as the best indicators to study cooperative behavior among players in the thermodynamic limit.
\end{abstract}

\maketitle

\section{\label{introsec}Introduction}
Within the framework of evolution, \textit{cooperation} is a very intriguing phenomenon where individuals (or, \textit{players}) of a population tend to forego their selfish interests for the benefit of others\cite{ref0}. Although this might not seem advantageous in the short term, the question of why and how cooperative behavior even arises among individuals of a population seems intriguing. Many theories have been proposed to explain how and why cooperation arises even if the Nash equilibrium strategy is \textit{defection}. In game theory, the equilibrium strategy that leads to the least loss or maximum gain for all players is defined as the \textit{Nash equilibrium} \cite{ref0, ref1, ref3, ref4}. In social dilemma games, like the \textit{two}-players Prisoner's dilemma, the Nash equilibrium strategy is \textit{defection}. However, as shown in Ref.~\cite{ref15}, for \textit{repeated} Prisoner's dilemma, i.e., both players interact among themselves multiple times, cooperation arises due to the adoption of \textit{tit-for-tat} strategies by the players, and this cooperative behavior sustains because of \textit{direct reciprocity} \cite{ref0} among the players. These indicate that cooperative populations eventually endure, and our modern-day society is a perfect example of that, where individuals with differing opinions and thought processes tend to cooperate, to a certain extent, in the context of religion, language, and society\cite{ref0}. A variety of mechanisms in addition to \textit{Direct reciprocity}, like \textit{Indirect reciprocity; Spatial selection; Group selection} and \textit{Kin selection} (see, Ref.~\cite{ref0}), exist which aid in understanding the origins of cooperative behavior among individuals, indicating that cooperation is the inevitable result of evolution \cite{ref0, ref7}. {Further, new mechanisms like  migration and aspiration have been shown to aid in coperative behavior too, see Refs.~\cite{ref-ref}}.

In this paper, on the other hand, we aim to understand the emergence of cooperative behavior not in an evolutionary setting (i.e., \textit{repeated games}) but rather in a \textit{one-shot} context, analytically. Specifically, we compare three analytical methods used to study cooperative behavior among players in the thermodynamic limit: \textit{Nash equilibrium mapping} (NEM), \textit{Aggregate selection} (AS), and \textit{Darwinian selection} (DS), with a numerical \textit{Agent based method} (ABM), which are all based on the \textit{1D}-Ising chain. Aggregate selection and Darwinian selection were previously called \textit{Hamiltonian dynamics} (HD) and \textit{Darwinian evolution} (DE), respectively, in Ref.~\cite{ref5}. However, the names \textit{Darwinian evolution} and especially \textit{Hamiltonian dynamics} are misnomers since in both Refs.~\cite{ref4, ref5}, the focus is on \textit{one-shot} games played in the thermodynamic limit. Thus, neither dynamics nor evolution is involved in either of these models. Therefore, we rename HD as \textit{Aggregate selection} (AS) and DE as \textit{Darwinian selection} (DS), respectively. Traditionally, cooperative behavior among players in social dilemmas is studied using numerical methods (see, Refs.~\cite{ref15, ref7}) like \textit{iterative methods}, \textit{replicator dynamics}, \textit{linear programming}. These methods are \textit{dynamical} in nature, i.e., they have time-dependency. However, we adopt an analytical non-dynamical approach to find exact expressions for the different indicators of cooperative behavior, which further helps us determine the fraction of the population cooperating in the thermodynamic limit. In our work, we find that the Nash equilibrium can be directly determined via the game payoff matrix in the thermodynamic limit and involves an exactly solvable model in Statistical Mechanics. All three analytical methods, i.e., NEM, AS, and DS, which we compare, are based on the $1D$-Ising model (see Ref.~\cite{ref14}).

{Before, we go any further, we clarify one question which may arise from the preceeding discussion: ``why do we need to understand emergence of cooperation in one-shot rather than evolutionary context''? The reason is in the evolutionary context (i.e., with repeated games) emergence of cooperation is mainly tackled numerically via agent based models, as analytical solutions are rare to find in the thermodynamic limit. However, in the thermodynamic limit for the one-shot case analytical models are available\cite{ref1,ref5}. The purpose of this paper is on how the three different analytical models used to analyze cooperation in one-shot games in the thermodynamic limit: Nash equilibrium mapping(NEM), Darwinian selection(DS) and Agreegate selection (AS) fare against each other when compared to a numerical agent based model. On why these three only because, till date these are the only three available. To compare these analytical models we utilize five indicators: game magnetization, game susceptibility, game correlation, game payoff capacity and average payoff. }

Our primary goal is to determine which model most closely approximates the players' behavior in the limit of thermodynamics (i.e., \textit{infinite player}). One of us has previously worked on this topic by considering game magnetization $(\mu_g)$ and individual player's average payoff $(\langle \Lambda_1 \rangle)$ as indicators of the emergence of cooperation (see, Refs.~\cite{ref1, ref3, ref4}). In this paper, we look at three new indicators: \textit{game susceptibility}, \textit{correlation} and \textit{payoff capacity}, for cooperative behavior and compare our results to that of $\mu_g$ and $\langle \Lambda_1 \rangle$, and determine which is the best indicator to study cooperative behavior among players in the thermodynamic limit. In a symmetric \textit{two-player, two-strategy} social dilemma, as the name suggests, two players have two different strategies $\mathcal{S}_1$ and $\mathcal{S}_2$ available to themselves, and they can opt for either of the two available strategies which might lead to identical or different outcomes (or, \textit{payoffs}) for both of them. The game payoffs $(\mathcal{A, B, C, D})$ are associated with pairs of strategies ($\mathcal{S}_1, \mathcal{S}_1$), ($\mathcal{S}_2, \mathcal{S}_1$), ($\mathcal{S}_1, \mathcal{S}_2$) and ($\mathcal{S}_2, \mathcal{S}_2$), where the \textit{left} element of the strategy pair corresponds to the strategy adopted by \textit{Player-1} and the \textit{right} element of the strategy pair corresponds to the strategy adopted by \textit{Player-2}, respectively. The \textit{game susceptibility} gives us the difference between the rate of change in the fraction of players choosing \textit{Cooperation} strategy (i.e., cooperators) and the fraction of players choosing \textit{Defection} strategy (i.e., defectors) owing to a modification in payoffs. On the other hand, \textit{correlation} gives us the degree of correlation between the strategies of two players at two different sites. Correlation, as an indicator for cooperation among players, was previously utilized in quantum Prisoner's dilemma to show the tuning of global correlations via local entanglement (see, Ref.~\cite{ref11}), which is an exciting model to understand Type-\textit{II} superconducting behavior. The \textit{payoff capacity}, which is analogous to specific heat, at constant volume, for the Ising model, corresponds to the number of player payoff changes when the \textit{noise} increases by a unit. The \textit{randomness} in the player's strategy selection is attributed to the \textit{noise} (or, \textit{selection pressure}) of the environment.

{Another question which may arise from the preceeding paragraph is \lq\lq{}why are new cooperation indicators needed?\rq\rq{} Now since we are looking to analyze cooperative behavior, it behooves us to find what is the best measure of coperation. Generally in both analytical and numerical models the net magnetization, i.e., difference between cooperators and defectors is taken as a good measure of the onset of coperation and what parameters enhance or reduce it. However,  in statistical mechanics via magnetic systems there are many ways apart from magnetization, to measure the response of a system to an external field or temperature, like susceptibility, specific heat, correlation and internal energy. What we do in this paper is to find analogs of each of these statistical mechanics measures and check which among them can best represent coperation in a game theoretic context. This the second aim of the work, first being a check on which analytical model is best when compared to a  numerical agent based model.}

To develop the mathematical formalism for the analytical methods, we use a $1D$-Ising chain, considering nearest neighbor interaction, where the sites represent the players and the spin states at each site correspond to strategies \cite{ref1, ref3, ref5}. {Further, the choice of indicators is driven by the mapping, since its a mapping to a 1D spin -$1/2$   Ising model. These five indicators are analogs of  thermodynamical quantities measured in the Ising model. However, in game theoretic works they have not had much resonance apart from the magnetization. We in this work show that each of these indicators can provide signatures of cooperative behavior in the thermodynamic limit. We infact show that the average payoff and the game payoff capacity are best indicators to measure cooperative behavior. The five indicators exhaust all possible thermodynamic meaures of the Ising model. }

{The detailed description and the calculations related to the different methods are mentioned in  Sec.~\ref{theory}. In the case of NEM, see Sec.\ref{sub-nem} and Refs.~\cite{ref1, ref3}, we map the game payoffs to the Hamiltonian of the $1D$-Ising chain with nearest neighbor interaction (the ID Ising model is introduced in Sec.~\ref{ising}), and for AS (in Sec.~\ref{sub-hdm}) and DS (in Sec.~\ref{sub-dem}), we adopt the formalism given in Ref.~\cite{ref5}. We introduce the Agent based model in Sec.~\ref{sub-abm}. We study two social dilemmas: The Hawk-Dove game(HDG) (see, Appendix~\ref{hdg}) and the Public goods game(PGG) (see, Sec.~\ref{pgg}) and find that for HDG, the results obtained for the game susceptibility in Sec.~\ref{sus-hdg}, correlation in Sec.~\ref{corr-hdg}, and payoff capacity  in Sec.~\ref{paycap-hdg} via NEM and ABM, were in good agreement with one another. On the contrary, we observed that DS gave indifferent results for all three indicators in question in HDG. For PGG, the results obtained via NEM and ABM for all three indicators, i.e., game susceptibility in Sec.\ref{sus-pgg}, correlation in Sec.~\ref{corr-pgg}, and payoff capacity in Sec.~\ref{paycap-pgg}, exactly match, whereas the results of AS did not. DS gave the same result as NEM and ABM for game susceptibility and correlation, but for payoff capacity, we observe that DS in the \textit{infinite noise} limit gives incompatible results.}

Next, we studied the correlation variation with the inter-site distance $j$ and found the results obtained via NEM and ABM for PGG in Sec.~\ref{corr-pgg} and HDG in Sec.~\ref{corr-hdg} were in agreement with each other. We have discussed more about them in the later Secs.~[\ref{sus-pgg-analysis}, \ref{corr-pgg-analysis},\ref{sus-hdg-analysis}, \ref{corr-hdg-analysis}]. After analyzing the results obtained via the analytical methods for the indicators in question, we see that at the thermodynamic (or, \textit{infinite player}) limit, cooperative behavior among the players emerges, contrary to what was previously thought impossible in \textit{two-player} games. Additionally, Nowak's study of cooperative behavior in repeated games with finite populations, using numerical dynamics (see Ref.~\cite{ref15}), also suggests the same. We can confirm that in \textit{one-shot} games, NEM is the sole reliable analytical method that should be used to study the emergence of cooperation among players in the thermodynamic limit. Finally, we compare in Tables~\ref{tab:my-table2} and \ref{tab:my-table1} the results of game magnetization and individual players' average payoff (from Ref.~\cite{ref4}) with our results for game susceptibility, correlation, and payoff capacity and find that both individual player's average payoff and payoff capacity serve as the best indicators to study cooperative behavior among players in the thermodynamic limit. The following section~\ref{theory} discusses the mathematical framework related to NEM in Sec.~\ref{sub-nem}, AS in Sec.~\ref{sub-hdm}, and DS in Sec.~\ref{sub-dem}. Later, we will discuss the algorithm in Sec.~\ref{sub-abm} related to ABM and apply these methods to both games.

%In Ref.~\cite{ref0}, the authors put forward five mechanisms: \textit{Direct reciprocity, Indirect reciprocity, Spatial selection, Group selection} and \textit{Kin selection}, that contribute to the emergence and sustainability of cooperation among individuals in a population. In Direct reciprocity, cooperation thrives when individuals interact repeatedly, fostering mutual benefits over time. Meanwhile, in Indirect reciprocity, \textit{Reputation} (see, Ref.~\cite{ref0}) of the individual matters, i.e., individuals are more likely to cooperate if they have a positive reputation based on past interactions. In Spatial selection, cooperation is favored in structured populations where individuals interact more frequently with nearby neighbors, whereas, in Group selection, cooperation is promoted when groups with more cooperators outperform those with fewer, leading to the spread of cooperative traits at the group level. Finally, in Kin selection, cooperation is explained by genetic relatedness, as individuals may be more inclined to cooperate with close relatives who share more genes. These five principles highlight the factors that promote and sustain cooperation in evolutionary contexts.

\section{\label{theory}Theory}
Here, we will discuss the three distinct analytical techniques used to explore social dilemmas in the thermodynamic limit: \textit{Nash equilibrium mapping} (NEM), \textit{Darwinian selection} (DS), and \textit{Aggregate selection} (AS) and also the numerical \textit{Agent based method} (ABM). Developed in close analogy to the \textit{1D}-Ising chain, these techniques are not based on the dynamical evolution of strategies, but rather on equilibrium statistical mechanics (see, Refs.~\cite{ref2, ref5}). This entails focusing on the thermodynamic limit of a \textit{one-shot game}. 

The idea of applying equilibrium statistical mechanics in the thermodynamic limit of games was introduced in Ref.~\cite{ref5}, where the authors discussed AS (previously called HD) and DS (previously called DE), respectively. It has been shown that with NEM \cite{ref1, ref3}, we get much better results than AS, and also, in many cases, it gave better results than DS (see, Ref.~\cite{ref4}). These are discussed in Refs.~\cite{ref1, ref3, ref4, ref6} where the authors have considered $\mu_g$ and $\langle \Lambda_1 \rangle$ as indicators. 

\subsection{\label{ising} 1D Ising Model}
For a $N$-site, \textit{1D}-Ising chain with nearest neighbour interactions, we write the Hamiltonian as,
\begin{equation}
    H = -\mathcal{J} \sum_{i=1}^{N} \mathfrak{s}_i \mathfrak{s}_{i+1} - \mathfrak{h}\sum_{i=1}^{N} \mathfrak{s}_i ,
    \label{eq1}
\end{equation}
where $\mathcal{J}$ denotes the coupling via which the nearest neighbour spins interact, $\mathfrak{h}$ is the externally applied uniform magnetic field, and $\mathfrak{s}_i = \pm 1$ denotes the \textit{two}-spin states ($\uparrow$ and $\downarrow$) at the $i^{th}$ site respectively. For the given $H$ in Eq.~(\ref{eq1}), we write the partition function $\zeta$ as\cite{ref8},
\begin{equation}
    \zeta = \sum_{\mathfrak{s}_1, \mathfrak{s}_2,...,\mathfrak{s}_N} e^{\beta (\mathcal{J} \sum_{i=1}^{N} \mathfrak{s}_i \mathfrak{s}_{i+1} + \frac{\mathfrak{h}}{2}\sum_{i=1}^{N} (\mathfrak{s}_i + \mathfrak{s}_{i+1}))},
    \label{eq2}
\end{equation}
with $\beta = \frac{1}{k_B T}$ representing the inverse of the temperature, $k_B$ being the \textit{Boltzmann constant}. Solving Eq.~(\ref{eq2}) using transfer matrix $\Gamma$ \cite{ref8}, we define,
\begin{gather}
    \Gamma(\mathfrak{s}, \mathfrak{s}') = \langle \mathfrak{s} | \Gamma |\mathfrak{s}' \rangle = e^{\beta (\mathcal{J} \mathfrak{s}\mathfrak{s}' + \frac{\mathfrak{h}}{2} (\mathfrak{s}+\mathfrak{s}'))}, \nonumber\\
    \text{or,}~ \Gamma = 
    \begin{bmatrix}
        \Gamma_{+,+} & \Gamma_{+,-}\\
        \Gamma_{-,+} & \Gamma_{-,-}
    \end{bmatrix}
    = \begin{bmatrix}
        e^{\beta (\mathcal{J}+\mathfrak{h})} & e^{-\beta \mathcal{J}}\\
        e^{-\beta \mathcal{J}} & e^{\beta (\mathcal{J}-\mathfrak{h})}
    \end{bmatrix},
    \label{eq3}
\end{gather}
where, $\Gamma_{\pm, \pm} \equiv \Gamma_{\mathfrak{s} = \pm 1,\mathfrak{s} = \pm 1}$, and $\zeta$ in terms of the elements of $\Gamma$ is,
\begin{equation}
    \zeta = \sum_{\mathfrak{s}_1, \mathfrak{s}_2,..., \mathfrak{s}_N} \Gamma(\mathfrak{s}_1, \mathfrak{s}_2) \Gamma(\mathfrak{s}_2, \mathfrak{s}_3)... \Gamma(\mathfrak{s}_N, \mathfrak{s}_1),
    \label{eq4}
\end{equation}
where, we have assumed a periodic boundary condition, i.e., $\mathfrak{s}_{N+1} = \mathfrak{s}_1$. The transfer matrix $\Gamma$ in Eq.~(\ref{eq3}) has the following property~\cite{ref8},
\begin{gather}
    \sum_{\mathfrak{s}_{i+1}} \Gamma(\mathfrak{s}_i, \mathfrak{s}_{i+1})\Gamma(\mathfrak{s}_{i+1}, \mathfrak{s}_{i+2}) = \Gamma^2 (\mathfrak{s}_i, \mathfrak{s}_{i+2}),
    \nonumber\\
    \text{or,}~ \zeta = \text{Tr}(\Gamma^N).
    \label{eq5}
\end{gather}
The sum of the eigenvalues of $\Gamma$ is Tr($\Gamma$). Supposing $(\Omega_{-},~\Omega_{+})$ are two eigenvalues of the square-matrix $\Gamma$. This implies $(\Omega_{-}^{N},~\Omega_{+}^{N})$ are the two eigenvalues of $\Gamma^N$. Thus, $\zeta = \text{Tr}(\Gamma^N) = \Omega_{-}^{N} + \Omega_{+}^{N}$, and we have the eigenvalues (of the $\Gamma$ matrix) in terms of the Ising parameters as \cite{ref8},
\begin{equation}
    \Omega_{\pm} = e^{-\beta \mathcal{J}} [e^{2\beta \mathcal{J}}\text{cosh}(\beta \mathfrak{h}) \pm \sqrt{1+e^{4\beta \mathcal{J}}\text{sinh}^2 (\beta \mathfrak{h})}],
    \label{eq6}
\end{equation}
where, $\Omega_{+}>\Omega_{-}$. The partition function $\zeta$, then is,
\begin{equation}
    \zeta = \text{Tr}(\Gamma^N) = \Omega_{-}^{N} + \Omega_{+}^{N} = \Omega_{+}^{N}\bigg[ 1 + \bigg(\dfrac{\Omega_-}{\Omega_+}\bigg)^N  \bigg].
    \label{eq7}
\end{equation}
From Eq.~(\ref{eq6}), we notice that $\Omega_{+}>\Omega_{-}$ and in the thermodynamic limit, i.e., $N\rightarrow \infty$, $(\Omega_-/\Omega_+)^N \approx 0$. Thus, the partition function, in $N\rightarrow\infty$ limit, effectively reduces to,
\begin{equation}
    \zeta = \Omega_{+}^N = [e^{\beta \mathcal{J}}\text{cosh}(\beta \mathfrak{h}) + e^{-\beta \mathcal{J}}\sqrt{1+e^{4\beta \mathcal{J}}\text{sinh}^2 (\beta \mathfrak{h})}]^N.
    \label{eq8}
\end{equation}
From partition function $\zeta$ in Eq.~(\ref{eq8}), we can calculate \textit{five} quantities that define our understanding of the \textit{1D}-Ising chain, and these are,\\
\textbf{1. Magnetization:} The free energy per spin ($\mathcal{G}$) calculated from the partition function $\zeta$ is,
\begin{equation}
    \mathcal{G} = -\dfrac{k_B T}{N} \ln{\zeta}\bigg|_{N\rightarrow \infty} = -k_B T \ln{\Omega_+}.
    \label{eq9}
\end{equation}
The average magnetization in the $1D$-Ising spin chain is,
\begin{equation}
    \mu = -\dfrac{\partial \mathcal{G}}{\partial \mathfrak{h}} = \dfrac{\text{sinh}(\beta \mathfrak{h})}{\sqrt{ \text{sinh}^2(\beta \mathfrak{h}) + e^{-4\beta \mathcal{J}}}}.
    \label{eq10}
\end{equation}
\textbf{2. Susceptibility:} From the expression of magnetization given in Eq.~(\ref{eq10}), we can calculate the magnetic susceptibility ($\Tilde{\chi}_\mathfrak{h}$) as \cite{ref8},
\begin{equation}
    \Tilde{\chi}_\mathfrak{h} = \dfrac{\partial \mu}{\partial \mathfrak{h}} = 2\dfrac{\partial \varrho_{\uparrow}}{\partial \mathfrak{h}},
    \label{eq11}
\end{equation}
where, magnetization $\mu$ gives us the difference between the fraction of $\uparrow$ and $\downarrow$ spins, i.e., $\mu = \varrho_{\uparrow} - \varrho_{\downarrow}$, and $\varrho_{\uparrow} + \varrho_{\downarrow} = 1$. Thus, $\mu = (2\varrho_{\uparrow} - 1)$, written in Eq.~(\ref{eq11}), replacing $\mu$.

We define $\Tilde{\chi}_\mathfrak{h}$ as a measure of the response of magnetization to an externally applied uniform magnetic field $\mathfrak{h}$. The average magnetization $\mu$, given in Eq.~(\ref{eq10}), can also be written as $\mu = \langle \Tilde{\mu} \rangle$, where $\Tilde{\mu} = \sum_{i=1}^{N} \mathfrak{s}_i$. Using the Hamiltonian given in Eq.~(\ref{eq1}), the partition function is written as \cite{ref12},
\begin{gather}
    \zeta = \sum_{\{\mathfrak{s}_i\}} e^{-\beta H(\{\mathfrak{s}_i\})} = \sum_{\{\mathfrak{s}_i\}} e^{\beta (\mathcal{J} \sum_{i=1}^{N} \mathfrak{s}_i \mathfrak{s}_{i+1} + \mathfrak{h}\sum_{i=1}^{N}\mathfrak{s}_i)}, \nonumber\\
    \text{or,}~~ \dfrac{\partial \zeta}{\partial \mathfrak{h}} = \sum_{\{\mathfrak{s}_i\}} e^{-\beta H(\{\mathfrak{s}_i\})} ~\beta~ \Tilde{\mu} .\label{eq12}\\
    \text{Thus,} ~\Tilde{\chi}_{\mathfrak{h}} = \dfrac{\partial \mu}{\partial \mathfrak{h}} = \dfrac{\partial \langle \Tilde{\mu} \rangle}{\partial \mathfrak{h}} = \dfrac{\partial}{\partial \mathfrak{h}} \bigg[ \dfrac{1}{\zeta}\sum_{\{\mathfrak{s}_i\}} \Tilde{\mu}~ e^{-\beta H(\{\mathfrak{s}_i\})}  \bigg] \nonumber\\
    = \bigg(-\dfrac{1}{\zeta^2} \dfrac{\partial \zeta}{\partial \mathfrak{h}} \bigg)\sum_{\{\mathfrak{s}_i\}} \Tilde{\mu}~ e^{-\beta H(\{\mathfrak{s}_i\})} + \dfrac{1}{\zeta} \sum_{\{\mathfrak{s}_i\}} \beta~\Tilde{\mu}^2~e^{-\beta H(\{\mathfrak{s}_i\})},  \nonumber\\
    = \bigg(-\dfrac{1}{\zeta^2} \sum_{\{\mathfrak{s}_i\}} e^{-\beta H(\{\mathfrak{s}_i\})} ~\beta~ \Tilde{\mu} \bigg)\sum_{\{\mathfrak{s}_i\}} \Tilde{\mu}~ e^{-\beta H(\{\mathfrak{s}_i\})} \nonumber\\
    + \dfrac{1}{\zeta} \sum_{\{\mathfrak{s}_i\}} \beta~\Tilde{\mu}^2~e^{-\beta H(\{\mathfrak{s}_i\})},\nonumber\\
    = -\beta\underbrace{\bigg[\dfrac{1}{\zeta} \sum_{\{\mathfrak{s}_i\}} \Tilde{\mu}~ e^{-\beta H(\{\mathfrak{s}_i\})}\bigg]^2}_{\langle \Tilde{\mu} \rangle^2} + \beta \underbrace{\bigg[ \dfrac{1}{\zeta} \sum_{\{\mathfrak{s}_i\}} \Tilde{\mu}^2~e^{-\beta H(\{\mathfrak{s}_i\})}\bigg]}_{\langle \Tilde{\mu}^2 \rangle}\nonumber\\
    \text{or,}~~ \Tilde{\chi}_\mathfrak{h} = \dfrac{\partial \mu}{\partial \mathfrak{h}} = \beta [  \langle \Tilde{\mu}^2 \rangle - \langle \Tilde{\mu} \rangle^2].
    \label{eq13}
\end{gather}
Normalizing the magnetic susceptibility $\Tilde{\chi}_\mathfrak{h}$, in Eqs.~(\ref{eq11}, \ref{eq13}), we have,
\begin{equation}
    \chi_\mathfrak{h} =\dfrac{\Tilde{\chi}_{\mathfrak{h}}}{\beta} = \dfrac{1}{\beta} \dfrac{\partial \mu}{\partial \mathfrak{h}} = [  \langle \Tilde{\mu}^2 \rangle - \langle \Tilde{\mu} \rangle^2].
    \label{eq13a}
\end{equation}
In Eq.~(\ref{eq13a}), we have shown that the magnetic susceptibility $\chi_{\mathfrak{h}}$ is related to the variance of magnetization.\\ 
\textbf{3. Correlation:} The expression for the spin-spin correlation can also be derived similarly. As discussed in Ref.~\cite{ref8}, we can express the average magnetization $\mu$ in terms of the transfer matrix $\Gamma$ as,
\begin{gather}
    \mu = \langle \mathfrak{s}_i \rangle = \dfrac{1}{\zeta} \sum_{\mathfrak{s}_1,..,\mathfrak{s}_N} \Gamma(\mathfrak{s}_1, \mathfrak{s}_2)..\mathfrak{s}_i \Gamma(\mathfrak{s}_{i}, \mathfrak{s}_{i+1}).. \Gamma(\mathfrak{s}_{N}, \mathfrak{s}_{1}),\nonumber\\
    = \dfrac{1}{\zeta} \text{Tr}(\mathfrak{s}_z \Gamma^N) = \dfrac{\text{sinh}(\beta \mathfrak{h})}{\sqrt{ \text{sinh}^2(\beta \mathfrak{h}) + e^{-4\beta \mathcal{J}}}},
    \label{eq14}
\end{gather}
where, $\mathfrak{s}_z = \pm 1$ depending on the spin at the $i^{th}$ site. Similarly, we can write the correlation for two spins located at the $i^{th}$ and $(i+j)^{th}$ sites as,
\begin{gather}
    \langle \mathfrak{s}_i \mathfrak{s}_{i+j}\rangle = \dfrac{\text{Tr}[\Gamma^{i-1}\mathfrak{s}_z \Gamma^j \mathfrak{s}_z \Gamma^{N-i-j+1}]}{\zeta}=\dfrac{\text{Tr}[\mathfrak{s}_z \Gamma^j \mathfrak{s}_z \Gamma^{N-j}]}{\zeta}.
    \label{eq15}
\end{gather}
Eq.~(\ref{eq15}) can be obtained, using the cyclic property of the trace. When $N\rightarrow\infty$, i.e., thermodynamic limit, $\zeta = \Omega_{+}^N$. We can calculate the eigenvectors $|\Omega_{\pm}\rangle$ of the transfer matrix, using the eigenvalue equations in Eq.~(\ref{eq6}), to get the correlation \cite{ref8} in the form:
\begin{gather}
    \langle \mathfrak{s}_i \mathfrak{s}_{i+j}\rangle = \frac{ \langle \Omega_{+} | \mathfrak{s}_z \Gamma^j \mathfrak{s}_z |\Omega_{+} \rangle \Omega_{+}^{N-j} + \langle \Omega_{-} | \mathfrak{s}_z \Gamma^j \mathfrak{s}_z |\Omega_{-} \rangle \Omega_{-}^{N-j} }{\Omega_{+}^N},\nonumber\\
    \text{or,}~ \langle \mathfrak{s}_i \mathfrak{s}_{i+j}\rangle = \dfrac{\langle \Omega_{+} | \mathfrak{s}_z \Gamma^j \mathfrak{s}_z |\Omega_{+} \rangle}{\Omega_{+}^j}.
    \label{eq16}
\end{gather}
In Eq.~(\ref{eq16}), the second term $(\langle \Omega_{-} | \delta_z \Gamma^j \delta_z |\Omega_{-} \rangle \Omega_{-}^{N-j})/\Omega_{+}^N$ vanishes since in the thermodynamic limit we have $(\Omega_{-}/ \Omega_{+})^N \rightarrow 0$. By calculating the eigenvectors of the transfer matrix~\cite{ref8}, we get correlation from Eq.~(\ref{eq16}) as,
\begin{equation}
    \mathfrak{c}_j = \langle \mathfrak{s}_i \mathfrak{s}_{i+j}\rangle = \cos^2 \varphi + \bigg(\dfrac{\Omega_{-}}{\Omega_{+}} \bigg)^j \sin^2 \varphi,
    \label{eq17}
\end{equation}
with $\cos^2 \varphi= \dfrac{\text{sinh}^2 (\beta \mathfrak{h})}{[\text{sinh}^2(\beta \mathfrak{h}) + e^{-4\beta \mathcal{J}}]} = \langle \mathfrak{s}_i\rangle^2$.\\
\textbf{4. Internal Energy:} From the partition function $\zeta$ defined in Eq.~(\ref{eq8}), we have the internal energy $\langle \mathbb{E} \rangle$ in terms of $\zeta$ as \cite{ref12},
\begin{gather}
    \langle\mathbb{E}\rangle = -\dfrac{\partial}{\partial \beta} \ln{\zeta} = -\dfrac{1}{\zeta} \dfrac{\partial \zeta}{\partial \beta}.
    \label{e1a}
\end{gather}
\textbf{5. Specific heat Capacity:} The specific heat capacity, at constant volume, $\Tilde{\mathbb{S}}_V$ is defined as the change in the internal energy $\langle \mathbb{E} \rangle$ with regards to a unit change in temperature $T$ \cite{ref12}, i.e.,
\begin{gather}
    \Tilde{\mathbb{S}}_V = \dfrac{1}{N} \dfrac{\text{d} \langle\mathbb{E}\rangle}{\text{d}T} = -\dfrac{\beta^2}{N} \dfrac{\text{d} \langle\mathbb{E}\rangle}{\text{d}\beta} \equiv \dfrac{\beta^2}{N} \dfrac{\partial}{\partial\beta} \bigg[\dfrac{1}{\zeta}\dfrac{\partial \zeta}{\partial \beta} \bigg].
    \label{e1}
\end{gather}
From Eq.~(\ref{e1}), we can relate $\Tilde{\mathbb{S}}_V$ to the variance of $\langle \mathbb{E} \rangle$ from the Hamiltonian given in Eq.~(\ref{eq1}) as,
\begin{gather}
    \zeta = \sum_{\{\mathfrak{s}_i\}} e^{-\beta H(\{\mathfrak{s}_i\})} = \sum_{\{\mathfrak{s}_i\}} e^{\beta [\mathcal{J} \sum_{i=1}^{N} \mathfrak{s}_i \mathfrak{s}_{i+1} + \mathfrak{h}\sum_{i=1}^{N}\mathfrak{s}_i]}, \nonumber\\
    \text{or,}~~\dfrac{\partial}{\partial \beta}\ln{\zeta} = \dfrac{1}{\zeta}\dfrac{\partial \zeta}{\partial \beta} = -\dfrac{1}{\zeta}\sum_{\{\mathfrak{s}_i\}} e^{-\beta H(\{\mathfrak{s}_i\})} ~H(\{\mathfrak{s}_i\}), \nonumber
\end{gather}
\begin{gather}
    \text{or,}~~ \dfrac{1}{\zeta}\dfrac{\partial \zeta}{\partial \beta} = -\dfrac{1}{\zeta}\sum_{\{\mathfrak{s}_i\}} e^{-\beta \mathbb{E}(\{\mathfrak{s}_i\})} ~\mathbb{E}(\{\mathfrak{s}_i\}).\label{e2}\\
    \text{Thus,} ~\Tilde{\mathbb{S}}_V = -\dfrac{\beta^2}{N} \dfrac{\partial }{\partial\beta}\bigg[\dfrac{1}{\zeta}\sum_{\{\mathfrak{s}_i\}} e^{-\beta \mathbb{E}(\{\mathfrak{s}_i\})} ~\mathbb{E}(\{\mathfrak{s}_i\})\bigg],\nonumber\\
    = -\dfrac{\beta^2}{N}\bigg\{\bigg[-\dfrac{1}{\zeta^2} \dfrac{\partial \zeta}{\partial \beta} \bigg]\sum_{\{\mathfrak{s}_i\}} e^{-\beta \mathbb{E}}~\mathbb{E} - \dfrac{1}{\zeta} \sum_{\{\mathfrak{s}_i\}} \mathbb{E}^2~e^{-\beta \mathbb{E}}\bigg\}, \nonumber\\
    = -\dfrac{\beta^2}{N}\bigg\{\bigg[\dfrac{1}{\zeta^2} \sum_{\{\mathfrak{s}_i\}} e^{-\beta \mathbb{E}}~ \mathbb{E} \bigg]\sum_{\{\mathfrak{s}_i\}} e^{-\beta \mathbb{E}}~\mathbb{E} - \dfrac{1}{\zeta} \sum_{\{\mathfrak{s}_i\}} \mathbb{E}^2~e^{-\beta \mathbb{E}}\bigg\},\nonumber\\
    \text{or,}~\Tilde{\mathbb{S}}_V = \dfrac{\beta^2}{N} \dfrac{\partial^2}{\partial\beta^2} \ln{\zeta} = \dfrac{\beta^2}{N}[\langle \mathbb{E}^2\rangle - \langle \mathbb{E}\rangle^2].
    \label{e3}
\end{gather}
From Eqs.~(\ref{e1},\ref{e3}), one can \textit{normalize} specific heat capacity $\Tilde{\mathbb{S}}_V$ as,
\begin{equation}
    \mathbb{S}_V = \dfrac{\Tilde{\mathbb{S}}_V}{\beta^2} = \dfrac{1}{N} \dfrac{\partial}{\partial\beta} \bigg[\dfrac{1}{\zeta}\dfrac{\partial \zeta}{\partial \beta} \bigg] = \dfrac{1}{N}[\langle \mathbb{E}^2\rangle - \langle \mathbb{E}\rangle^2]
    \label{e3a}
\end{equation}
In Eq.~(\ref{e3a}), we have shown that the specific heat capacity $\mathbb{S}_V$ for the Ising model is related to the variance of the internal energy $\langle \mathbb{E}\rangle$ of the $1D$-Ising chain.

In social dilemmas, we similarly define the \textit{game payoff capacity} $\mathbb{S}_g$, analogous to the specific heat capacity $\mathbb{S}_V$ of the Ising model, which gives us the net change in average payoff, for each player, with respect to a unit change in the selection pressure (or, \textit{noise}). A measure of the unpredictability in the player's strategy selection is defined as \textit{noise} and is expressed as $\beta~(= 1/k_B T)$, where $k_B$ is the \textit{Boltzmann constant}. For the Ising model, $T$ denotes the temperature of the $1D$-spin chain. Analogously, in social dilemmas, $T\rightarrow 0$ (or, $\beta \rightarrow \infty$) corresponds to \textit{zero noise} (ZN), i.e., no change in strategies adopted by the players, whereas, $T\rightarrow \infty$ (or, $\beta \rightarrow 0$) indicates \textit{infinite noise} (IN), i.e., complete unpredictability in strategy selection by the players.

When it comes to the Ising model, the \textit{equilibrium} is defined as the lowest energy configuration. However, in social dilemmas, especially \textit{two-player} games, players look for the \textit{Nash equilibrium}, which is defined as the maximum feasible payoff for both players, deviating from which yields a worse outcome for the other. Hence, in order to establish a correspondence between the \textit{Nash equilibrium} of social dilemmas and the \textit{equilibrium configuration} of the Ising model, we relate the negative of the payoffs to the energies. Thus, the average internal energy of the Ising model $\langle \mathbb{E}\rangle$ is related to the average game payoff $\langle \Lambda \rangle$ as $\langle \Lambda \rangle= -\langle \mathbb{E}\rangle$. 

\begin{figure*}[!ht]
    \centering
    \includegraphics[width = 0.88\textwidth]{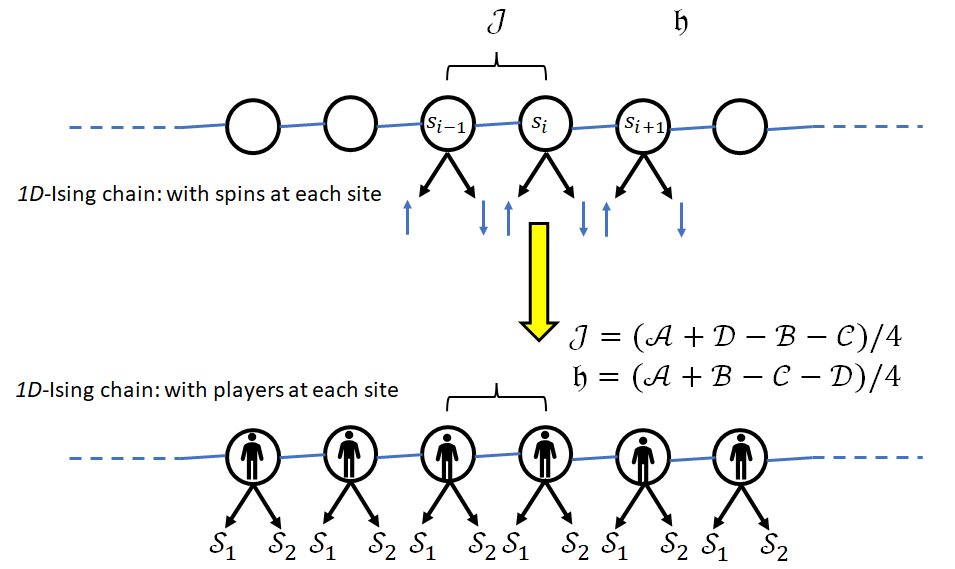}
    \caption{\centering{\textbf{NEM method}: Mapping a $1D$-Ising chain consisting of two-spins $(\uparrow, \downarrow)$ to a social dilemma game with available strategies $(\mathcal{S}_1, \mathcal{S}_2)$. The Ising parameters $(\mathcal{J}, \mathfrak{h})$ are expressed in terms of game payoffs $(\mathcal{A, B, C, D})$.}}
    \label{fig:0}
\end{figure*}

Both game susceptibility and payoff capacity have a multiplicative factor of $\beta$ and $\beta^2$, respectively (see, Eqs.~(\ref{eq13}, \ref{e3})), and since we will be comparing the three different analytical methods in the ZN and the IN limits, these factors can affect the results when we study their variation in the $\beta\rightarrow 0$ and $\beta\rightarrow \infty$ limits. Hence, we normalize both \textit{game susceptibility} and \textit{payoff capacity}, and along with \textit{correlation}, we work with these as the indicators of cooperative behaviour. Earlier works had game magnetization ($\mu_g$) and individual player's average payoff ($\langle \Lambda_1 \rangle$) as the indicators for cooperative behaviour (see, Ref.~\cite{ref4}).

\subsection{\label{sub-nem}Nash equilibrium mapping (NEM)}
NEM has been discussed elaborately in Refs.~\cite{ref1, ref2, ref3}, but still, we will briefly introduce the formalism in this section for the ease of the readers. In this method, we map a social dilemma game to a $1D$-Ising chain model with an infinite number of spin sites (see, Fig.~\ref{fig:0}). To begin with, we consider a two-strategy; two-player social dilemma game and map it to an Ising chain with two spin sites. We begin by writing the Hamiltonian for a $two$-site $1D$-Ising chain,
\begin{equation}
    H = -\mathcal{J}(\mathfrak{s}_1\mathfrak{s}_2 + \mathfrak{s}_2\mathfrak{s}_1) - \mathfrak{h}(\mathfrak{s}_1 + \mathfrak{s}_2) = \Delta_1 + \Delta_2.
    \label{eq18}
\end{equation}
Here, we have the energy for each spin site as,
\begin{equation}
    \Delta_1 = -\mathcal{J}\mathfrak{s}_1\mathfrak{s}_2 - \mathfrak{h}\mathfrak{s}_1,~~\text{and}~~\Delta_2 = -\mathcal{J}\mathfrak{s}_2\mathfrak{s}_1 - \mathfrak{h}\mathfrak{s}_2.
    \label{eq19}
\end{equation}
The total energy of the \textit{two-site} Ising chain is $\Delta= \Delta_1 +\Delta_2$. For any social dilemma, the Nash equilibrium involves finding the maximum feasible payoffs for the players. However, in a $1D$-Ising chain, we minimize the total $two$-site energy to obtain the equilibrium. This means that in order to find a correspondence between the Nash equilibrium of games and the equilibrium configuration of Ising model, we write the negative of the energy matrix, where each element ($-\Delta_i$) is associated with spin values of $\pm 1$ at each site of the Ising chain, since negative energy minimization implies payoff maximization. Thus,
\begin{equation}
\begin{footnotesize}
    -\Delta = 
    \left[\begin{array}{c|c c} 
    	 & \mathfrak{s}_2 = +1 & \mathfrak{s}_2 = -1\\ 
    	\hline 
    	\mathfrak{s}_1 = +1 & (\mathcal{J}+\mathfrak{h}), (\mathcal{J}+\mathfrak{h}) & -(\mathcal{J}-\mathfrak{h}), -(\mathcal{J}+\mathfrak{h})\\
        \mathfrak{s}_1 = -1 & -(\mathcal{J}+\mathfrak{h}), -(\mathcal{J}-\mathfrak{h}) & (\mathcal{J}-\mathfrak{h}), (\mathcal{J}-\mathfrak{h})
    \end{array}\right].
    \label{eq20}
\end{footnotesize} 
\end{equation}
If we consider a $two$-player \textit{symmetric} game (with $two$ available strategies $\mathcal{S}_1$ and $\mathcal{S}_2$), then we can write the payoff matrix ($\Lambda$) as,
\begin{equation}
    \Lambda = \left[\begin{array}{c|c c} 
    	 & \mathcal{S}_1 & \mathcal{S}_2\\ 
    	\hline 
    	\mathcal{S}_1 & \mathcal{A, A} & \mathcal{B, C}\\
        \mathcal{S}_2 & \mathcal{C, B} & \mathcal{D, D}
    \end{array}\right]. 
    \label{eq21}
\end{equation}
where, $(\mathcal{A,B,C,D})$ are game payoffs. In the \textit{two-strategy}, \textit{two-player} game, the game payoffs $(\mathcal{A,B,C,D})$ either fulfil the criterion: $\mathcal{A+D=B+C}$ (as in PGG) or don't (as in HDG). {When social dilemmas (like PGG) satisfy this payoff criterion, it reflects the idea that the total resource is available to all the players and the distribution of payoffs is a result of individual choices within that constraint.} For \textit{PGG}, the sum of payoffs for cooperators and defectors should be equal to the sum of payoffs for those who benefit from the public good, regardless of their contribution, whereas, for \textit{HDG}, there is no such requisite for the players. Both HDG and PGG are discussed in detail in Secs.~[\ref{hdg},~\ref{pgg}], respectively.

For the Hamiltonian $H$ given in Eq.~(\ref{eq18}), considering only energy of a single spin $i$ and by using a set of one-to-one linear transformations on $\Lambda$, that keeps the Nash equilibrium invariant (see, Refs.~\cite{ref1, ref2, ref4, ref9}),
\begin{equation}
    \mathcal{A}\rightarrow \frac{\mathcal{A}-\mathcal{C}}{2},~\mathcal{B} \rightarrow \frac{\mathcal{B}-\mathcal{D}}{2},~\mathcal{C}\rightarrow \frac{\mathcal{C}-\mathcal{A}}{2},~\mathcal{D}\rightarrow \frac{\mathcal{D}-\mathcal{B}}{2},
\end{equation}
so as to equate $\Lambda$ in Eq.~(\ref{eq21}) with $-\Delta$ in Eq.~(\ref{eq20}), we get the $(\mathcal{J},\mathfrak{h})$ parameters in terms of payoffs as,
\begin{equation}
    \mathfrak{h} = \frac{\mathcal{(A-C)+(B-D)}}{4},~ \text{and}~\mathcal{J} = \frac{\mathcal{(A-C)-(B-D)}}{4}.
    \label{eqpar}
\end{equation}
In NEM, we map the Ising parameters $(\mathcal{J}, \mathfrak{h})$ to the game parameters $(\mathcal{A,B,C,D})$ (see, Ref.~\cite{ref1}). The partition function $\zeta^{NEM}$ in terms of the game parameters $(\mathcal{A,B,C,D})$ is then,
\begin{gather}
    \zeta^{2-site~Ising} = e^{2\beta (\mathcal{J}+\mathfrak{h})} + e^{2\beta (\mathcal{J}-\mathfrak{h})} + 2e^{-2\beta\mathcal{J}},\nonumber\\
    \text{or,}~\zeta^{NEM} = e^{\beta\mathcal{(A-C)}} + e^{-\beta \mathcal{(B+D)}} + 2e^{\frac{\beta}{2} \mathcal{(B+C-A-D)}}.
    \label{eqpartnem}
\end{gather}
\textbf{Game magnetization:} As discussed in detail and derived in Refs.~\cite{ref1, ref2, ref4, ref11}, by using the payoff matrix $\Lambda$ defined in Eq.~(\ref{eq21}) and partition function $\zeta^{NEM}$, we have the game magnetization ($\mu_g^{NEM}$), in terms of the game parameters, 
\begin{equation}
    \mu_g^{NEM} = \dfrac{1}{\beta}\dfrac{\partial}{\partial \mathfrak{h}}\ln{\zeta^{NEM}} =\dfrac{\sinh{\beta \frac{\mathcal{(A-C)+(B-D)}}{4}}}{\mathfrak{Z}},
    \label{eq22}    %eq23
\end{equation}
where, $\mathfrak{Z} = \sqrt{\sinh^2{\beta \frac{\mathcal{(A-C)+(B-D)}}{4}} + e^{-4\beta \frac{\mathcal{(A-C)-(B-D)}}{4}}}$, respectively. In certain games, like PGG, the payoffs $(\mathcal{A, B, C, D})$ satisfy: $\mathcal{A+D} = \mathcal{B+C}$, and we have the game magnetization reducing to,
\begin{equation}
    \mu_g^{NEM} = \tanh{\beta \bigg(\dfrac{\mathcal{A-C}}{2}\bigg)}.
    \label{eq23}    %eq24
\end{equation}
In social dilemmas, the parameter $T$ (which denotes \textit{temperature} in the Ising model) is defined as a measure of the amount of \textit{randomness} (or, \textit{noise}) in the strategy selection by the players. For $T\rightarrow 0$ (or, $\beta \rightarrow \infty$), the system is noiseless and is at Nash equilibrium. When the system is noiseless, players' selection of strategies are not random. Whereas, for $T\rightarrow \infty$ (or, $\beta \rightarrow 0$), we have complete randomness in the selection of strategies (or, \textit{maximum noise}).\\
\textbf{Game susceptibility:} We have the expression for the susceptibility in Eq.~(\ref{eq13a}) and to derive the game susceptibility $\chi_\sigma^{NEM}$, we differentiate $\mu_g$ with the payoffs $\mathcal{A, B, C, D}$ and normalize by $\beta$, i.e.,
\begin{equation}
    \chi_\sigma^{NEM} = \dfrac{1}{\beta}\dfrac{\partial}{\partial \sigma}\mu_g^{NEM}~\text{, for}~\sigma\in \{\mathcal{A, B, C, D}\}.
    \label{eq24}    %eq25
\end{equation}
In Eq.~(\ref{eq24}), we are not explicitly writing the expression for the game susceptibility since in the $1D$-Ising chain, the magnetic susceptibility is defined as $\chi_\mathfrak{h} = \frac{1}{\beta} \frac{\partial \mu}{\partial \mathfrak{h}}$ (see, Eq.~(\ref{eq13a})), i.e., derivative with respect to the external magnetic field $\mathfrak{h}$. Similarly, for social dilemmas, the payoffs that appear only in the field factor $\mathfrak{h}$ are \textit{field payoffs} for which the game susceptibility exists. For payoffs that appear only in the coupling factor $\mathcal{J}$, i.e., the \textit{coupling payoffs}, or both in coupling and field factors, i.e., the \textit{mixed payoffs}, for them, the game susceptibility is not defined. For different games, we first define the corresponding game payoffs, and then we determine the game susceptibility with respect to those payoffs that do not appear in $\mathcal{J}$ but appear only in $\mathfrak{h}$, i.e., the \textit{field payoffs}. For example, in HDG~\cite{ref9}, the \textit{payoff matrix} $\Lambda^{H-D}$ is given as,
\begin{table}[H]
\centering
$\Lambda^{H-D}$ = 
\setlength{\extrarowheight}{2pt}
\begin{tabular}{*{4}{c|}}
    \multicolumn{2}{c}{} & \multicolumn{2}{c}{}\\\cline{3-4}
    \multicolumn{1}{c}{} &  & $\mathcal{S}_1 \equiv \mathbb{H}$  & $\mathcal{S}_2 \equiv \mathbb{D}$ \\\cline{2-4}
    \multirow{2}*{}  & $\mathcal{S}_1 \equiv \mathbb{H}$ & $(\mathfrak{-d, -d})$ & $(\mathfrak{r, -r})$ \\\cline{2-4}
    & $\mathcal{S}_2 \equiv \mathbb{D}$ & $(\mathfrak{-r, r})$ & $(0,0)$ \\\cline{2-4}
\end{tabular}
\end{table}
where, $\mathbb{H}$ and $\mathbb{D}$ denotes the \textit{Hawk} and \textit{Dove} strategy, respectively. A detailed overview of HDG is given in Sec.~\ref{hdg}, where $\mathfrak{d}$ and $\mathfrak{r}$ denotes the injury cost and the reward associated with the game. From $\Lambda^{H-D}$ and Eqs.~(\ref{eq21}, \ref{eqpar}), we have $\mathcal{J} = -\frac{\mathfrak{d}}{4}$ and $\mathfrak{h} = \frac{2\mathfrak{r}-\mathfrak{d}}{4}$. Here, we observe that $\mathcal{J}$ is dependent on $\mathfrak{d}$ and independent of $\mathfrak{r}$, while $\mathfrak{h}$ is dependent on both $\mathfrak{r}$ and $\mathfrak{d}$. Hence, we calculate the game susceptibility only with respect to the \textit{field payoff} $\mathfrak{r}$, as $\mathfrak{d}$ is the \textit{mixed payoff}, i.e., it appears in both $\mathcal{J}$ and $\mathfrak{h}$.
\\
\textbf{Game correlation:} To find the expression for the correlation $\langle \mathfrak{s}_i \mathfrak{s}_{i+j}\rangle$, which is analogous to the spin-spin correlation, we replace the expressions for $\mathcal{J}$ and $\mathfrak{h}$ in Eq.~(\ref{eq17}) with the game parameters,
\begin{equation}
    \mathfrak{c}_j^{NEM} = \langle \mathfrak{s}_i \mathfrak{s}_{i+j}\rangle = \cos^2 \varphi + \bigg(\dfrac{\Omega_{-}}{\Omega_{+}} \bigg)^j \sin^2 \varphi,
    \label{eq25}    %eq26
\end{equation}
where, we have $j$ as the \textit{distance} from the $i^{th}$ site and
\begin{gather}
    \cos^2 \varphi= \dfrac{\sinh^2{\beta \mathfrak{h}}}{\mathfrak{X}},~\Omega_{\pm} = e^{\beta \mathcal{J}} [\text{cosh}(\beta \mathfrak{h}) \pm \sqrt{\mathfrak{X}}],\label{eq26} \\  
    \text{with,}~\mathfrak{X} = \sinh^2{\beta \mathfrak{h}} + e^{-4\beta \mathcal{J}}.\label{eq27}
\end{gather}
\textbf{Individual player's average payoff:} In order to derive the expression for the individual player's average payoff $\langle \Lambda_1 \rangle^{NEM}$, we use the partition function $\zeta^{NEM}$. For the Hamiltonian $H$ given in Eq.~(\ref{eq18}), we have $\zeta^{NEM}$ from Eq.~(\ref{eqpartnem}) as,
\begin{gather}
    \zeta^{NEM} = e^{\beta\mathcal{(A-C)}} + e^{-\beta \mathcal{(B+D)}} + 2e^{\frac{\beta}{2} \mathcal{(B+C-A-D)}},\nonumber\\
    \text{Thus,}~~\langle \Lambda_1 \rangle^{NEM} = -\dfrac{1}{2}\langle \mathbb{E}\rangle^{NEM} = \dfrac{1}{2}\bigg[\dfrac{1}{\zeta^{NEM}}\dfrac{\partial \zeta^{NEM}}{\partial \beta} \bigg], \nonumber\\
    \text{or,}~\langle \Lambda_1 \rangle^{NEM}= \dfrac{1}{2\zeta^{NEM}}[\mathcal{(B+C-A-D)}e^{\frac{\beta}{2} \mathcal{(B+C-A-D)}}\nonumber\\
    + \mathcal{(A-C)}e^{\beta\mathcal{(A-C)}} - \mathcal{(B+D)}e^{-\beta \mathcal{(B+D)}}].
    \label{e4a}
\end{gather}
\textbf{Game payoff capacity:} To determine the expression for the game payoff capacity $\mathbb{S}_{g}^{NEM}$, we again start from the partition function $\zeta^{NEM}$. For the Hamiltonian $H$ given in Eq.~(\ref{eq18}), we have $\zeta^{NEM}$ from Eq.~(\ref{eqpartnem}) as,
\begin{gather}
    \zeta^{NEM} = e^{\beta\mathcal{(A-C)}} + e^{-\beta \mathcal{(B+D)}} + 2e^{\frac{\beta}{2} \mathcal{(B+C-A-D)}}.\nonumber
\end{gather}
From Eq.~(\ref{e3a}), we have,
\begin{gather}
    \mathbb{S}_{g}^{NEM} = \dfrac{1}{2} \dfrac{\partial}{\partial\beta} \bigg[\dfrac{1}{\zeta^{NEM}}\dfrac{\partial \zeta^{NEM}}{\partial \beta} \bigg].
    \label{e4}
\end{gather}

{This NEM method to study games in the thermodynamic limit, is applicable to any two-player two-strategy game, both cooperative like Public Goods game or non-cooperative like Hawk-Dove game or Prisoner\rq{}s dilemma game, since we are making an exact mapping to the spin-1/2 Ising model via Nash equilibrium. Thus be it the battle of sexes(BoS) or, game of chicken or, Volunteer\rq{}s dilemma, or the vaccination game, all these games can be understood in the thermodynamic limit via the Nash equilibrium mapping. In fact in a recent work, our group has worked on the Nash equilibrium mapping in the context of the vaccination game, in the thermodynamic limit, see Ref.~\cite{ref16}.}

\subsection{\label{sub-hdm}Aggregate selection (AS)}
The framework of AS is designed using an analogy with the $1D$-Ising chain, similar to that of NEM. This method was introduced in Ref.~\cite{ref5}, where it was addressed as HD (see, Sec.~\ref{introsec}), and the authors showed that the equilibrium fraction of cooperators is given by the expectation value of a thermal observable similar to magnetization (see, Refs.~\cite{ref4, ref5}). We will be using the same formalism to derive the expressions for the game susceptibility, the correlation, and the payoff capacity. In AS, we relate the players' strategies to the spin states, and we represent the former as \textit{ket vectors}~\cite{ref5}. We denote the spin upstate with $|0\rangle = [1~~0]^{\text{T}}$, where `T' is the \textit{transpose} of the row vector, and we relate it to the cooperate strategy, whereas, the orthogonal spin downstate is denoted by $|1\rangle = [0~~1]^{\text{T}}$ and we relate it to the defect strategy. We can express the state $|\nu \rangle$ corresponding to a $N$-site system as,
\begin{equation}
    |\nu \rangle = |\gamma_1\rangle \otimes |\gamma_2\rangle \otimes |\gamma_3 \rangle ...\otimes |\gamma_N\rangle,
    \label{eq28}
\end{equation}
where, $|\gamma_i\rangle$ denotes the spin/strategy vector associated with the player at the $i^{th}$ site, and $\gamma_i \in \{0,1\} ~\forall~ i\in \{1, 2,..., N\}$. We consider a periodic boundary condition, i.e., $\gamma_{N+1} = \gamma_1$.

For a symmetric two-player game, we can write the payoff matrix $\Lambda_1$ for one of the players, let's say the row player, as,
\begin{equation}
    \Lambda_1 = \left[\begin{array}{c|c c} 
    	 & \mathcal{S}_1 & \mathcal{S}_2\\ 
    	\hline 
    	\mathcal{S}_1 & \mathcal{A} & \mathcal{B}\\
        \mathcal{S}_2 & \mathcal{C} & \mathcal{D}
    \end{array}\right].
    \label{eq29}    %eq29
\end{equation}
Similarly, we write the payoff matrix for the column player. Considering Eq.~(\ref{eq29}), we define the Hamiltonian using the energy matrix $\Delta$, which is the negative of $\Lambda_1$ (similar to NEM), thus,
\begin{equation}
    \Delta = \begin{bmatrix}
        \Delta_{00} & \Delta_{01}\\
        \Delta_{10} & \Delta_{11}
    \end{bmatrix}
    =
    \begin{bmatrix}
        -\mathcal{A} & -\mathcal{B}\\
        -\mathcal{C} & -\mathcal{D}
    \end{bmatrix}.
    \label{eq30}    %eq25
\end{equation}
According to Ref.~\cite{ref5}, we can write the Hamiltonian by considering nearest neighbour interactions among the players as,
\begin{equation}
    H = \sum_{i=1}^{N} \sum_{\{\gamma_{i},\gamma_{i+1}\}} \Delta_{\gamma_i \gamma_{i+1}} \mathbb{P}^{(i)}_{\gamma_i} \otimes \mathbb{P}^{(i+1)}_{\gamma_{i+1}},
    \label{eq31}       
\end{equation}
where, $\{\gamma_{i},\gamma_{i+1}\}\in \{0,1\}$, $\mathbb{P}_0 = |0 \rangle \langle 0|$ and $\mathbb{P}_1 = |1 \rangle \langle 1|$ are the \textit{projection} operators, $\forall~\gamma_i \in \{0,1\},~i\in \{1, 2,..., N\}$. As discussed in Refs.~\cite{ref4, ref5}, $\mathbb{P}_{\gamma_i}^{(i)}$ (or, $\mathbb{P}_{\gamma_{i+1}}^{(i+1)}$) denotes projection operator for cooperation ($\uparrow$-spin) when $\gamma_i=0$ (or, $\gamma_{i+1} =0$), while projection operator for defection ($\downarrow$-spin) when $\gamma_i=1$ (or, $\gamma_{i+1} =1$), at the $i^{th}$ (or, $(i+1)^{th}$) spin site, with $(\gamma_i,\gamma_{i+1}) \in \{0,1\}$, which denote the indices of the energy matrix $\Delta$. In Ref.~\cite{ref5}, the authors have used AS to calculate the game magnetization, which is used as an indicator of cooperation among the players in the thermodynamic limit. The problems associated with this method are discussed elaborately in Refs.~\cite{ref1, ref3, ref4}. 

In Ref.~\cite{ref4}, the authors have shown that AS fails in comparison to the other analytical methods when calculating the game magnetization as well as the individual player's average payoff. For our case, we will use AS to determine the three indicators in question, i.e., the game susceptibility, the correlation, and the payoff capacity, and we will compare it with the other methods. As derived in Ref.~\cite{ref4}, we have the AS partition function $\zeta^{AS}$ as,
\begin{equation}
\small{
    \zeta^{AS} = \text{Tr}(e^{-\beta H})=\sum_{|\nu \rangle} \langle \nu| e^{-\beta H}|\nu \rangle = (e^{\beta\mathcal{A}} + e^{\beta\mathcal{D}})^N.}
    \label{hdm-partition}
\end{equation}

\textbf{Game magnetization:} As discussed in detail and derived in Refs.~\cite{ref4, ref5}, one can calculate the average magnetization per player as,
\begin{equation}
    \mu_g^{AS} = \langle \hat{\mathcal{M}_z}\rangle_\beta = \dfrac{1}{N\zeta^{AS}} \sum_{|\nu\rangle}\langle \nu| \hat{\mathcal{M}_z} e^{-\beta H} |\nu\rangle ,
    \label{eq32}
\end{equation} 
where, the summation over $|\nu \rangle$ indicates the determination of the total magnetization of the $N$-site Ising chain, $\zeta^{AS} = (e^{\beta\mathcal{A}} + e^{\beta\mathcal{D}})^N$ is the partition function for the Hamiltonian $H$ and $\hat{\mathcal{M}_z}$ is defined as the \textit{order parameter}. $\hat{\mathcal{M}_z}$ is expressed as $\hat{\mathcal{M}_z} = \sum_{i}\hat{\mathcal{M}_z}^{(i)} =\sum_{i} (\mathbb{P}_0^{(i)} - \mathbb{P}_1^{(i)})$, $\forall~i\in \{1, 2,..., N\}$, and its expectation value is equivalent to the game magnetization $\mu_g^{NEM}$ obtained in NEM. 

From Refs.~\cite{ref4, ref5}, AS is only valid for games whose payoffs satisfy the criterion: $\mathcal{A+D=B+C}$. If the criterion of $\mathcal{A+D=B+C}$ is dissatisfied, the sole method for determining the parameters will be via a numerical approach, which will be in clear violation of the fact that AS is an analytical method \cite{ref4}. As shown in Ref.~\cite{ref3}, for $N$-players, we have the RHS of Eq.~(\ref{eq32}) as,
\begin{gather}
    \dfrac{\sum_{|\nu\rangle}\langle \nu| \hat{\mathcal{M}_z} e^{-\beta H} |\nu\rangle}{N\zeta^{AS}} = \dfrac{\cancel{N}(e^{\beta \mathcal{A}} - e^{\beta \mathcal{D}})}{\cancel{N}\zeta^{AS}(e^{\beta \mathcal{A}} + e^{\beta \mathcal{D}})^{1-N}},
    \label{eq33}\\
    \text{or,}~\mu_g^{AS} = \dfrac{e^{\beta \mathcal{A}} - e^{\beta \mathcal{D}}}{e^{\beta \mathcal{A}} + e^{\beta \mathcal{D}}} = \tanh\bigg[{\beta\bigg(\dfrac{\mathcal{A}-\mathcal{D}}{2}\bigg)}\bigg].
    \label{eq34}
\end{gather}
\textbf{Game susceptibility:} We can calculate the game susceptibility, from Eqs.~(\ref{eq13a}, \ref{eq34}), as,
\begin{equation}
    \chi_\sigma^{AS} = \dfrac{1}{\beta} \dfrac{\partial}{\partial \sigma} \mu_g^{AS},~\text{where $\sigma \in \{\mathcal{A}, \mathcal{D}\}$.} 
    \label{ne46}
\end{equation}
In Eq.~(\ref{ne46}), we are not explicitly writing the expression for the game susceptibility since game susceptibility is only defined for \textit{field payoffs}, and they are specific for each game. Hence, for different games, we first define the corresponding game payoffs, and then we determine the game susceptibility with respect to the \textit{field payoffs}.\\
\textbf{Game correlation:} To calculate the correlation, we consider the fact that $\hat{\mathcal{M}_z}^{(i)}$ is nothing but the $\mathfrak{s}_z$ (Pauli $z$-matrix) operator acting on the $i^{th}$ site and it gives the spin/strategy at that site. We thus have,
\begin{gather}
    \mathfrak{c}_j^{AS} = \langle \mathfrak{s}_i \mathfrak{s}_{i+j}\rangle = \langle \hat{\mathcal{M}_z}^{(i)} \hat{\mathcal{M}_z}^{(i+j)} \rangle_{\beta}, \nonumber\\
    \text{or,}~~ \mathfrak{c}_j^{AS} = \dfrac{\sum_{|\nu \rangle}\langle \nu| \hat{\mathcal{M}_z}^{(i)}\hat{\mathcal{M}_z}^{(i+j)} e^{-\beta H} |\nu \rangle}{N\zeta^{AS}},\nonumber\\
    = \dfrac{1}{N\zeta^{AS}} \sum_{|\nu\rangle}\langle \nu| \{\mathbb{P}_0^{(i)} - \mathbb{P}_1^{(i)}\}  \{\mathbb{P}_0^{(i+j)} - \mathbb{P}_1^{(i+j)}\} e^{-\beta H} |\nu\rangle, \nonumber\\
    = \sum_{|\nu \rangle}\bigg[\dfrac{\langle \nu| \mathbb{P}_0^{(i)} \mathbb{P}_0^{(i+j)}e^{-\beta H} |\nu\rangle + \langle \nu| \mathbb{P}_1^{(i)} \mathbb{P}_1^{(i+j)}e^{-\beta H} |\nu\rangle}{N\zeta^{AS}} \nonumber\\
     - \dfrac{\langle \nu| \mathbb{P}_0^{(i)} \mathbb{P}_1^{(i+j)}e^{-\beta H} |\nu\rangle + \langle \nu| \mathbb{P}_1^{(i)} \mathbb{P}_0^{(i+j)}e^{-\beta H} |\nu\rangle}{N\zeta^{AS}}\bigg] ,\nonumber\\
    = \dfrac{1}{\cancel{N}\zeta^{AS}} \cancel{N}(e^{\beta \mathcal{A}} + e^{\beta \mathcal{D}})^{N-2}[e^{2\beta \mathcal{A}} + e^{2\beta \mathcal{D}} - 2 e^{\beta (\mathcal{A+D})}], \nonumber\\
    = \bigg[\dfrac{e^{\beta \mathcal{A}} - e^{\beta \mathcal{D}}}{e^{\beta \mathcal{A}} + e^{\beta \mathcal{D}}} \bigg]^2 = \tanh^2{\bigg[\beta\bigg(\dfrac{\mathcal{A}-\mathcal{D}}{2}\bigg)\bigg]} = \langle \hat{\mathcal{M}_z}^{(i)} \rangle_{\beta}^2 .\nonumber\\
    \text{Thus,}~\mathfrak{c}_j^{AS} = \langle \hat{\mathcal{M}_z}^{(i)} \hat{\mathcal{M}_z}^{(i+j)} \rangle_{\beta} = \tanh^2{\bigg[\beta\bigg(\dfrac{\mathcal{A}-\mathcal{D}}{2}\bigg)\bigg]} .
    \label{eq35}
\end{gather}
We observe from Eq.~(\ref{eq35}) that the correlation is independent of the distance $j$ and is equal to the square of the average magnetization per player. This is also evident from the fact that in Refs.~\cite{ref4, ref5}, the authors define a $K$-matrix with elements $K_{\gamma \gamma'} = e^{-\beta \Delta_{\gamma\gamma'}}$, and for games that satisfy the criterion: $\mathcal{A+D = B+C}$, the two eigenvalues of the $K$-matrix are $\Omega_- = 0$ and $\Omega_+ = e^{\beta \mathcal{A}} + e^{\beta \mathcal{D}}$, with $\Omega_- < \Omega_+$ from Eq.~(\ref{eq6}). We see that $\frac{\Omega_{-}}{\Omega_{+}} = 0$ and substituting in Eq.~(\ref{eq17}), we have the correlation as the square of the average magnetization per player, i.e., $\langle \hat{\mathcal{M}_z}^{(i)}\rangle_{\beta}$ for AS, which is independent of the distance $j$. We will use this expression derived in Eq.~(\ref{eq35}) in the later sections when we apply it to different social dilemma games.\\
\textbf{Individual player's average payoff:} In order to derive the expression for the individual player's average payoff $\langle \Lambda_1 \rangle^{AS}$, we need the partition function $\zeta^{AS} = (e^{\beta\mathcal{A}} + e^{\beta\mathcal{D}})^N$ and from Eq.~(\ref{e1a}), we have, 
\begin{gather}
    \langle \Lambda_1 \rangle^{AS} = \bigg[\dfrac{1}{N\zeta^{AS}}\dfrac{\partial \zeta^{AS}}{\partial \beta} \bigg] = \dfrac{\mathcal{A}e^{\beta\mathcal{A}} + \mathcal{D}e^{\beta\mathcal{D}}}{e^{\beta\mathcal{A}} + e^{\beta\mathcal{D}}},
    \label{e5a}
\end{gather}
where, $(\mathcal{A, D})$ are the payoffs associated with a social dilemma that satisfies the criterion: $\mathcal{A+D=B+C}$.\\
\textbf{Game payoff capacity:} To determine the expression for the game payoff capacity $\mathbb{S}_{g}^{AS}$, we again start from the partition function $\zeta^{AS} = (e^{\beta\mathcal{A}} + e^{\beta\mathcal{D}})^N$ and from Eq.~(\ref{e3a}), we have, 
\begin{gather}
    \mathbb{S}_{g}^{AS} = \dfrac{1}{N} \dfrac{\partial}{\partial\beta} \bigg[\dfrac{1}{\zeta^{AS}}\dfrac{\partial \zeta^{AS}}{\partial \beta} \bigg] = \dfrac{(\mathcal{A-D})^2 e^{\beta(\mathcal{A+D})}}{(e^{\beta\mathcal{A}}+\mathrm{e}^{\beta\mathcal{D}})^2},
    \label{e5}
\end{gather}
where, $(\mathcal{A, D})$ are the payoffs associated with a social dilemma that satisfies the criterion: $\mathcal{A+D=B+C}$.

\subsection{\label{sub-dem}Darwinian selection (DS)}
In DS, just one player of interest, i.e., the \textit{principal} player, pursues the maximum gain without considering the payoff of the other nearest interacting players. In Ref.~\cite{ref5}, this method, where it was addressed as DE (see, Sec.~\ref{introsec}), gave a better result than AS when considering game magnetization as an indicator. In DS, the aim is to minimize the principal player's energy (i.e., maximizing the principal player's gain), whereas, in AS, we deal with maximizing the total payoff in order to maximize the cumulative gain for all players. For determining the analytical expression for the game susceptibility, we begin with the AS Hamiltonian given in Eq.~(\ref{eq31}),
\begin{equation}
    H = \sum_{i=1}^{N} \sum_{\{\gamma_{i},\gamma_{i+1}\}} \Delta_{\gamma_i \gamma_{i+1}} \mathbb{P}^{(i)}_{\gamma_i} \otimes \mathbb{P}^{(i+1)}_{\gamma_{i+1}},
\end{equation}
where, $\{\gamma_{i},\gamma_{i+1}\}\in \{0,1\}$, $\mathbb{P}_0 = |0 \rangle \langle 0|$ and $\mathbb{P}_1 = |1 \rangle \langle 1|$ are the \textit{projection} operators, $\forall~\gamma_i \in \{0,1\},~\text{and}~ i\in \{1, 2,..., N\}$. $\mathbb{P}_{\gamma_i}^{(i)}$ (or, $\mathbb{P}_{\gamma_{i+1}}^{(i+1)}$) denotes projection operator for cooperation ($\uparrow$-spin) when $\gamma_i=0$ (or, $\gamma_{i+1} =0$), while projection operator for defection ($\downarrow$-spin) when $\gamma_i=1$ (or, $\gamma_{i+1} =1$), at the $i^{th}$ (or, $(i+1)^{th}$) spin site, with $\{\gamma_i,\gamma_{i+1}\} \in \{0,1\}$, which denote the indices of the energy matrix $\Delta$.

For DS, the effective Hamiltonian is \cite{ref5}, 
\begin{equation}
    H^{(1)} = \sum_{\{\gamma_{1},\gamma_{2}\}} \Delta_{\gamma_1 \gamma_2} \mathbb{P}^{(1)}_{\gamma_1} \otimes \mathbb{P}^{(2)}_{\gamma_{2}},
    \label{eq36}
\end{equation}
where, $\{\gamma_{1},\gamma_{2}\}\in \{0,1\}$, and we consider the principal player to be on the site with index $i=1$. Since we are considering nearest neighbour interaction, we get the Hamiltonian, in Eq.~(\ref{eq36}), with interaction between the sites $1,2$. In Eq.~(\ref{eq36}), $\{\gamma_1, \gamma_2\} \in \{0,1\}$ corresponds to $\{|\gamma_1\rangle, |\gamma_2\rangle\} \in \{ |0\rangle, |1\rangle \}$, and $|\gamma_1\rangle,~ |\gamma_2\rangle$ denote the strategies opted by the principal player and its nearest neighbour, respectively. In this case, we need to minimize the energy (or, maximize the payoff) of the principal player at spin site $1$ and the state of the \textit{two}-player game is given as $|\nu \rangle = |\gamma_1 \gamma_2\rangle$ respectively. As derived in Ref.~\cite{ref4}, we have the DS partition function $\zeta^{DS}$ as,
\begin{gather}
    \zeta^{DS} = \sum_{|\nu \rangle} \langle \nu| e^{-\beta H^{(1)}}|\nu \rangle = \sum_{\{\gamma_1, \gamma_2\}} \langle \gamma_1 \gamma_2| e^{-\beta H^{(1)}}|\gamma_1 \gamma_2 \rangle, \nonumber\\
    \text{or,}~\zeta^{DS} = \sum_{\{\gamma_1, \gamma_2\}} e^{-\beta \Delta_{\gamma_1\gamma_2}} =e^{\beta \mathcal{A}} + e^{\beta \mathcal{B}} + e^{\beta \mathcal{C}} + e^{\beta \mathcal{D}}.
    \label{dem-partition}
\end{gather}
\textbf{Game magnetization:} As discussed in Ref.~\cite{ref5} and derived in Section~(\textbf{II D}) of Ref.~\cite{ref4}, using the order parameter, i.e., $\hat{\mathcal{M}_z}^{(1)} = \mathbb{P}_0^{(1)} - \mathbb{P}_1^{(1)}$, defined for the principal player at the site with index $i=1$, one can get the expression for average magnetization of the principal player $\langle \hat{\mathcal{M}_z}^{(1)}\rangle_{\beta}$ as,
\begin{gather}
    \mu_g^{DS} = \langle \hat{\mathcal{M}_z}^{(1)}\rangle_{\beta} = \dfrac{1}{\zeta^{DS}} \sum_{|\nu \rangle} \langle \nu| \hat{\mathcal{M}_z}^{(1)} e^{-\beta H^{(1)}}|\nu \rangle, \nonumber\\
    = \dfrac{1}{\zeta^{DS}} \sum_{\{\gamma_1, \gamma_2\}} \langle \gamma_1 \gamma_2| (\mathbb{P}_0^{(1)} - \mathbb{P}_1^{(1)}) e^{-\beta H^{(1)}}|\gamma_1 \gamma_2 \rangle, \label{mgds}
\end{gather}
In Eq.~(\ref{mgds}), we have two terms: $\langle \gamma_1 \gamma_2| \mathbb{P}_0^{(1)} e^{-\beta H^{(1)}}|\gamma_1 \gamma_2 \rangle$ and $\langle \gamma_1 \gamma_2|  \mathbb{P}_1^{(1)} e^{-\beta H^{(1)}}|\gamma_1 \gamma_2 \rangle$, in the RHS, with the summation over $\{\gamma_1, \gamma_2\} \in \{0,1\}$ and both projection operators $\mathbb{P}_0^{(1)} = |0\rangle\langle 0|,~\mathbb{P}_1^{(1)} = |1\rangle\langle 1|$ acting on the site with index $i=1$. Now, $\mathbb{P}_0^{(1)}|\gamma_1=0\rangle |\gamma_2 \rangle = |\gamma_1=0\rangle |\gamma_2 \rangle$, and $\mathbb{P}_1^{(1)}|\gamma_1=1\rangle |\gamma_2 \rangle = |\gamma_1=1\rangle |\gamma_2 \rangle,~\forall~\gamma_2\in \{0,1\}$. 

Hence, $\langle \gamma_1 \gamma_2| \mathbb{P}_0^{(1)} e^{-\beta H^{(1)}}|\gamma_1 \gamma_2 \rangle = \sum_{\gamma_2 \in \{0,1\}}e^{-\beta \Delta_{0\gamma_2}}$, for $\gamma_1 = 0$ and $\gamma_2\in \{0,1\}$, and $\langle \gamma_1 \gamma_2| \mathbb{P}_1^{(1)} e^{-\beta H^{(1)}}|\gamma_1 \gamma_2 \rangle = \sum_{\gamma_2 \in \{0,1\}}e^{-\beta \Delta_{1\gamma_2}}$, for $\gamma_1 = 1$ and $\gamma_2\in \{0,1\}$. We now have the RHS in Eq.~(\ref{mgds}), with the expression of $\zeta^{DS}$ in Eq.~(\ref{dem-partition}), as,
\begin{gather}
    \mu_g^{DS} = \dfrac{ \sum_{\gamma_2 \in \{0,1\}}e^{-\beta \Delta_{0\gamma_2}} - \sum_{\gamma_2 \in \{0,1\}} e^{-\beta \Delta_{1\gamma_2}}}{e^{\beta \mathcal{A}} + e^{\beta \mathcal{B}} + e^{\beta \mathcal{C}} + e^{\beta \mathcal{D}}}, \nonumber\\
    \text{or,}~\mu_g^{DS}=\dfrac{e^{\beta \mathcal{A}} + e^{\beta \mathcal{B}} - e^{\beta \mathcal{C}} - e^{\beta \mathcal{D}}}{e^{\beta \mathcal{A}} + e^{\beta \mathcal{B}} + e^{\beta \mathcal{C}} + e^{\beta \mathcal{D}}}.
    \label{eq37}
\end{gather}
The formalism to derive this expression is similar to that of AS. Now, DS is applicable to all games, irrespective of whether or not the payoff criterion: $\mathcal{A+D=B+C}$ is fulfilled. If the payoffs fulfil the criterion: $\mathcal{A+D=B+C}$, $\mu_g^{DS}$, from Eq.~(\ref{eq37}) reduces to,
\begin{equation}
    \mu_g^{DS} = \langle \hat{\mathcal{M}_z}^{(1)}\rangle_{\beta}  = \tanh\bigg[{\beta\bigg(\dfrac{\mathcal{A-C}}{2}\bigg)}\bigg].
    \label{eq38}
\end{equation}
In Ref.~\cite{ref4}, it was shown by one of us that when the payoffs satisfy the criterion: $\mathcal{A+D=B+C}$, $\mu_g^{NEM}=\mu_g^{DS}$, whereas, they differ significantly when $\mathcal{A+D\neq B+C}$. \\
\textbf{Game susceptibility:} We can calculate the game susceptibility, from Eqs.~(\ref{eq13a}, \ref{eq37}), as,
\begin{equation}
    \chi_\sigma^{DS} = \dfrac{1}{\beta}\dfrac{\partial}{\partial \sigma}\mu_g^{DS}, ~\text{where $\sigma \in \{\mathcal{A, B, C, D}\}$}.
    \label{ne54}
\end{equation} 
In Eq.~(\ref{ne54}), we are not explicitly determining the expression for the game susceptibility since game susceptibility is only defined for \textit{field payoffs}, and they are specific for each game. Hence, for different games, we first define the corresponding game payoffs, and then we determine the game susceptibility with respect to the \textit{field payoffs}.\\ 
\textbf{Game correlation:} Considering the fact that $\hat{\mathcal{M}_z}^{(1)}$ is the Pauli $\mathfrak{s}_z$-operator acting on the $1^{st}$ site, which gives the spin/strategy for that site, we calculate the correlation as follows, we choose \textit{two} principal sites/players and for that, we need to modify the Hamiltonian $H^{(1)}$ in Eq.~(\ref{eq36}) to include two sites and their nearest neighbour interaction. Let us consider now the Hamiltonian with two principal players/sites, i.e., $1^{st}$-site and the $j^{th}$-site, so we have,
\begin{equation}
    H^{(1,j)} = \sum_{\substack{\{\gamma_j,\gamma_{j+1}\\ \gamma_1,\gamma_2\}}} [\Delta_{\gamma_1\gamma_2} \mathbb{P}_{\gamma_1}^{(1)} \otimes \mathbb{P}_{\gamma_2}^{(2)} + \Delta_{\gamma_j \gamma_{j+1}} \mathbb{P}_{\gamma_j}^{(j)} \otimes \mathbb{P}_{\gamma_{j+1}}^{(j+1)}].
    \label{eq39}
\end{equation}
where, $\{\gamma_1,\gamma_2, \gamma_j,\gamma_{j+1}\}\in \{0,1\}$.
We need to minimize the energy (or, maximize the payoff) of both the principal players at sites $\{1,j\}$, and the state of the \textit{two}-player game at both sites is given as $|\nu \rangle = |\gamma_1 \gamma_2 \gamma_j \gamma_{j+1}\rangle$, with $\{|\gamma_1\rangle, |\gamma_2\rangle, |\gamma_j\rangle, |\gamma_{j+1}\rangle\} \in \{|0\rangle, |1\rangle\}$ being the strategies of the respective player. The partition function for a two-site DS method ($\zeta^{DS}_2$) is,
\begin{gather}
    \zeta^{DS}_2 = \text{Tr}(e^{-\beta H^{(1,j)}}) = \sum_{|\nu\rangle} \langle \nu| e^{-\beta H^{(1,j)}}|\nu\rangle, \nonumber\\
    \zeta^{DS}_2 = \sum_{|\gamma_1 \gamma_2 \gamma_j \gamma_{j+1}\rangle} \langle \gamma_1 \gamma_2 \gamma_j \gamma_{j+1}| e^{-\beta H^{(1,j)}}|\gamma_1 \gamma_2 \gamma_j \gamma_{j+1}\rangle, \nonumber \\
    \text{or,}~~\zeta^{DS}_2 = \sum_{\{\gamma_1, \gamma_2\}} e^{-\beta \Delta_{\gamma_1 \gamma_2}}\cdot \sum_{\{\gamma_j, \gamma_{j+1}\}} e^{-\beta \Delta_{\gamma_j \gamma_{j+1}}} .\nonumber\\
    \text{Thus,}~~ \zeta^{DS}_2 = (e^{\beta \mathcal{A}} + e^{\beta \mathcal{B}} + e^{\beta \mathcal{C}} + e^{\beta \mathcal{D}})^2,
    \label{eq40}
\end{gather}
where, we have $\gamma_i \in \{0,1\}~ \forall ~i \in \{1,2,j,j+1\}$. From Eq.~(\ref{eq40}), we notice that the two-site DS partition function $\zeta^{DS}_2$ is simply the square of the single-site DS partition function $\zeta^{DS}$ in Eq.~(\ref{dem-partition}), since the two principal players do not directly interact among themselves, and hence, the \textit{joint} two-site DS partition function is simply the product of the individual principal player's partition functions. 

\begin{table*}[!ht]
\renewcommand{\arraystretch}{3.5}
\centering
\resizebox{\textwidth}{!}{
\begin{tabular}{|cl|c|c|c|}
\hline
\multicolumn{2}{|c|}{\textit{\textbf{}}} & $\textbf{NEM}^{*}$ & \textbf{DS} & $\textbf{AS}^{\#}$ \\ \hline
\multicolumn{2}{|c|}{{$\mathbf{H}$}} & $H = -\mathcal{J} \sum_{i=1}^{N} \mathfrak{s}_i \mathfrak{s}_{i+1} - \mathfrak{h}\sum_{i=1}^{N} \mathfrak{s}_i$ & $H^{(1)} = \sum_{\gamma,\gamma' = 0}^{1} \Delta_{\gamma\gamma'} \mathbb{P}_{\gamma}^{(1)} \otimes \mathbb{P}_{\gamma'}^{(2)}$ & $H = \sum_{i=1}^{N} \sum_{\gamma,\gamma'=0}^{1} \Delta_{\gamma\gamma'} \mathbb{P}^{(i)}_{\gamma} \otimes \mathbb{P}^{(i+1)}_{\gamma'}$ \\ \hline
\multicolumn{2}{|c|}{{$\bm{\zeta}$}} & $\zeta^{NEM} = e^{\beta\mathcal{(A-C)}} + e^{-\beta \mathcal{(B+D)}} + 2e^{\frac{\beta}{2} \mathcal{(B+C-A-D)}}$ & $\zeta^{DS} = e^{\beta \mathcal{A}} + e^{\beta \mathcal{B}} + e^{\beta \mathcal{C}} + e^{\beta \mathcal{D}}$ & $\zeta^{AS} = (e^{\beta\mathcal{A}} + e^{\beta\mathcal{D}})^N$ \\ \hline
\multicolumn{2}{|c|}{{$\bm{\langle \Lambda_1 \rangle}$}} &                               \begin{tabular}{@{}c@{}}
    $\langle \Lambda_1 \rangle^{NEM}= \dfrac{1}{2\zeta^{NEM}}[{\mathcal{(B+C-A-D)}e^{\frac{\beta}{2} \mathcal{(B+C-A-D)}}}$ \\
    $+ {\mathcal{(A-C)}e^{\beta\mathcal{(A-C)}} - \mathcal{(B+D)}e^{-\beta \mathcal{(B+D)}}}]$ 
\end{tabular}
& $\langle \Lambda_1 \rangle^{DS}= \dfrac{\mathcal{A}e^{\beta\mathcal{A}} + \mathcal{B}e^{\beta\mathcal{B}} + \mathcal{C}e^{\beta\mathcal{C}} + \mathcal{D}e^{\beta\mathcal{D}}}{\zeta^{DS}}$ & $\langle \Lambda_1 \rangle^{AS}= \dfrac{\mathcal{A}e^{\beta\mathcal{A}} + \mathcal{D}e^{\beta\mathcal{D}}}{e^{\beta\mathcal{A}} + e^{\beta\mathcal{D}}}$ \\ \hline
\multicolumn{2}{|c|}{{$\bm{\mu_g}$}} & $\mu_g^{NEM} = \dfrac{e^{2\beta \mathcal{J}}\sinh{\beta \mathfrak{h}}}{\sqrt{1 + e^{4\beta\mathcal{J}}\sinh^2{\beta \mathfrak{h}}}}$ & $\mu_g^{DS} = \dfrac{e^{\beta \mathcal{A}} + e^{\beta \mathcal{B}} - e^{\beta \mathcal{C}} - e^{\beta \mathcal{D}}}{\zeta^{DS}}$ & $\mu_g^{AS} = \tanh\bigg[{\beta\bigg(\dfrac{\mathcal{A-D}}{2}\bigg)}\bigg]$ \\ \hline
\multicolumn{2}{|c|}{{$\bm{\chi_\sigma}$}} & $\chi_\sigma^{NEM} = \dfrac{1}{\beta}\dfrac{\partial \mu_g^{NEM}}{\partial \sigma},~\forall~\sigma\in \{\mathcal{A, B, C, D}\}$ & $\chi_\sigma^{DS} = \dfrac{1}{\beta}\dfrac{\partial \mu_g^{DS}}{\partial \sigma},~\forall~\sigma\in \{\mathcal{A, B, C, D}\}$ & $\chi_\sigma^{AS} = \dfrac{1}{\beta}\dfrac{\partial \mu_g^{AS}}{\partial \sigma},~\forall~\sigma \in \{\mathcal{A, D}\}$ \\ \hline
\multicolumn{2}{|c|}{{$\bm{\mathfrak{c}_j}$}} & $\mathfrak{c}_j^{NEM} = \cos^2 \varphi + \bigg(\dfrac{\Omega_{-}}{\Omega_{+}} \bigg)^j \sin^2 \varphi$ & {$\mathfrak{c}_j^{DS} = \bigg[\dfrac{e^{\beta \mathcal{A}} + e^{\beta \mathcal{B}} - e^{\beta \mathcal{C}} - e^{\beta \mathcal{D}}}{\zeta^{DS}}\bigg]^2$} & $\mathfrak{c}_j^{AS} = \tanh^2{\bigg[\beta\bigg(\dfrac{\mathcal{A-D}}{2}\bigg)\bigg]}$ \\ \hline
\multicolumn{2}{|c|}{{$\bm{\mathbb{S}_g}$}} & $\mathbb{S}_{g}^{NEM} = \dfrac{1}{2} \dfrac{\partial}{\partial\beta} \bigg[\dfrac{1}{\zeta^{NEM}}\dfrac{\partial }{\partial \beta}\zeta^{NEM} \bigg]$ & 
    \begin{tabular}{@{}c@{}}
        $\mathbb{S}_{g}^{DS}= \bigg(\dfrac{\mathcal{A}^2 e^{\beta\mathcal{A}} + \mathcal{B}^2 e^{\beta\mathcal{B}} + \mathcal{C}^2 e^{\beta\mathcal{C}} + \mathcal{D}^2 e^{\beta\mathcal{D}}}{\zeta^{DS}}\bigg)$ \\ 
        $- \bigg(\dfrac{\mathcal{A}e^{\beta\mathcal{A}} + \mathcal{B}e^{\beta\mathcal{B}} + \mathcal{C}e^{\beta\mathcal{C}} + \mathcal{D}e^{\beta\mathcal{D}}}{\zeta^{DS}}\bigg)^2$
    \end{tabular} 
    & $\mathbb{S}_{g}^{AS} = \dfrac{(\mathcal{A-D})^2 e^{\beta(\mathcal{A+D})}}{(e^{\beta\mathcal{A}}+\mathrm{e}^{\beta\mathcal{D}})^2}$ \\ \hline
\end{tabular}}
\caption{\centering{Here, \textbf{H} is the Hamiltonian, $\bm{\zeta}$ is the Partition function, $\bm{\langle \Lambda_1\rangle}$ is the average payoff/player, $\bm{\mu_g}$ is the game magnetization, $\bm{\chi_{\sigma}}$ is the game susceptibility, $\bm{\mathfrak{c}_j}$ is the correlation and $\bm{\mathbb{S}_g}$ is the game payoff capacity.${}^*$For \textbf{NEM}: $\mathfrak{h} = \frac{\mathcal{(A-C)+(B-D)}}{4}$, $\mathcal{J} = \frac{\mathcal{(A-C)-(B-D)}}{4}$, $\cos^2 \varphi= \dfrac{\sinh^2{\beta \mathfrak{h}}}{\sinh^2{\beta \mathfrak{h}} + e^{-4\beta \mathcal{J}}} = 1 - \sin^2 \varphi$, and $\Omega_{\pm} = e^{-\beta \mathcal{J}} [e^{2\beta \mathcal{J}}\text{cosh}(\beta \mathfrak{h}) \pm \sqrt{1+e^{4\beta \mathcal{J}}\text{sinh}^2 (\beta \mathfrak{h})}]$. ${}^{\#}$\textbf{AS} is only applicable to games that satisfy the payoffs criterion: $\mathcal{A+D = B+C}$. \textbf{NOTE:} $\mu_g$ and $\langle\Lambda_1\rangle$ have been dealt with in Ref.~\cite{ref4}}}
\label{table-analytical}
\hrule
\end{table*}

The correlation, $\mathfrak{c}_j^{DS}=\langle \mathfrak{s}_1 \mathfrak{s}_{j}\rangle \equiv \langle \hat{\mathcal{M}_z}^{(1)} \hat{\mathcal{M}_z}^{(j)} \rangle_{\beta}$, is calculated in a similar way to that of AS, and we have,
\begin{gather}
    \mathfrak{c}_j^{DS} = \langle \mathfrak{s}_1 \mathfrak{s}_{j}\rangle = \dfrac{\sum_{|\nu \rangle}\langle \nu| \hat{\mathcal{M}_z}^{(1)}\hat{\mathcal{M}_z}^{(j)} e^{-\beta H^{(1,j)}} |\nu \rangle}{2\zeta^{DS}},\nonumber\\
    = \dfrac{1}{2\zeta^{DS}} \sum_{|\nu \rangle}\langle \nu| \{\mathbb{P}_0^{(1)} - \mathbb{P}_1^{(1)}\}  \{\mathbb{P}_0^{(j)} - \mathbb{P}_1^{(j)}\} e^{-\beta H^{(1,j)}} |\nu \rangle, \nonumber\\
    \text{or,}~~ \mathfrak{c}_j^{DS} = \dfrac{1}{2\zeta^{DS}} \bigg[\sum_{|\nu \rangle}\{\langle \nu| \mathbb{P}_0^{(1)} \mathbb{P}_0^{(j)}e^{-\beta H^{(1,j)}} |\nu \rangle \nonumber\\
    + \langle \nu| \mathbb{P}_1^{(1)} \mathbb{P}_1^{(j)}e^{-\beta H^{(1,j)}} |\nu \rangle - \langle \nu| \mathbb{P}_0^{(1)} \mathbb{P}_1^{(j)}e^{-\beta H^{(1,j)}} |\nu \rangle \nonumber\\
    - \langle \nu| \mathbb{P}_1^{(1)} \mathbb{P}_0^{(j)}e^{-\beta H^{(1,j)}} |\nu \rangle \} \bigg],\nonumber\\
    = \dfrac{1}{2\zeta^{DS}}\bigg\{\sum_{\gamma_2}e^{-\beta \Delta_{0\gamma_2}}\cdot\sum_{\gamma_{j+1}}e^{-\beta \Delta_{0\gamma_{j+1}}} + \sum_{\gamma_2}e^{-\beta \Delta_{1\gamma_2}}\nonumber\\
     \sum_{\gamma_{j+1}}e^{-\beta \Delta_{1\gamma_{j+1}}}  - \sum_{\gamma_2}e^{-\beta \Delta_{0\gamma_2}}\cdot
    \sum_{\gamma_{j+1}}e^{-\beta \Delta_{1\gamma_{j+1}}} \nonumber
\end{gather}
\begin{gather}
    - \sum_{\gamma_2}e^{-\beta \Delta_{1\gamma_2}}\cdot \sum_{\gamma_{j+1}}e^{-\beta \Delta_{0\gamma_{j+1}}}  \bigg\},  \nonumber\\
    \text{or,}~ \mathfrak{c}_j^{DS} = \langle \hat{\mathcal{M}_z}^{(1)} \hat{\mathcal{M}_z}^{(j)} \rangle_{\beta}  = \dfrac{1}{\cancel{2}\zeta^{DS}}\cancel{2}[(e^{\beta \mathcal{C}} + e^{\beta \mathcal{D}})^2 \nonumber\\
    + (e^{\beta \mathcal{A}} + e^{\beta \mathcal{B}})^2 - 2(e^{\beta \mathcal{C}} + e^{\beta \mathcal{D}})(e^{\beta \mathcal{A}} + e^{\beta \mathcal{B}})]. \nonumber\\
    \text{Thus,}~ \mathfrak{c}_j^{DS} = \langle \mathfrak{s}_1 \mathfrak{s}_{j}\rangle =\bigg[ \dfrac{e^{\beta \mathcal{A}} + e^{\beta \mathcal{B}} - e^{\beta \mathcal{C}} - e^{\beta \mathcal{D}}}{e^{\beta \mathcal{A}} + e^{\beta \mathcal{B}} + e^{\beta \mathcal{C}} + e^{\beta \mathcal{D}}} \bigg]^2.
    \label{eq41}
\end{gather}
In Eq.~(\ref{eq41}), we find that the correlation is independent of the distance between the sites, similar to the case of AS, and we will further discuss this in the later sections.\\
\textbf{Individual player's average payoff:} In order to determine the expression for the individual player's average payoff $\langle \Lambda_1 \rangle^{DS}$, we use the expression of the partition function, i.e., $\zeta^{DS} = (e^{\beta\mathcal{A}} + e^{\beta\mathcal{B}} + e^{\beta\mathcal{C}} + e^{\beta\mathcal{D}})$ (see, Eq.~(\ref{dem-partition}) and Ref.~\cite{ref4}) and from Eq.~(\ref{e1a}), we have, 
\begin{gather}
    \langle \Lambda_1 \rangle^{DS} = \dfrac{\mathcal{A}e^{\beta\mathcal{A}} + \mathcal{B}e^{\beta\mathcal{B}} + \mathcal{C}e^{\beta\mathcal{C}} + \mathcal{D}e^{\beta\mathcal{D}}}{e^{\beta\mathcal{A}} + e^{\beta\mathcal{B}} + e^{\beta\mathcal{C}} + e^{\beta\mathcal{D}}},
    \label{e6a}
\end{gather}
where, $(\mathcal{A, B, C, D})$ are the payoffs of the social dilemma.\\
\textbf{Game payoff capacity:} To determine the expression for the payoff capacity $\mathbb{S}_{g}^{DS}$ for games, we use the expression for the partition function, i.e., $\zeta^{DS} = e^{\beta\mathcal{A}} + e^{\beta\mathcal{B}} + e^{\beta\mathcal{C}} + e^{\beta\mathcal{D}}$, and from Eq.~(\ref{e3a}), we have, 
\begin{gather}
    \mathbb{S}_{g}^{DS} = \dfrac{\partial}{\partial\beta} \bigg[\dfrac{1}{\zeta^{DS}}\dfrac{\partial \zeta^{DS}}{\partial \beta} \bigg], \nonumber\\
    \text{or,}~\mathbb{S}_{g}^{DS}= \dfrac{1}{\zeta^{DS}}\bigg[(\mathcal{A}^2 e^{\beta\mathcal{A}} + \mathcal{B}^2 e^{\beta\mathcal{B}} + \mathcal{C}^2 e^{\beta\mathcal{C}} + \mathcal{D}^2 e^{\beta\mathcal{D}})\nonumber\\
    -\dfrac{1}{\zeta^{DS}}(\mathcal{A}e^{\beta\mathcal{A}} + \mathcal{B}e^{\beta\mathcal{B}} + \mathcal{C}e^{\beta\mathcal{C}} + \mathcal{D}e^{\beta\mathcal{D}})^2\bigg].
    \label{e6}
\end{gather}
where, $(\mathcal{A, B, C, D})$ are the payoffs of the social dilemma. All three analytical methods that we discussed in this section (summarized in Table~\ref{table-analytical}) are based on the $1D$-Ising chain, where we map the spin sites to the players and the spins ($\uparrow, \downarrow$) to the two strategies ($\mathcal{S}_1, \mathcal{S}_2$) available to the players.

\subsection{\label{sub-abm}Agent based method (ABM)}
ABM is a numerical technique often used to study social dilemmas in the thermodynamic limit. For our case, we consider a thousand players, and these players are placed on a $1D$-Ising chain where we consider nearest neighbour interaction and periodic boundary condition (see, Refs.~\cite{ref5, ref7}). The energy matrix $\Delta$ in Eq.~(\ref{eq30}) gives the energy of each site in the Ising chain. We now update the strategy choice of the players by running a conditional loop a million times, i.e., a thousand strategy runs per player on average. The basic structure of the algorithm (based on the \textit{Metropolis algorithm}\cite{ref13}) for determining magnetization and the average game payoff is well explained in Ref.~\cite{ref4} (which was co-authored by one of us), but for our case, we need to determine the game susceptibility, the correlation as well as the payoff capacity. So, here is a brief outline of the modified algorithm:
\begin{enumerate}
    \item Allocate a random strategy of \textit{0} (strategy $\mathcal{S}_1$: say \textit{defection}) or \textit{1} (strategy $\mathcal{S}_2$: say \textit{cooperation}) to each player located at each site on the $1D$-Ising chain. These values are mapped to the (Hawk/Dove) strategies in the HDG, or to (Contribute/Defect) strategies in  PGG.
    \item Arbitrarily select a player of interest to determine its specific strategy as well as the strategy of the nearest neighbour. Based on these strategies, the energy (or, $-ve$ payoff) $\Delta$ of the player of interest is determined. The energy (or, $-ve$ payoff) of the player of interest is calculated for the two possibilities, i.e., if it had chosen the alternative strategy while keeping the strategy of its nearest neighbour the same or chosen the same strategy as its nearest neighbour. 
    \item The difference of energy ($d\Delta$) is calculated for the player of interest, for the two different possibilities, and the current strategy of the player of interest is flipped based on whether the Fermi function $(e^{\beta \cdot d\Delta}+1)^{-1} > 0.5$, else, if Fermi function $(e^{\beta \cdot d\Delta}+1)^{-1}<0.5$, it is not flipped~\cite{ref5, ref7}. 
    \item Three conditions now arise depending on the indicators:
    \begin{itemize}
        \item For \textit{game susceptibility}: Determine the difference between the fraction of players choosing to cooperate and the fraction of players choosing to defect. This gives the average magnetization at each site, for each cycle. Then calculate the variance of the game magnetization. Multiply this with the factor $\frac{1}{2}$, or $-\frac{1}{4}$ (depending on the game payoff), to get the game susceptibility. You will see in Eqs.~(\ref{eq47}, \ref{eq61}, \ref{eq61cost}, \ref{eq61punish}) how we are getting an extra multiplicative factor of $\frac{1}{2}$, or $-\frac{1}{4}$ for both HDG and PGG.
        \item For \textit{correlation}: We consider two players of interest, and hence, we slightly modify the first 4 steps to include the condition of two players of interest. After the individual spin-flipping operations for both the randomly chosen players of interest, if both of them have the same strategy, then a correlation value of $+1$ is added to the total correlation, whereas, if both the strategies are different, then a correlation value of $-1$ is added to the total correlation. 
        \item For \textit{game payoff capacity}: For a given payoff, we determine the total energy and the square of the total energy for all the players in the Ising chain. Then, we calculate the average of both the total energy (which is equivalent to the negative of the average payoff) as well as the square of the total energy and determine the game payoff capacity from the relation given in Eq.~(\ref{e3a}). 
    \end{itemize} 
    \item Go to step 2 and repeat this process a \textit{thousand} times.
\end{enumerate}
This algorithm can be better understood after looking at the Python code, for both HDG and PGG (see,  Supplementary Material\cite{supp}). We notice that the probability of strategy flipping decreases when the energy difference $d\Delta$ increases, since, our main aim is to avail the maximum payoff, and this is possible only when our system achieves equilibrium, i.e., the minimum possible energy configuration. A detailed overview of ABM is also given in Refs.~\cite{ref4, ref5, ref7}. 

When we study models involving players who can interact among themselves via Ising-type coupling, the parameter $T$ (or, $\beta$) plays an important role in measuring the noise/uncertainty. We can also interpret $\beta = \frac{1}{k_B T}$ as the intensity of selection (as shown in Ref.\cite{ref7}) where, for $\beta\ll 1$, we have random strategy selection, whereas, for $\beta\gg 1$, there is no randomness in strategy selection. We also see that for increasing payoff, the transition probability for changing strategies asymptotically goes from $0\rightarrow 1$ as $\beta$ goes from $0 \rightarrow \infty$~\cite{ref7}.

\section{\label{pgg}The Public Goods Game (PGG)}
PGG~\cite{ref9} is a social dilemma that is similar to Prisoner's Dilemma. This game involves players who have two options, either to \textit{cooperate/contribute} (denoted by $\mathbb{C}$) a set number of tokens, of cost $\mathfrak{t}$, into a public pot or \textit{defect/free-ride} (denoted by $\mathbb{D}$). The total amount of tokens collected is treated as a \textit{public good} which can be shared equally by all the players. The payoffs for the \textit{cooperators} ($P_\mathfrak{C}$) and the \textit{defectors} ($P_\mathfrak{D}$) are given as \cite{ref3},
\begin{equation}
    P_\mathfrak{D} = \dfrac{k n_{\mathfrak{C}} \mathfrak{t}}{N}, ~~P_\mathfrak{C} = P_\mathfrak{D}-\mathfrak{t},
    \label{eq53}
\end{equation}
where, $\mathfrak{t}$ is the cost, $N$ is the number of players, $k$ is the multiplication factor of the public good (with $k>1$) and $n_{\mathfrak{C}}$ is the number of cooperators respectively. {The rationale for this specific form of payoff matrix for PGG~\cite{ref3} is as follows: Consider a group of $N$ players with $n_c$ cooperators (and obviously, $N - n_c$ defectors). One considers only two levels of investment: zero, corresponding to defectors withholding their money, or a fixed amount $c$ denoting the cooperators contribution. The value of the public good is determined via multiplication factor $r$ of the common pool. For mutual cooperation to perform better than mutual defection $r > 1$ must hold. However, if $r > N$ the social dilemma raised by the PGG is relaxed as each invested dollar has a positive net return. At the same time, the higher gain of defectors is preserved regardless of the group composition.}

For a two-player game, we have $N=2$, and when both of them choose $\mathbb{C}$-strategy, $P_\mathfrak{C} = \mathfrak{t}(k - 1)$, since, $n_{\mathfrak{C}} = N = 2$. When both of them choose $\mathbb{D}$-strategy, $P_\mathfrak{D} = 0$. However, when one player chooses $\mathbb{D}$-strategy and the other player chooses  $\mathbb{C}$-strategy (i.e., $n_{\mathfrak{C}} = 1,~N=2$), then the defector gets $P_{\mathfrak{D}}= \frac{k\mathfrak{t}}{2}$, and the cooperator gets $P_{\mathfrak{C}}= [\frac{k\mathfrak{t}}{2} - \mathfrak{t}]$. If we denote the cooperator's payoff (for $n_{\mathfrak{C}} = N = 2$), $P_\mathfrak{C} = \mathfrak{t}(k - 1) = 2\mathfrak{r}$, for $\mathfrak{r}$ being the \textit{reward}, then we have $[\frac{k\mathfrak{t}}{2} - \mathfrak{t}] = [\mathfrak{r}-\frac{\mathfrak{t}}{2}]$ and $\frac{k\mathfrak{t}}{2} = [\mathfrak{r}+\frac{\mathfrak{t}}{2}]$, respectively, with reward $\mathfrak{r}>0$ and cost $\mathfrak{t}>0$. If we introduce punishment ($\mathfrak{p}$) for the defectors, then the defector's overall payoff gets reduced by $\mathfrak{p}$. Here, we consider that $\mathfrak{p}>0$ and we have the \textit{two-player} $(N=2)$ payoff matrix $\Lambda$ as,
\begin{table}[H]
\centering
$\Lambda$ = 
\setlength{\extrarowheight}{5pt}
\begin{tabular}{*{4}{c|}}
    \multicolumn{2}{c}{} & \multicolumn{2}{c}{}\\\cline{3-4}
    \multicolumn{1}{c}{} & $\bm{N=2}$ & $\mathcal{S}_1 \equiv \mathbb{C}$  & $\mathcal{S}_2 \equiv \mathbb{D}$ \\\cline{2-4}
    \multirow{2}*{}  & $\mathcal{S}_1 \equiv \mathbb{C}$ & $(2\mathfrak{r},2\mathfrak{r})$ & $(\mathfrak{r}-\frac{\mathfrak{t}}{2}, \mathfrak{r} + \frac{\mathfrak{t}}{2} -\mathfrak{p})$ \\\cline{2-4}
    & $\mathcal{S}_2 \equiv \mathbb{D}$ & $(\mathfrak{r} + \frac{\mathfrak{t}}{2} -\mathfrak{p}, \mathfrak{r} - \frac{\mathfrak{t}}{2})$ & $(-\mathfrak{p},-\mathfrak{p})$ \\\cline{2-4}
\end{tabular}
\end{table}
We notice that PGG fulfils the payoff criterion: $\mathcal{A+D=B+C}$ for both $\mathfrak{p}=0$ as well as $\mathfrak{p}\neq 0$. This means that we can apply AS to this game. For a single player (the \textit{row} player), the payoff matrix can be written as,
\begin{equation}
    \Lambda_1 = \begin{bmatrix}
            2\mathfrak{r} & \mathfrak{r}-\frac{\mathfrak{t}}{2}\\
            \mathfrak{r} + \frac{\mathfrak{t}}{2} -\mathfrak{p} & -\mathfrak{p} 
          \end{bmatrix}
          \equiv \begin{bmatrix}
            \mathcal{A} & \mathcal{B}\\
            \mathcal{C} & \mathcal{D} 
        \end{bmatrix}.
          \label{eq54}
\end{equation}
We will be using the payoff matrix $\Lambda_1$ given in Eq.~(\ref{eq54}) for further calculations related to PGG.

\subsection{\label{sus-pgg}Game susceptibility}
There are three parameters in PGG: reward ($\mathfrak{r}$), cost ($\mathfrak{t}$) and punishment ($\mathfrak{p}$). We will calculate the \textit{reward susceptibility} with respect to $\mathfrak{r}$, the \textit{cost susceptibility} with respect to $\mathfrak{t}$, and the \textit{punishment susceptibility} with respect to $\mathfrak{p}$. In the Ising model, thermodynamic susceptibility is defined as the change in magnetization as a response to a unit change in the external magnetic field $\mathfrak{h}$, and it is independent of the coupling constant $\mathcal{J}$. However, in social dilemmas,  we work with game payoffs, instead of Ising parameters, and in \textit{NEM}, we linearly map the Ising parameters to the game payoffs. For PGG, in NEM, $\mathcal{J} = 0$ and $\mathfrak{h} = \frac{2\mathfrak{r}-\mathfrak{t} + 2\mathfrak{p}}{4}$ (from Eq.~(\ref{eq54})). We see that $\mathcal{J}$ is equal to \textit{zero} and independent of $(\mathfrak{r}, \mathfrak{t}, \mathfrak{p})$. Hence, the susceptibilities corresponding to the \textit{field payoffs} $\mathfrak{r}, \mathfrak{t}$ and $\mathfrak{p}$, can be calculated. We also consider $\mathfrak{t}>0, \mathfrak{r}>0,$ and $\mathfrak{p}>0$.

\subsubsection{\underline{NEM}}
For calculating the \textit{reward, cost} and \textit{punishment susceptibilities}, using \textit{Nash equilibrium mapping}, we write the game magnetization $\mu_g^{NEM}$ in terms of $\mathfrak{r}$, $\mathfrak{t}$ and $\mathfrak{p}$. From Eqs.~(\ref{eq23}, \ref{eq54}), $\mu_g^{NEM}$ in terms of the game parameters is given as,
\begin{equation}
    \mu_g^{NEM} = \tanh{\beta \bigg(\dfrac{2\mathfrak{r}-\mathfrak{t}+2\mathfrak{p}}{4}\bigg)}.
    \label{eq55}
\end{equation}
When $T\rightarrow 0$ (or, $\beta \rightarrow \infty$), i.e., ZN limit, we find $\mu_g^{NEM} \rightarrow +1$, for $2\mathfrak{r}>(\mathfrak{t}-2\mathfrak{p})$, i.e., every player \textit{cooperates}, and $\mu_g^{NEM} \rightarrow -1$, for $2\mathfrak{r}<(\mathfrak{t}-2\mathfrak{p})$, i.e., every player \textit{defects}. When $T\rightarrow \infty$ (or, $\beta \rightarrow 0$), i.e., IN limit, $\mu_g^{NEM} \rightarrow 0$ since the players choose their strategies randomly, resulting in roughly the same number of defectors and cooperators. From Eq.~(\ref{eq55}), we have the reward susceptibility as,
\begin{gather}
    \chi_{\mathfrak{r}}^{NEM} = \dfrac{1}{\beta}\dfrac{\partial}{\partial \mathfrak{r}}\mu_g^{NEM} =  \dfrac{1}{2}\text{sech}^2\bigg[\dfrac{\beta}{4}(2\mathfrak{r}-\mathfrak{t}+2\mathfrak{p})\bigg].
    \label{eq56}
\end{gather}
In Eq.~(\ref{eq56}), when $T\rightarrow 0$ (or, $\beta \rightarrow \infty$), i.e., ZN limit, $\chi_\mathfrak{r}^{NEM} \rightarrow 0,~\text{when}~2\mathfrak{r}\neq (\mathfrak{t} - 2\mathfrak{p})$, since, in this case, the rate of turnover from \textit{Defectors} (with strategy $\mathbb{D}$) to \textit{Cooperators} (with strategy $\mathbb{C}$) or vice-versa vanishes. When $\beta$ is finite, for increasing values of $\mathfrak{r}$, with fixed $(\mathfrak{t,p})$, the rate of change of the number of players choosing to \textit{cooperate} increases as $\mathfrak{r}$ approaches $(\mathfrak{t}-2\mathfrak{p})$ and then the rate decreases again as $\mathfrak{r}$ increases since most of the players have changed from $\mathbb{D}$-strategy to $\mathbb{C}$-strategy. When $T\rightarrow \infty$ (or, $\beta \rightarrow 0$), i.e., IN limit, $\chi_\mathfrak{r}^{NEM} \rightarrow \frac{1}{2}, \forall~\mathfrak{r},$ due to randomness in strategy selection. In IN limit, the rate of turnover from defectors to cooperators and vice-versa averages out to $\frac{1}{2}$. This can be verified from Eq.~(\ref{eq55}), where the \textit{first}-order correction (denoted by $\mu_g^{{NEM}{(1)}}$) of game magnetization $\mu_g^{NEM}$, Taylor expanded around $\beta$, is given as, $\mu_g^{{NEM}{(1)}} = \frac{\beta(2\mathfrak{r}-\mathfrak{t} + 2\mathfrak{p})}{4}$, and this gives us the \textit{zeroth}-order reward susceptibility correction, in the IN (or, $\beta\rightarrow 0$) limit, as,
\begin{equation}
    \lim_{\beta\rightarrow 0}\chi_\mathfrak{r}^{NEM(0)} = \lim_{\beta\rightarrow 0}\frac{1}{\cancel{\beta}} \frac{\partial}{\partial \mathfrak{r}}\underbrace{\frac{\cancel{\beta}(2\mathfrak{r}-\mathfrak{t} + 2\mathfrak{p})}{4}}_{=\mu_g^{NEM(1)}} =\frac{1}{2}.
\end{equation}
The \textit{first} and \textit{higher} order terms in the Taylor expansion of $\chi_\mathfrak{r}^{NEM}$ (see, Eq.~(\ref{eq56})) about $\beta$ , in the IN limit, vanishes.

One thing to note is that the reward susceptibility $\chi_\mathfrak{r}^{NEM}$ is always positive, which indicates that the rate of the turnover from Defectors (with strategy $\mathbb{D}$) to Cooperators (with strategy $\mathbb{C}$) is more than vice-versa. This implies that for increasing reward $\mathfrak{r}$, there is always a greater fraction of players shifting to $\mathbb{C}$-strategy than to $\mathbb{D}$-strategy.

From Eq.~(\ref{eq55}), we have the cost susceptibility in terms of $(\mathfrak{r, t, p})$ as:
\begin{gather}
    \chi_{\mathfrak{t}}^{NEM} = \dfrac{1}{\beta}\dfrac{\partial}{\partial \mathfrak{t}}\mu_g^{NEM} =  -\dfrac{1}{4}\text{sech}^2\bigg[\dfrac{\beta}{4}(2\mathfrak{r}-\mathfrak{t}+2\mathfrak{p})\bigg].
    \label{eqcostpggnem}
\end{gather}
In Eq.~(\ref{eqcostpggnem}), when $T\rightarrow 0$ (or, $\beta \rightarrow \infty$), i.e., ZN limit, $\chi_\mathfrak{t}^{NEM} \rightarrow 0,$ when $\mathfrak{t}\neq (2\mathfrak{r} + 2\mathfrak{p})$, since, in this case, the rate of turnover from \textit{Cooperators} (with strategy $\mathbb{C}$) to \textit{Defectors} (with strategy $\mathbb{D}$) or vice-versa vanishes. It is evident that when $\beta$ is finite, for increasing values of $\mathfrak{t}$, with fixed $(\mathfrak{r,p})$, the rate of change of the number of players shifting to \textit{defection} will increase as $\mathfrak{t}$ approaches $2(\mathfrak{r}+\mathfrak{p})$ and then the rate decreases again as $\mathfrak{t}$ increases since most of the players have changed from $\mathbb{C}$-strategy to $\mathbb{D}$-strategy. When $T\rightarrow \infty$ (or, $\beta \rightarrow 0$), i.e., IN limit, $\chi_\mathfrak{t}^{NEM} \rightarrow -\frac{1}{4},$ $\forall~\mathfrak{t},$ due to complete randomness in strategy selection. This can also be verified from Eq.~(\ref{eq55}), where the \textit{first}-order correction (denoted by $\mu_g^{{NEM}{(1)}}$) of game magnetization $\mu_g^{NEM}$, Taylor expanded around $\beta$, is given as, $\mu_g^{{NEM}{(1)}} = \frac{\beta(2\mathfrak{r}-\mathfrak{t} + 2\mathfrak{p})}{4}$, and this gives us the \textit{zeroth}-order cost susceptibility correction, in the IN (or, $\beta\rightarrow 0$) limit, as,
\begin{equation}
    \lim_{\beta\rightarrow 0}\chi_\mathfrak{t}^{NEM(0)} = \lim_{\beta\rightarrow 0}\frac{1}{\cancel{\beta}} \frac{\partial}{\partial \mathfrak{t}}\underbrace{\frac{\cancel{\beta}(2\mathfrak{r}-\mathfrak{t} + 2\mathfrak{p})}{4}}_{=\mu_g^{NEM(1)}} =-\frac{1}{4}.
\end{equation}
The \textit{first} and \textit{higher} order terms in the Taylor expansion of $\chi_\mathfrak{t}^{NEM}$ (see, Eq.~(\ref{eqcostpggnem})) about $\beta$ , in the IN limit, vanishes.

The cost susceptibility $\chi_\mathfrak{t}^{NEM}$ is always negative, which indicates that the rate of the turnover from Defectors (with strategy $\mathbb{D}$) to Cooperators (with strategy $\mathbb{C}$) is less than vice-versa. This implies that for increasing cost $\mathfrak{t}$, there is always a greater fraction of players shifting to $\mathbb{D}$-strategy than to $\mathbb{C}$-strategy.

Finally, from Eq.~(\ref{eq55}), we have the punishment susceptibility as,
\begin{gather}
    \chi_{\mathfrak{p}}^{NEM} = \dfrac{1}{\beta}\dfrac{\partial}{\partial \mathfrak{p}}\mu_g^{NEM} =  \dfrac{1}{2}\text{sech}^2\bigg[\dfrac{\beta}{4}(2\mathfrak{r}-\mathfrak{t}+2\mathfrak{p})\bigg].
    \label{eqpunishpgg}
\end{gather}
In Eq.~(\ref{eqpunishpgg}), when $T\rightarrow 0$ (or, $\beta \rightarrow \infty$), i.e., ZN limit, $\chi_\mathfrak{p}^{NEM} \rightarrow 0,~\text{when}~2\mathfrak{p}\neq (\mathfrak{t} - 2\mathfrak{r})$, since, in this case, the rate of turnover from \textit{Defectors} (with strategy $\mathbb{D}$) to \textit{Cooperators} (with strategy $\mathbb{C}$) or vice-versa vanishes. When $\beta$ is finite, for increasing values of $\mathfrak{p}$, with fixed $(\mathfrak{r,t})$, the rate of change of the number of players choosing to \textit{cooperate} increases as $\mathfrak{p}$ approaches $(\mathfrak{t}-2\mathfrak{r})$ and then the rate decreases again as the $\mathfrak{p}$ increases since most of the players have changed from $\mathbb{D}$-strategy to $\mathbb{C}$-strategy. When $T\rightarrow \infty$ (or, $\beta \rightarrow 0$), i.e., IN limit, $\chi_\mathfrak{p}^{NEM} \rightarrow \frac{1}{2}, \forall~\mathfrak{p},$ due to randomness in strategy selection. In IN limit, the rate of turnover from defectors to cooperators and vice-versa averages out to $\frac{1}{2}$. Similar to the case of reward susceptibility $\chi_\mathfrak{r}^{NEM}$, this can be verified from Eq.~(\ref{eq55}), where the \textit{first}-order correction (denoted by $\mu_g^{{NEM}{(1)}}$) of game magnetization $\mu_g^{NEM}$, Taylor expanded around $\beta$, is given as, $\mu_g^{{NEM}{(1)}} = \frac{\beta(2\mathfrak{r}-\mathfrak{t} + 2\mathfrak{p})}{4}$, and this gives us the \textit{zeroth}-order punishment susceptibility correction, in the IN (or, $\beta\rightarrow 0$) limit, as,
\begin{equation}
    \lim_{\beta\rightarrow 0}\chi_\mathfrak{p}^{NEM(0)} = \lim_{\beta\rightarrow 0}\frac{1}{\cancel{\beta}} \frac{\partial}{\partial \mathfrak{p}}\underbrace{\frac{\cancel{\beta}(2\mathfrak{r}-\mathfrak{t} + 2\mathfrak{p})}{4}}_{=\mu_g^{NEM(1)}} =\frac{1}{2}.
\end{equation}
The \textit{first} and \textit{higher} order terms in the Taylor expansion of $\chi_\mathfrak{p}^{NEM}$ (see, Eq.~(\ref{eqpunishpgg})) about $\beta$ , in the IN limit, vanishes.

The punishment susceptibility $\chi_\mathfrak{p}^{NEM}$ is always positive, which indicates that the rate of the turnover from Defectors (with strategy $\mathbb{D}$) to Cooperators (with strategy $\mathbb{C}$) is more than vice-versa. This implies that for increasing punishment $\mathfrak{p}$, there is always a greater fraction of players shifting to $\mathbb{C}$-strategy than to $\mathbb{D}$-strategy.

\subsubsection{\underline{AS}}
For calculating the \textit{reward, cost} and \textit{punishment susceptibilities}, using the \textit{Aggregate selection} method, we write the game magnetization $\mu_g^{AS} = \langle \hat{\mathcal{M}_z} \rangle_\beta$ in terms of $\mathfrak{r}$, $\mathfrak{t}$ and $\mathfrak{p}$. From Eqs.~(\ref{eq34}, \ref{eq54}), the AS game magnetization in terms of the game parameters is given as,
\begin{equation}
    \mu_g^{AS} = \langle \hat{\mathcal{M}_z}\rangle_\beta = \tanh{\beta \bigg(\dfrac{2\mathfrak{r}+\mathfrak{p}}{2}\bigg)}.
    \label{eq57}
\end{equation}
In Eq.~(\ref{eq57}), we notice that there is no $\mathfrak{t}$-dependency in $\mu_g^{AS}$, so $\chi_{\mathfrak{t}}^{AS} = 0,~\forall~\mathfrak{t}, \beta,$ which clearly shows the failure of AS while determining the cost susceptibility for PGG. When $T\rightarrow 0$ (or, $\beta \rightarrow \infty$), i.e., ZN limit, we find $\mu_g^{AS} \rightarrow +1$, since $(2\mathfrak{r}+\mathfrak{p})>0$ and in this case, all players \textit{cooperate} (or, \textit{provide}). The main aim of AS is to minimize the system’s total energy in order to maximize the cumulative payoff for all the players, and for $\beta \rightarrow \infty$, each and every player provides, leading to maximum payoffs. When $T\rightarrow \infty$ (or, $\beta \rightarrow 0$), i.e., IN limit, $\mu_g^{AS} \rightarrow 0$ since the players choose their strategies randomly, resulting in roughly the same number of \textit{Cooperators} and \textit{Defectors}. From Eq.~(\ref{eq57}), we have the reward susceptibility $\chi_{\mathfrak{r}}^{AS}$ as,
\begin{gather}
    \chi_{\mathfrak{r}}^{AS} = \dfrac{1}{\beta}\dfrac{\partial}{\partial \mathfrak{r}}\mu_g^{AS} =\text{sech}^2 \bigg[ \dfrac{\beta}{2}(2\mathfrak{r}+\mathfrak{p})\bigg].
    \label{eq58}
\end{gather}
In Eq.~(\ref{eq58}), when $T\rightarrow 0$ (or, $\beta \rightarrow \infty$), i.e., ZN limit, $\chi_{\mathfrak{r}}^{AS} \rightarrow 0$ since in this case, all players choose to \textit{Cooperate} (i.e., $\mu_g^{AS} = +1$) and the rate of change of strategy from \textit{Defection} to \textit{Cooperation} vanishes. When $T\rightarrow \infty$ (or, $\beta \rightarrow 0$), i.e., IN limit, $\chi_{\mathfrak{r}}^{AS} \rightarrow 1,$ $\forall~\mathfrak{r},$ because of randomness in strategy selection. This can be verified from Eq.~(\ref{eq57}), where the \textit{first}-order correction (denoted by $\mu_g^{{AS}{(1)}}$) of game magnetization $\mu_g^{AS}$, Taylor expanded around $\beta$, is given as, $\mu_g^{{AS}{(1)}} = \frac{\beta(2\mathfrak{r}+\mathfrak{p})}{2}$, and this gives us the \textit{zeroth}-order reward susceptibility correction, in the IN (or, $\beta\rightarrow 0$) limit, as,
\begin{equation}
    \lim_{\beta\rightarrow 0}\chi_\mathfrak{r}^{AS(0)} = \lim_{\beta\rightarrow 0}\frac{1}{\cancel{\beta}} \frac{\partial}{\partial \mathfrak{r}}\underbrace{\frac{\cancel{\beta}(2\mathfrak{r} + \mathfrak{p})}{2}}_{=\mu_g^{AS(1)}} = 1.
\end{equation}
The \textit{first} and \textit{higher} order terms in the Taylor expansion of $\chi_\mathfrak{r}^{AS}$ (see, Eq.~(\ref{eq58})) about $\beta$, in the IN limit, vanishes. The reward susceptibility $\chi_{\mathfrak{r}}^{AS}$ is always positive, which indicates that the rate of turnover from \textit{Defectors} (with strategy $\mathbb{D}$) to \textit{Cooperators} (with strategy $\mathbb{C}$) is more than vice-versa. This implies that for increasing reward $\mathfrak{r}$, there is always a greater fraction of players shifting to $\mathbb{C}$-strategy than to $\mathbb{D}$-strategy.

Similarly, we have the punishment susceptibility $\chi_{\mathfrak{p}}^{AS}$, from Eq.~(\ref{eq57}), as,
\begin{gather}
    \chi_{\mathfrak{p}}^{AS} = \dfrac{1}{\beta}\dfrac{\partial}{\partial \mathfrak{p}}\mu_g^{AS} =\dfrac{1}{2}\text{sech}^2 \bigg[ \dfrac{\beta}{2}(2\mathfrak{r}+\mathfrak{p})\bigg].
    \label{eq58punishment}
\end{gather}
In Eq.~(\ref{eq58punishment}), when $T\rightarrow 0$ (or, $\beta \rightarrow \infty$), i.e., ZN limit, $\chi_{\mathfrak{p}}^{AS} \rightarrow 0$ since in this case, all players choose to \textit{Cooperate} (i.e., $\mu_g^{AS} = +1$) and the rate of change of strategy from \textit{Cooperation} to \textit{Defection} or vice-versa vanishes. When $T\rightarrow \infty$ (or, $\beta \rightarrow 0$), i.e., IN limit, $\chi_{\mathfrak{p}}^{AS} \rightarrow \frac{1}{2},$ $\forall~\mathfrak{p},$ because of the randomness in strategy selection. This can also be verified from Eq.~(\ref{eq57}), where the \textit{first}-order correction (denoted by $\mu_g^{{AS}{(1)}}$) of game magnetization $\mu_g^{AS}$, Taylor expanded around $\beta$, is given as, $\mu_g^{{AS}{(1)}} = \frac{\beta(2\mathfrak{r}+\mathfrak{p})}{2}$, and this gives us the \textit{zeroth}-order punishment susceptibility correction, in the IN (or, $\beta\rightarrow 0$) limit, as,
\begin{equation}
    \lim_{\beta\rightarrow 0}\chi_\mathfrak{p}^{AS(0)} = \lim_{\beta\rightarrow 0}\frac{1}{\cancel{\beta}} \frac{\partial}{\partial \mathfrak{p}}\underbrace{\frac{\cancel{\beta}(2\mathfrak{r} + \mathfrak{p})}{2}}_{=\mu_g^{AS(1)}} = \frac{1}{2}.
\end{equation}
The \textit{first} and \textit{higher} order terms in the Taylor expansion of $\chi_\mathfrak{p}^{AS}$ (see, Eq.~(\ref{eq58punishment})) about $\beta$, in the IN limit, vanishes. The punishment susceptibility $\chi_{\mathfrak{p}}^{AS}$ is always positive, which indicates that the rate of turnover from \textit{Defectors} (with strategy $\mathbb{D}$) to \textit{Cooperators} (with strategy $\mathbb{C}$) is more than vice-versa. This implies that for increasing punishment $\mathfrak{p}$, there is always a greater fraction of players shifting to $\mathbb{C}$-strategy than to $\mathbb{D}$-strategy. The results of the game susceptibility via NEM and AS will be discussed in Sec.~\ref{sus-pgg-analysis}, where we will also compare the results obtained via DS and ABM.

\subsubsection{\underline{DS}}
For calculating the reward, cost and punishment susceptibilities, using the \textit{Darwinian selection} method, we write the game magnetization $\mu_g^{DS} = \langle \hat{\mathcal{M}_z}^{(1)}\rangle_\beta$ in terms of $\mathfrak{r}$, $\mathfrak{t}$ and $\mathfrak{p}$. From Eqs.~(\ref{eq38}, \ref{eq54}), the DS game magnetization in terms of the game parameters is,
\begin{equation}
    \mu_g^{DS} = \tanh{\bigg[\dfrac{\beta}{4}({2\mathfrak{r}-\mathfrak{t}+2\mathfrak{p}})\bigg]}.
    \label{eq59}
\end{equation}
This is identical to the NEM expression (see, Eq.~(\ref{eq55})), and hence they have the same characteristics in the ZN and the IN limits. From Eq.~(\ref{eq59}), we have the reward susceptibility as,
\begin{gather}
    \chi_{\mathfrak{r}}^{DS} =\dfrac{1}{\beta}\dfrac{\partial}{\partial \mathfrak{r}} \mu_g^{DS} = \dfrac{1}{2}\text{sech}^2\bigg[\dfrac{\beta}{4}(2\mathfrak{r}-\mathfrak{t}+2\mathfrak{p})\bigg].
    \label{eq60}
\end{gather}
This is also the same expression obtained in the case of NEM (see, Eq.~(\ref{eq56})), and hence they have the same characteristics in the ZN and the IN limits. Similarly, from Eq.~(\ref{eq59}), we have the cost susceptibility as,
\begin{gather}
    \chi_{\mathfrak{t}}^{DS} = \dfrac{1}{\beta}\dfrac{\partial}{\partial\mathfrak{t}}\mu_g^{DS} =-\dfrac{1}{4}\text{sech}^2\bigg[\dfrac{\beta}{4}(2\mathfrak{r}-\mathfrak{t}+2\mathfrak{p})\bigg].
    \label{eqcostpggdem}
\end{gather}
which is also the same expression obtained for cost susceptibility in the case of NEM (see, Eq.~(\ref{eqcostpggnem})), and hence they have the same characteristics in the ZN and the IN limits. 

\begin{figure*}[!ht]
    \centering
    \begin{subfigure}[b]{0.86\columnwidth}
        \centering
        \includegraphics[width = \textwidth]{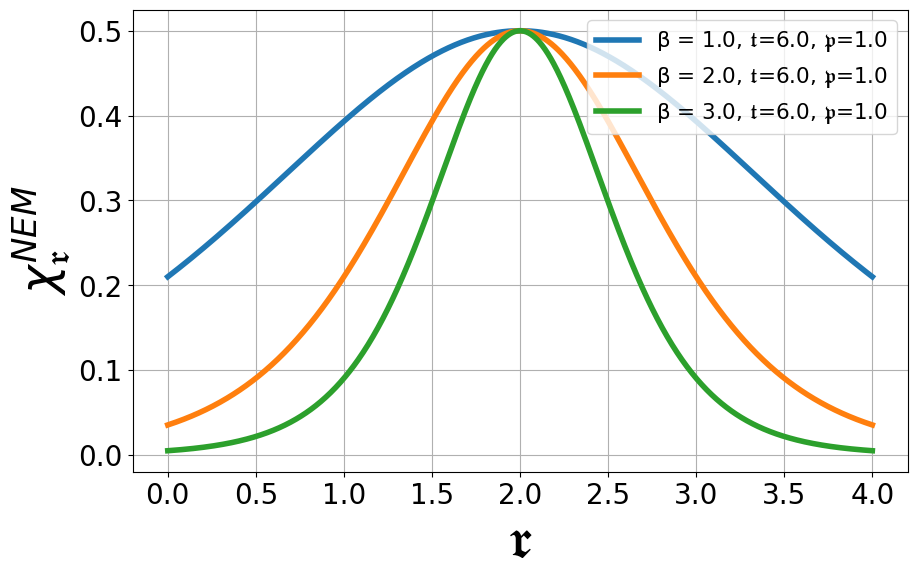}
        \caption{$\chi_\mathfrak{r}^{NEM}$ vs $\mathfrak{r}$}
        \label{fig3a}
    \end{subfigure}
    \begin{subfigure}[b]{0.86\columnwidth}
        \centering
        \includegraphics[width=\textwidth, height=4.7cm]{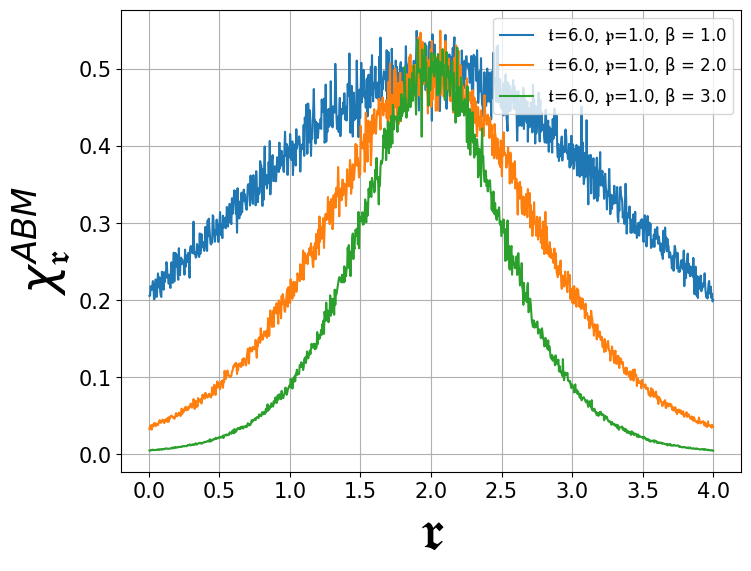}
        \caption{$\chi_\mathfrak{r}^{ABM}$ vs $\mathfrak{r}$}
        \label{fig3b}
    \end{subfigure}
    \begin{subfigure}[b]{0.86\columnwidth}
        \centering
        \includegraphics[width = \textwidth]{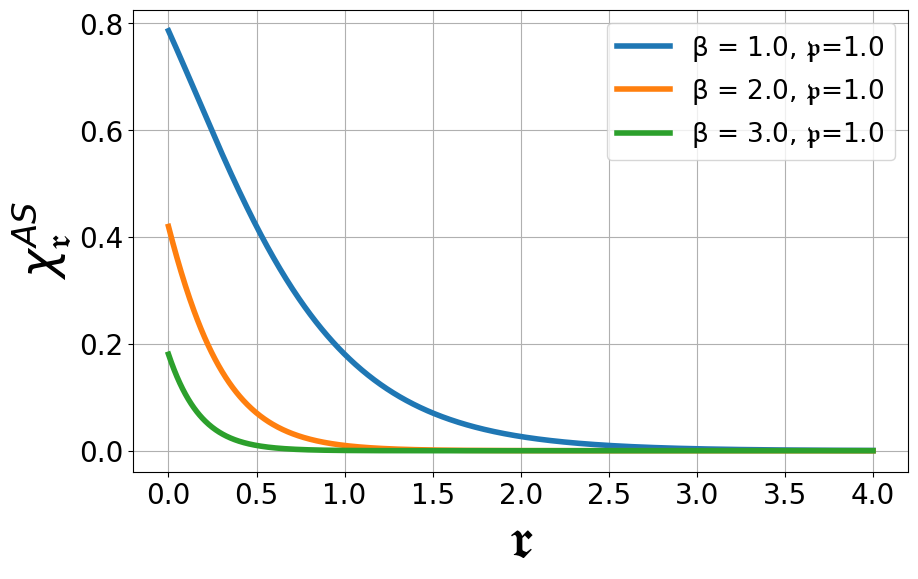}
        \caption{$\chi_\mathfrak{r}^{AS}$ vs $\mathfrak{r}$}
        \label{fig3c}
    \end{subfigure}
    \begin{subfigure}[b]{0.86\columnwidth}
        \centering
        \includegraphics[width = \textwidth]{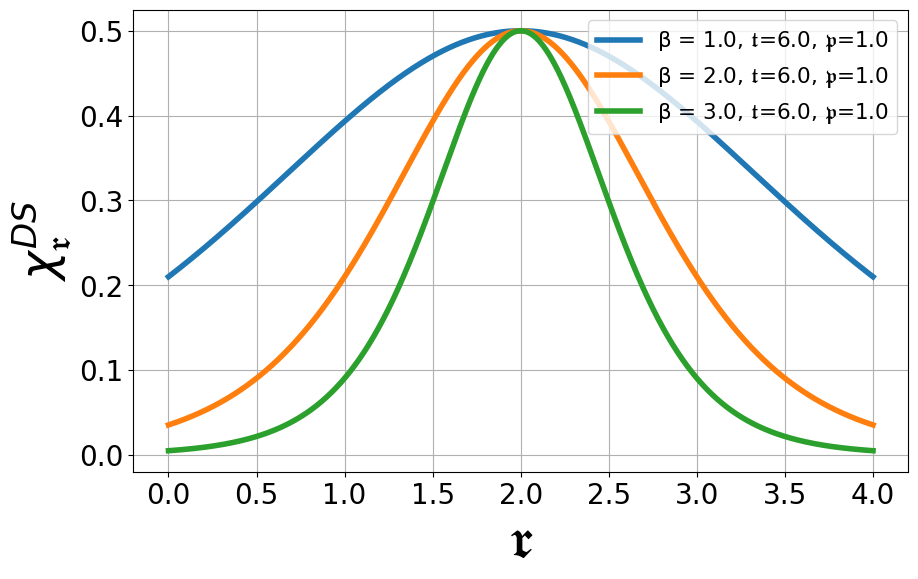}
        \caption{$\chi_\mathfrak{r}^{DS}$ vs $\mathfrak{r}$}
        \label{fig3d}
    \end{subfigure}
    \caption{\centering{\textbf{Reward susceptibility} $\chi_\mathfrak{r}$ vs \textbf{reward} $\mathfrak{r}$ for \textbf{cost} $\mathfrak{t}=6.0$ and \textbf{punishment} $\mathfrak{p}=1.0$ via NEM, AS, DS and ABM in PGG. { Cooperation (C) becomes the dominant strategy for the condition $2\mathfrak{r} >
(\mathfrak{t} - 2\mathfrak{p})$, with NEM, DS and ABM agreeing on the Nash equilibrium at $2\mathfrak{r}=\mathfrak{t}-2\mathfrak{p}$, while AS is an outlier.}}}
    \label{fig:3}
\end{figure*}

From Eq.~(\ref{eq59}), we have the punishment susceptibility as,
\begin{gather}
    \chi_{\mathfrak{p}}^{DS} = \dfrac{1}{\beta}\dfrac{\partial}{\partial \mathfrak{p}} \mu_g^{DS} = \dfrac{1}{2}\text{sech}^2\bigg[\dfrac{\beta}{4}(2\mathfrak{r}-\mathfrak{t}+2\mathfrak{p})\bigg].
    \label{eqpunishpggdem}
\end{gather}
which is also the same expression obtained for punishment susceptibility in the case of NEM (see, Eq.~(\ref{eqpunishpgg})), and hence they have the same characteristics in the ZN and the IN limits. In both finite and limiting values of $\beta$, the results from DS and AS were significantly different (see, Figs.~\ref{fig:3} and \ref{fig3punishment}), where we see a discrepancy in the Nash equilibrium predicted by DS and AS. For PGG, the game susceptibility expressions obtained via NEM and DS were identical, since, in the thermodynamic limit, the average magnetization of the player of interest in DS matches the average magnetization of all players in NEM. In AS, we do not even have an analytical expression for the cost susceptibility since the game magnetization $\mu_g^{AS}$ (see, Eq.~(\ref{eq57})) is independent of the cost parameter $\mathfrak{t}$. This inconsistency in the results of DS and AS will be discussed in Sec.~\ref{sus-pgg-analysis}.

\subsubsection{\underline{ABM}}
In Eq.~(\ref{eq13a}), we have shown that the magnetic susceptibility for the $1D$-Ising chain is related to the variance of the magnetization. Since there exists a one-to-one mapping between the Ising parameters $(\mathcal{J}, \mathfrak{h})$ and the game parameters $(\mathcal{A, B, C, D})$, one can determine the game susceptibility from the variance of the game magnetization $\mu_g$. For PGG, the payoffs $(\mathcal{A, B, C,D})$ associated to the game fulfil the criterion: $\mathcal{A+D=B+C}$, where $\mathcal{A}= 2\mathfrak{r},~\mathcal{B}=\mathfrak{r} - \mathfrak{t}/2,~\mathcal{C}=-\mathfrak{r} + \mathfrak{t}/2 - \mathfrak{p},~\mathcal{D}=-\mathfrak{p}$. So, $(\mathcal{J}, \mathfrak{h})$ in terms of the payoffs are,
\begin{gather}
    \mathcal{J} = \bigg[\dfrac{\mathcal{(A-C)-(B-D)}}{4}\bigg] = 0,\nonumber\\
    \mathfrak{h} = \bigg[\dfrac{\mathcal{(A-C)+(B-D)}}{4}\bigg] = \dfrac{2\mathfrak{r}+ 2\mathfrak{p}-\mathfrak{t}}{4}.
    \label{neweq3}
\end{gather}
From Eq.~(\ref{neweq3}), we can write the Hamiltonian $H$ in terms of the payoffs as,
\begin{gather}
    H = -\mathcal{J} \sum_{i=1}^{N} \mathfrak{s}_i \mathfrak{s}_{i+1} - \mathfrak{h}\sum_{i=1}^{N} \mathfrak{s}_i,\nonumber\\
    \text{or,}~H = 0 - \dfrac{(2\mathfrak{r}-\mathfrak{t} + 2\mathfrak{p})}{4}\sum_{i=1}^{N} \mathfrak{s}_i\nonumber = - \dfrac{(2\mathfrak{r}-\mathfrak{t} + 2\mathfrak{p})}{4} \Tilde{\mu}_g\nonumber\\
    \text{So,}~~\zeta  = \sum_{\{\mathfrak{s}_i\}} e^{-\beta H(\{\mathfrak{s}_i\})} = \sum_{\{\mathfrak{s}_i\}} e^{\beta [ \frac{(2\mathfrak{r}-\mathfrak{t} + 2\mathfrak{p})}{4}\sum_{i=1}^{N}\mathfrak{s}_i]}, \nonumber\\
    \dfrac{\partial \zeta}{\partial \mathfrak{r}} = \sum_{\{\mathfrak{s}_i\}} e^{-\beta H(\{\mathfrak{s}_i\})} ~\frac{\beta}{2} \Tilde{\mu}_g,~\dfrac{\partial \zeta}{\partial \mathfrak{t}} = -\sum_{\{\mathfrak{s}_i\}} e^{-\beta H(\{\mathfrak{s}_i\})} ~\frac{\beta}{4}\Tilde{\mu}_g .\nonumber
\end{gather}
\begin{gather}
    \text{or,} ~\chi_{\mathfrak{r}} = \dfrac{1}{\beta}\dfrac{\partial \mu_g}{\partial \mathfrak{r}} = \dfrac{1}{\beta}\dfrac{\partial \langle \Tilde{\mu}_g \rangle}{\partial \mathfrak{r}} = \dfrac{\partial}{\beta\partial \mathfrak{r}} \bigg[ \dfrac{1}{\zeta}\sum_{\{\mathfrak{s}_i\}} \Tilde{\mu}_g~ e^{-\beta H(\{\mathfrak{s}_i\})}  \bigg], \nonumber\\
    = -\dfrac{1}{\beta\zeta^2} \dfrac{\partial \zeta}{\partial \mathfrak{r}}\sum_{\{\mathfrak{s}_i\}} \Tilde{\mu}_g~ e^{-\beta H(\{\mathfrak{s}_i\})} + \dfrac{1}{\zeta} \sum_{\{\mathfrak{s}_i\}} \dfrac{1}{2}~\Tilde{\mu}_g^2~e^{-\beta H(\{\mathfrak{s}_i\})},  \nonumber\\
    = \bigg[-\dfrac{1}{\zeta^2} \sum_{\{\mathfrak{s}_i\}} e^{-\beta H(\{\mathfrak{s}_i\})} ~\dfrac{1}{2}~ \Tilde{\mu}_g \bigg]\sum_{\{\mathfrak{s}_i\}} \Tilde{\mu}_g~ e^{-\beta H(\{\mathfrak{s}_i\})} \nonumber\\
    + \dfrac{1}{\zeta} \sum_{\{\mathfrak{s}_i\}} \dfrac{1}{2}~\Tilde{\mu}_g^2~e^{-\beta H(\{\mathfrak{s}_i\})},\nonumber\\
    = -\dfrac{1}{2}\underbrace{\bigg[\dfrac{1}{\zeta} \sum_{\{\mathfrak{s}_i\}} \Tilde{\mu}_g~ e^{-\beta H(\{\mathfrak{s}_i\})}\bigg]^2}_{\langle \Tilde{\mu}_g \rangle^2} + \dfrac{1}{2} \underbrace{\bigg[ \dfrac{1}{\zeta} \sum_{\{\mathfrak{s}_i\}} \Tilde{\mu}_g^2~e^{-\beta H(\{\mathfrak{s}_i\})}\bigg]}_{\langle \Tilde{\mu}_g^2 \rangle}\nonumber\\
    \text{or,}~~ \chi_{\mathfrak{r}} = \dfrac{1}{\beta}\dfrac{\partial \mu_g}{\partial \mathfrak{r}} = \dfrac{1}{2} [  \langle \Tilde{\mu}_g^2 \rangle - \langle \Tilde{\mu}_g \rangle^2].
    \label{neweq4}
\end{gather}
As $\Tilde{\mu}_g = \mu_g^{ABM}$, we have,
\begin{equation}
    \chi_{\mathfrak{r}}^{ABM} = \dfrac{1}{2}[\langle (\mu_g^{ABM})^2 \rangle - \langle \mu_g^{ABM} \rangle^2],
    \label{eq61}
\end{equation}
where, for a particular configuration, $\mu_g^{ABM}$ denotes the game magnetization obtained for the system, each time the conditional loop runs for a particular value of $(\mathfrak{r}, \beta)$. 

Following the same formalism, for the cost susceptibility $\chi_{\mathfrak{t}}^{ABM}$, we have,
\begin{equation}
    \chi_{\mathfrak{t}}^{ABM} =  -\dfrac{1}{4}[\langle (\mu_g^{ABM})^2 \rangle - \langle \mu_g^{ABM} \rangle^2],
    \label{eq61cost}
\end{equation}
where, for a particular configuration, $\mu_g^{ABM}$ denotes the game magnetization obtained for the system, each time the conditional loop runs for a particular value of $(\mathfrak{t}, \beta)$. Similarly, for the punishment susceptibility $\chi_{\mathfrak{p}}^{ABM}$, we have,
\begin{equation}
    \chi_{\mathfrak{p}}^{ABM} =  \dfrac{1}{2}[\langle (\mu_g^{ABM})^2 \rangle - \langle \mu_g^{ABM} \rangle^2],
    \label{eq61punish}
\end{equation}
where, for a particular configuration, $\mu_g^{ABM}$ denotes the game magnetization obtained for the system, each time the conditional loop runs for a particular value of $(\mathfrak{p}, \beta)$. The algorithm used in ABM is described in Sec.~\ref{sub-abm}, where we consider the Energy matrix $\Delta = -\Lambda_1$ (or, $-ve$ of payoff matrix, see, Eq.~(\ref{eq54})). So,
\begin{equation}
    \Delta = \begin{bmatrix}
            -2\mathfrak{r} & -\mathfrak{r}+\frac{\mathfrak{t}}{2}\\
            -\mathfrak{r} - \frac{\mathfrak{t}}{2} +\mathfrak{p} & \mathfrak{p} 
        \end{bmatrix}.
        \label{eq62}
\end{equation}
When $T\rightarrow 0$ (or, $\beta \rightarrow \infty$), i.e., ZN limit, $\chi_{\mathfrak{r}}^{ABM} \rightarrow 0$ when $2\mathfrak{r}\neq (\mathfrak{t} - 2\mathfrak{p})$, i.e., the rate of turnover from \textit{Defectors} to \textit{Cooperators} and vice-versa vanishes. When $T\rightarrow \infty$ (or, $\beta \rightarrow 0$), i.e. IN limit, $\chi_{\mathfrak{r}}^{ABM} \rightarrow \frac{1}{2}$, i.e., when the randomness in strategy selection is maximum, in the IN limit, the rate of turnover from defectors to cooperators and vice-versa averages out to $\frac{1}{2}$. In the IN limit, $\langle \mu_g^{ABM} \rangle \rightarrow 0$, since the players choose their strategies randomly, leading to an equiprobable selection of $\mathbb{C}$ and $\mathbb{D}$-strategies. However, in the same $\beta \rightarrow 0$ limit, $\langle (\mu_g^{ABM})^2 \rangle \rightarrow 1$, and this leads to a value of $\frac{1}{2}$ for the reward susceptibility (from Eq.~(\ref{eq61})), in the IN limit. The reward susceptibility $\chi_{\mathfrak{r}}^{ABM}$ is always positive, which indicates that the rate of switching from \textit{Defectors} (with strategy $\mathbb{D}$) to \textit{Cooperators} (with strategy $\mathbb{C}$) is greater than vice-versa. This implies that for increasing reward $\mathfrak{r}$, the rate at which players switch to $\mathbb{C}$-strategy is always greater than the rate at which players switch to $\mathbb{D}$-strategy. 

Similarly, for $\chi_{\mathfrak{t}}^{ABM}$, when $T\rightarrow 0$ (or, $\beta \rightarrow \infty$), i.e., ZN limit, $\chi_{\mathfrak{t}}^{ABM} \rightarrow 0$ when $\mathfrak{t}\neq (2\mathfrak{r} + 2\mathfrak{p})$, i.e., in this case, the rate of switching strategies from \textit{Cooperate} to \textit{Defect} and vice-versa vanishes. When $T\rightarrow \infty$ (or, $\beta \rightarrow 0$), i.e. IN limit, $\chi_{\mathfrak{t}}^{ABM} \rightarrow -\frac{1}{4},$ due to maximum randomness in strategy selection. In the IN limit, $\langle \mu_g^{ABM} \rangle \rightarrow 0$, since the players choose their strategies randomly, leading to an equiprobable selection of $\mathbb{C}$ and $\mathbb{D}$-strategies. However, in the same $\beta \rightarrow 0$ limit, $\langle (\mu_g^{ABM})^2 \rangle \rightarrow 1$, and this leads to a value of $-\frac{1}{4}$ for the cost susceptibility (from Eq.~(\ref{eq61cost})), in the IN limit. The cost susceptibility $\chi_{\mathfrak{t}}^{ABM}$ is always negative, which indicates that the rate of switching strategies from $\mathbb{C}$ to $\mathbb{D}$ is greater than vice-versa. This implies that for increasing cost $\mathfrak{t}$, the rate at which players switch to $\mathbb{C}$-strategy is always lesser than the rate at which players switch to $\mathbb{D}$-strategy. 

Finally, for $\chi_{\mathfrak{p}}^{ABM}$, when $T\rightarrow 0$ (or, $\beta \rightarrow \infty$), i.e., ZN limit, $\chi_{\mathfrak{p}}^{ABM} \rightarrow 0$ when $2\mathfrak{p}\neq (\mathfrak{t} - 2\mathfrak{r})$, i.e., in this case, the rate of turnover from \textit{Defectors} to \textit{Cooperators} and vice-versa vanishes. When $T\rightarrow \infty$ (or, $\beta \rightarrow 0$), i.e. IN limit, $\chi_{\mathfrak{p}}^{ABM} \rightarrow \frac{1}{2},$ due to maximum randomness in strategy selection, as in the IN limit, the rate of turnover from defectors to cooperators and vice-versa averages out to $\frac{1}{2}$. In the IN limit, $\langle \mu_g^{ABM} \rangle \rightarrow 0$, since the players choose their strategies randomly, leading to an equiprobable selection of $\mathbb{C}$ and $\mathbb{D}$-strategies. However, in the same $\beta \rightarrow 0$ limit, $\langle (\mu_g^{ABM})^2 \rangle \rightarrow 1$, and this leads to a value of $\frac{1}{2}$ for the punishment susceptibility (from Eq.~(\ref{eq61punish})), in the IN limit. Here also, the punishment susceptibility $\chi_{\mathfrak{p}}^{ABM}$ is always positive, which indicates that the rate of switching from Defectors (with strategy $\mathbb{D}$) to Cooperators (with strategy $\mathbb{C}$) is greater than vice-versa. This implies that for increasing punishment $\mathfrak{p}$, the rate at which players switch from $\mathbb{D}$ to $\mathbb{C}$-strategy is always greater than vice-versa. 

\begin{figure*}
    \centering
    \begin{subfigure}[b]{0.66\columnwidth}
        \centering
        \includegraphics[width = \textwidth]{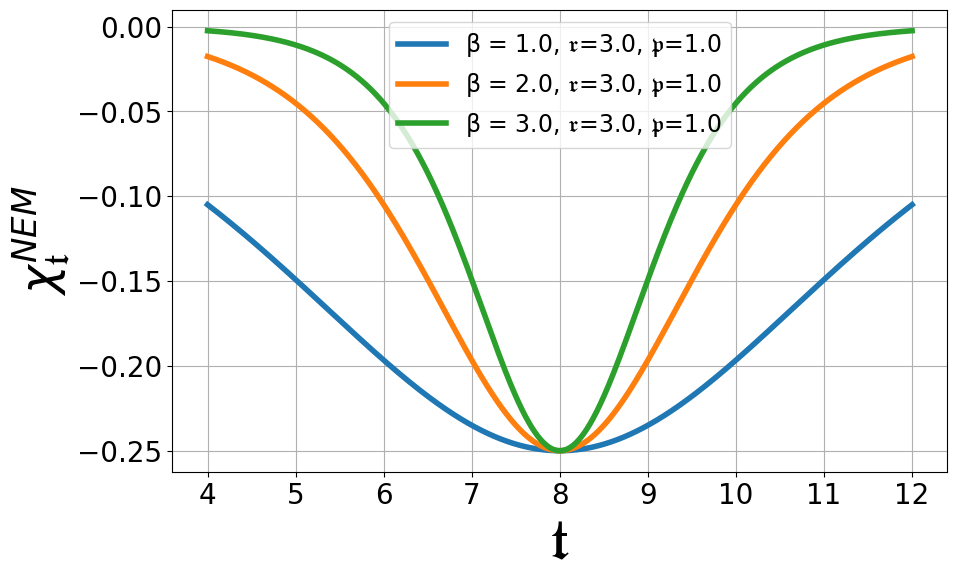}
        \caption{$\chi_\mathfrak{t}^{NEM}$ vs $\mathfrak{t}$}
        \label{fig3anew}
    \end{subfigure}
    \begin{subfigure}[b]{0.7\columnwidth}
        \centering
        \includegraphics[width=5.8cm,height=3.52cm]{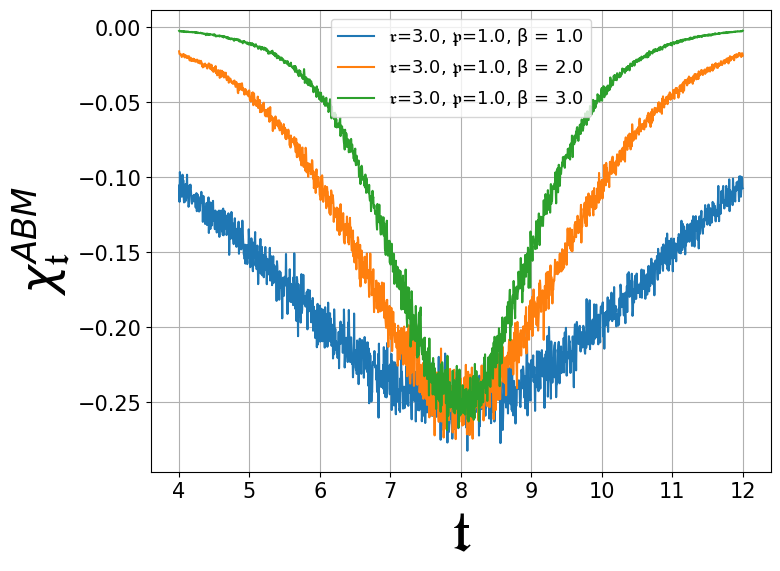}
        \caption{$\chi_\mathfrak{t}^{ABM}$ vs $\mathfrak{t}$}
        \label{fig3bnew}
    \end{subfigure}
    \begin{subfigure}[b]{0.66\columnwidth}
        \centering
        \includegraphics[width = \textwidth]{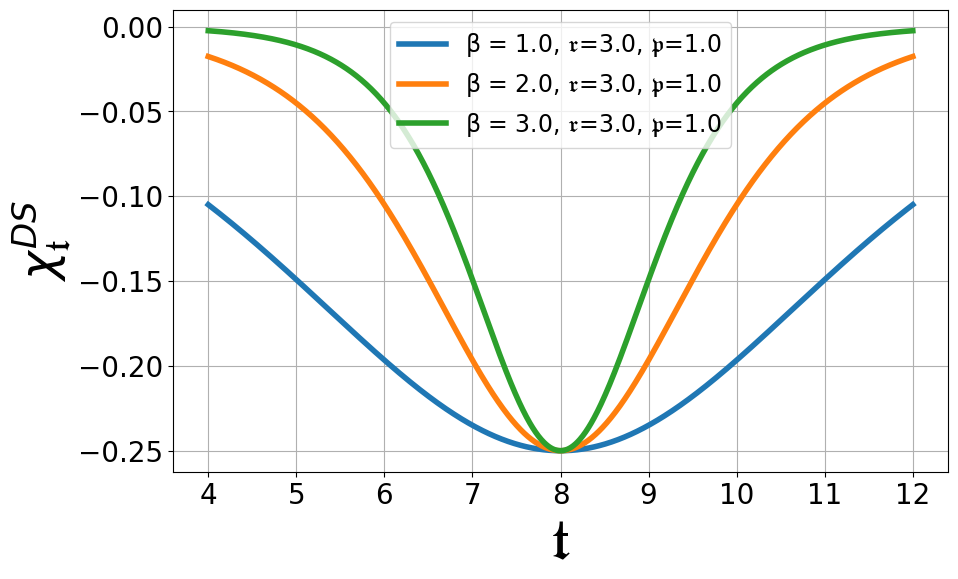}
        \caption{$\chi_\mathfrak{t}^{DS}$ vs $\mathfrak{t}$}
        \label{fig3cnew}
    \end{subfigure}
    \caption{\centering{\textbf{Cost susceptibility} $\chi_\mathfrak{t}$ vs \textbf{cost} $\mathfrak{t}$ for \textbf{reward} $\mathfrak{r}=3.0$ and \textbf{punishment} $\mathfrak{p}=1.0$ via NEM, DS and ABM in PG G. { For $ t < (2r + 2p)$, even though Cooperation remains the dominant strategy,
we observe that for increasing cost $t$, the rate of shifting to Defection always exceeds the rate of shifting to Cooperation. Cost susceptibility for AS model vanishes for all cost values, therefore its not shown. }}}
    \label{fig:3new}
\end{figure*} 

\begin{figure*}
    \centering
    \begin{subfigure}[b]{0.96\columnwidth}
        \centering
        \includegraphics[width = \textwidth]{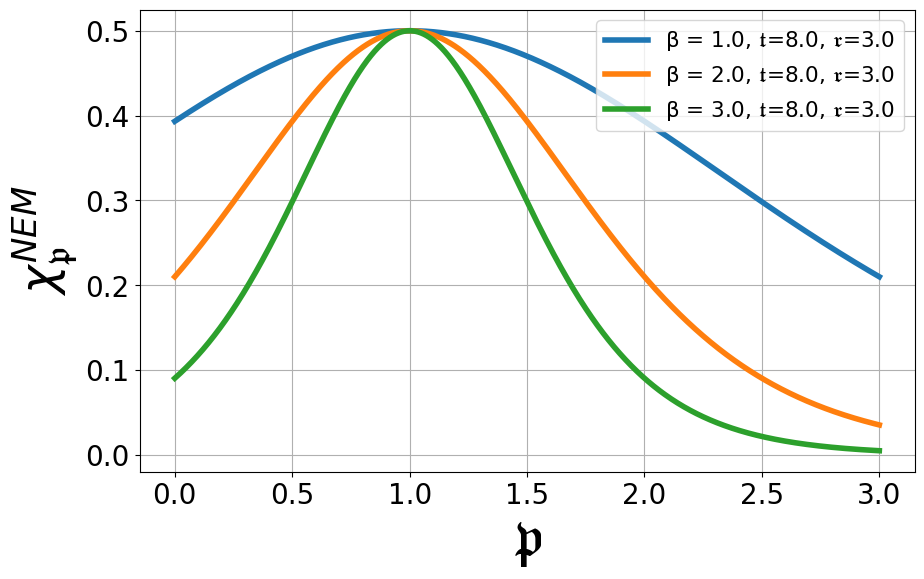}
        \caption{$\chi_\mathfrak{p}^{NEM}$ vs $\mathfrak{p}$}
        \label{fig3a-p}
    \end{subfigure}
    \begin{subfigure}[b]{0.96\columnwidth}
        \centering
        \includegraphics[width=\textwidth, height=5.15cm]{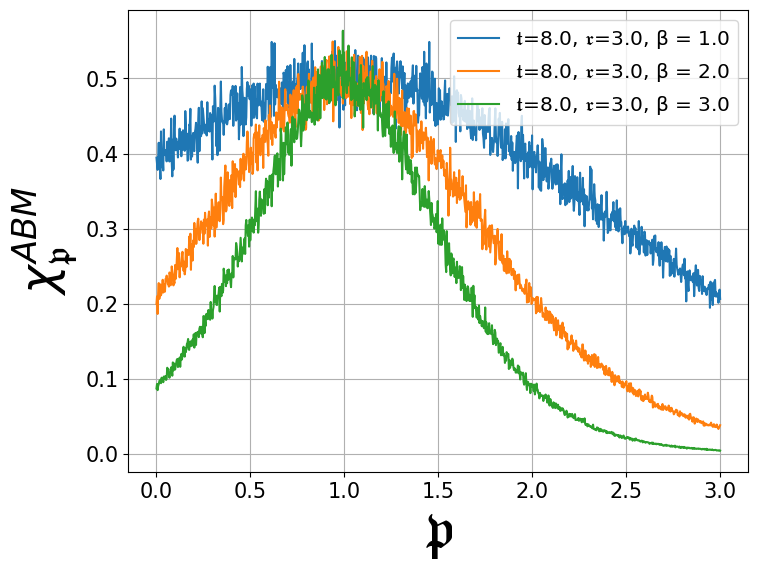}
        \caption{$\chi_\mathfrak{p}^{ABM}$ vs $\mathfrak{p}$}
        \label{fig3b-p}
    \end{subfigure}
    \begin{subfigure}[b]{0.96\columnwidth}
        \centering
        \includegraphics[width = \textwidth]{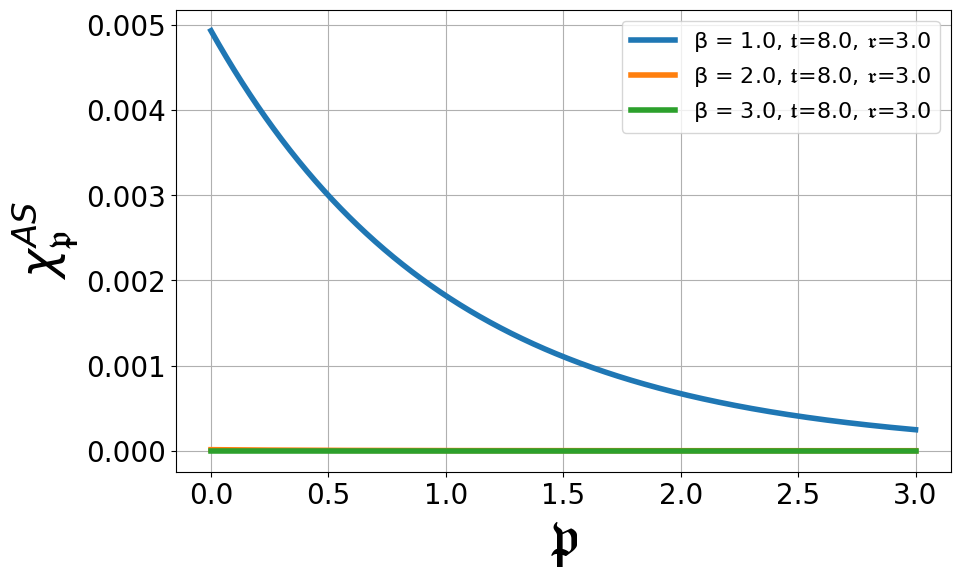}
        \caption{$\chi_\mathfrak{p}^{AS}$ vs $\mathfrak{p}$}
        \label{fig3c-p}
    \end{subfigure}
    \begin{subfigure}[b]{0.96\columnwidth}
        \centering
        \includegraphics[width = \textwidth]{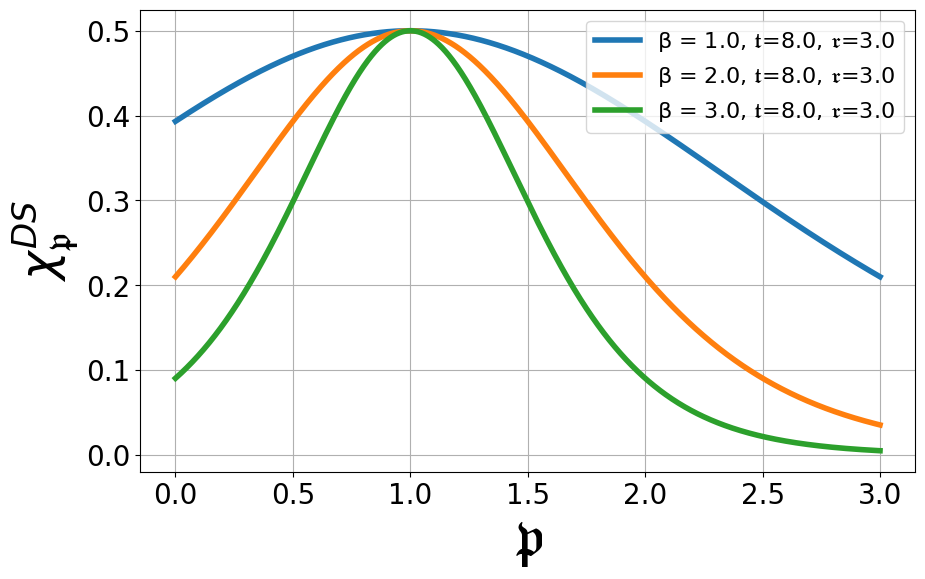}
        \caption{$\chi_\mathfrak{p}^{DS}$ vs $\mathfrak{p}$}
        \label{fig3d-p}
    \end{subfigure}
    \caption{\centering{\textbf{Punishment susceptibility} $\chi_\mathfrak{p}$ vs \textbf{punishment} $\mathfrak{p}$ for \textbf{cost} $\mathfrak{t}=8.0$ and \textbf{reward} $\mathfrak{r}=3.0$ via NEM, AS, DS and ABM in PGG. { For $2\mathfrak{p}<(\mathfrak{t}-2\mathfrak{r})$, even though \textit{Defection} remains  dominant, for increasing punishment $\mathfrak{p}$, the rate of shifting to Cooperation always exceeds the rate of shifting to Defection for all models apart from AS. }}}
    \label{fig3punishment}
\end{figure*} 
\subsubsection{\label{sus-pgg-analysis}Analysis of game susceptibility for PGG}
Here, we will discuss the results obtained for the \textit{reward, cost} and \textit{punishment susceptibilities} via NEM, AS, DS and ABM, respectively. For $\chi_{\mathfrak{r}}$, as shown in Fig.~\ref{fig:3}, we observe that for all four models, Cooperation $(\mathbb{C})$ becomes the dominant strategy for the condition $2\mathfrak{r}>(\mathfrak{t}-2\mathfrak{p})$, but the switch from \textit{Defection} to \textit{Cooperation} is always greater than vice-versa $\forall~\mathfrak{r}$. For $2\mathfrak{r}<(\mathfrak{t}-2\mathfrak{p})$, even though \textit{Defection} remains the dominant strategy, we observe that for increasing reward $\mathfrak{r}$, the rate at which players switch to $\mathbb{C}$-strategy always exceeds the rate at which players switch to $\mathbb{D}$-strategy. For increasing values of $\mathfrak{r}$, we observe an initial increase in the rate of change of strategies from $\mathbb{D}$ to $\mathbb{C}$ when $\mathfrak{r}$ approaches $\frac{1}{2}(\mathfrak{t} - 2\mathfrak{p})$, reaching a maximum at $2\mathfrak{r}=\mathfrak{t}-2\mathfrak{p}$, and then the rate again decreases as $2\mathfrak{r}$ becomes greater than $(\mathfrak{t} - 2\mathfrak{p})$. The results obtained via NEM, DS and ABM follow a very similar trend wherein all have the same inflexion point at $2\mathfrak{r}=\mathfrak{t}-2\mathfrak{p}$, which indicates the \textit{Nash equilibria}, whereas, AS gives incorrect result where it depicts the inflexion point at $\mathfrak{r}\rightarrow 0$.

For $\chi_{\mathfrak{t}}$, we have the results from NEM, DS and ABM, since in the case of AS, $\mu_g^{AS}$ had no $\mathfrak{t}$-dependency, so $\chi_{\mathfrak{t}}^{AS} = 0,~\forall~\mathfrak{t}$, which is clearly an incorrect result. As shown in Fig.~\ref{fig:3new}, we observe that for all three models, Defection $(\mathbb{D})$ becomes the dominant strategy for the condition $\mathfrak{t}>(2\mathfrak{r}+2\mathfrak{p})$, but the rate of shifting from $\mathbb{C}$ to $\mathbb{D}$ is greater than vice-versa $\forall~\mathfrak{t}$. For $\mathfrak{t}<(2\mathfrak{r}+2\mathfrak{p})$, even though \textit{Cooperation} remains the dominant strategy, we observe that for increasing cost $\mathfrak{t}$, the rate of shifting to $\mathbb{D}$-strategy always exceeds the rate of shifting to $\mathbb{C}$-strategy. However, for $\chi_{\mathfrak{t}}$, in NEM, DS and ABM, for increasing values of $\mathfrak{t}$, we observe an initial increase in the rate of change of strategies from $\mathbb{C}$ to $\mathbb{D}$ when $\mathfrak{t}$ approaches $2(\mathfrak{r} + \mathfrak{p})$, reaching a maximum at $\mathfrak{t}=2(\mathfrak{r}+\mathfrak{p})$, and then the rate again decreases as $\mathfrak{t}$ becomes greater than $2(\mathfrak{r} + \mathfrak{p})$. The results obtained via NEM, DS and ABM follow again a very similar trend wherein all have the same inflexion point at $\mathfrak{t}=2(\mathfrak{r}+\mathfrak{p})$, which indicates the \textit{Nash equilibria}. 

Finally, for $\chi_{\mathfrak{p}}$, as shown in Fig.~\ref{fig3punishment}, we observe that for all four models, Cooperation $(\mathbb{C})$ becomes the dominant strategy for the condition $2\mathfrak{p}>(\mathfrak{t}-2\mathfrak{r})$, but the rate of switching from $\mathbb{D}$ to $\mathbb{C}$ is greater than vice-versa $\forall~\mathfrak{p}$. For $2\mathfrak{p}<(\mathfrak{t}-2\mathfrak{r})$, even though \textit{Defection} remains the dominant strategy, we observe that for increasing punishment $\mathfrak{p}$, the rate of shifting to $\mathbb{C}$-strategy always exceeds the rate of shifting to $\mathbb{D}$-strategy. For increasing values of $\mathfrak{p}$, we observe an initial increase in the rate of change of strategies from $\mathbb{D}$ to $\mathbb{C}$ when $\mathfrak{p}$ approaches $\frac{1}{2}(\mathfrak{t} - 2\mathfrak{r})$, reaching a maximum at $2\mathfrak{p}=\mathfrak{t}-2\mathfrak{r}$, and then the rate again decreases as $2\mathfrak{p}$ becomes greater than $(\mathfrak{t} - 2\mathfrak{r})$. The results obtained via NEM, DS and ABM follow a very similar trend wherein all have the same inflexion point at $2\mathfrak{p}=\mathfrak{t}-2\mathfrak{r}$, which indicates the \textit{Nash equilibria}, whereas, AS gives incorrect result where it depicts the inflexion point at $\mathfrak{p}\rightarrow 0$.

For games, like PGG, where the payoffs $(\mathcal{A, B, C, D})$ fulfil the criterion: $\mathcal{A+D=B+C}$, the Nash equilibrium mapping coincides with the player of interest concept of DS.

\subsection{\label{corr-pgg}Correlation}
In this section, we will discuss the correlation for the PGG and its variation with game parameters $\mathfrak{r}$ and $\mathfrak{t}$. As mentioned earlier, we have discussed the different methods to calculate correlation in Sec.~\ref{theory}.

\subsubsection{\underline{NEM}}
Using \textit{Nash equilibrium mapping}, the expression for the correlation between the strategies of the players at $i^{th}$ and $(i+j)^{th}$ sites is given as,
\begin{equation}
    \mathfrak{c}_j^{NEM} = \langle \mathfrak{s}_i \mathfrak{s}_{i+j} \rangle = \cos^2 \varphi + \bigg(\dfrac{\Omega_{-}}{\Omega_{+}} \bigg)^j \sin^2 \varphi,
    \label{eq63}
\end{equation}
where, $j$ is the \textit{distance} from $i^{th}$ site and from Eqs.~(\ref{eq26}, \ref{eq27}), we have $(\cos^2\varphi,~ \sin^2\varphi,~\Omega_{\pm})$ in terms of the reward $\mathfrak{r}$, cost $\mathfrak{t}$ and punishment $\mathfrak{p}$,
\begin{gather}
    \cos^2\varphi = \dfrac{\mathfrak{T}-1}{\mathfrak{T}},~\Omega_{\pm} = \cosh\bigg[\dfrac{\beta(2\mathfrak{r}-\mathfrak{t}+2\mathfrak{p})}{4}\bigg] \pm \sqrt{\mathfrak{T}},\nonumber\\
    \text{where,}~ \mathfrak{T} = 1 + \sinh^2\bigg[\dfrac{\beta(2\mathfrak{r}-\mathfrak{t}+2\mathfrak{p})}{4}\bigg].
    \label{eq64}
\end{gather}
In PGG, when $T\rightarrow 0$ (or, $\beta \rightarrow \infty$), i.e., ZN limit, $\mathfrak{c}_j^{NEM} \rightarrow +1,$ for $2\mathfrak{r}\neq (\mathfrak{t} - 2\mathfrak{p})$ (from Eq.~(\ref{eq63})), indicating \textit{positive} correlation. This indicates that the strategies of all the players are the same (either all \textit{Cooperate} or all \textit{Defect}), and they are perfectly correlated. This can be verified by Taylor expanding, up to \textit{first}-order, the expression of $\mathfrak{c}_j^{NEM}$ in Eq.~(\ref{eq63}) about $\frac{1}{\beta}$. 

\begin{figure*}[!ht]
    \centering
    \begin{subfigure}[b]{\textwidth}
        \centering
        \includegraphics[width = 0.9\textwidth]{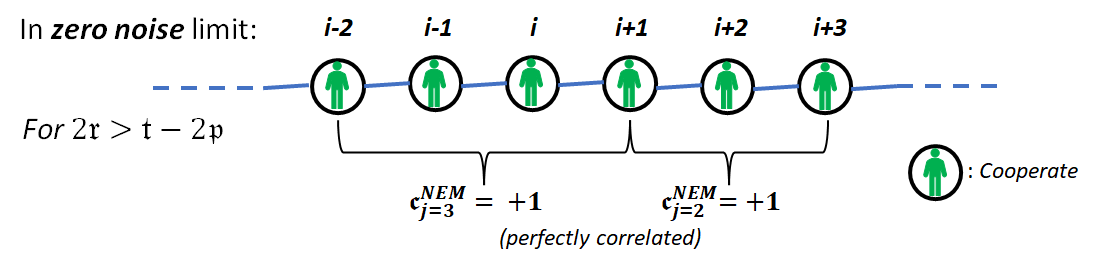}
    \label{fig:pgg-zero-coop}
    \end{subfigure}
    \begin{subfigure}[b]{\textwidth}
        \centering
        \includegraphics[width = 0.9\textwidth]{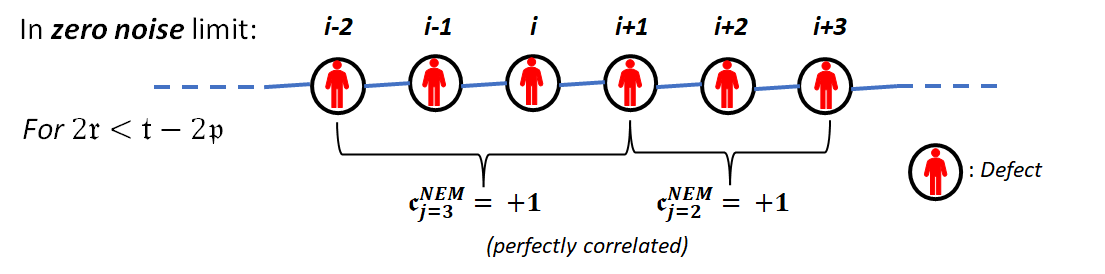}
    \label{fig:pgg-zero-defect}
    \end{subfigure}
    \caption{\centering{\textbf{NEM}: For PGG, in the ZN limit, if $2\mathfrak{r} < (\mathfrak{t}-2\mathfrak{p})$, all players choose to \textit{Defect}, whereas, if $2\mathfrak{r} > (\mathfrak{t}-2\mathfrak{p})$, then all players choose to \textit{Cooperate}.}}
    \label{pgg-zero}
\end{figure*}

An illustration of this case is shown in Fig.~\ref{pgg-zero}, where we see that in the ZN limit, for $2\mathfrak{r}<(\mathfrak{t}-2\mathfrak{p})$, all players choose to \textit{Defect}, owing to the relatively high cost and low reward associated with the game. The players do not modify their strategies, due to the absence of \textit{noise}, and hence, they show maximum correlation. Similarly, in the same ZN limit, for $2\mathfrak{r}>(\mathfrak{t}-2\mathfrak{p})$, all players choose to \textit{Cooperate}, owing to the relatively high reward now associated with the game. Here too, the players do not modify their strategies, due to the absence of \textit{noise}, and hence, they show maximum correlation. Both situations correspond to the best feasible payoff, i.e., \textit{Nash equilibrium}, in the ZN limit. When $T\rightarrow 0$ (or, $\beta \rightarrow \infty$), the \textit{two-player} pure Nash equilibrium, i.e., $(\mathbb{C}, \mathbb{C})$ for $2\mathfrak{r}>(\mathfrak{t}-2\mathfrak{p})$ and $(\mathbb{D}, \mathbb{D})$ for $2\mathfrak{r}<(\mathfrak{t}-2\mathfrak{p})$, becomes the Nash equilibrium for all the players in the thermodynamic limit, as illustrated in Fig.~\ref{pgg-zero}. When $T\rightarrow \infty$ (or, $\beta \rightarrow 0$), i.e., IN limit, $\mathfrak{c}_j^{NEM} \rightarrow 0$ since the players choose their strategies randomly, leading to the absence of any correlation.

\subsubsection{\underline{AS}}
Using \textit{Aggregate selection} method, the correlation (from Eq.~(\ref{eq35})) between strategies of the players at $i^{th}$ site and $(i+j)^{th}$ site, in terms of game parameters $\mathfrak{r},~\mathfrak{t},~\text{and}~\mathfrak{p}$, is,
\begin{equation}
    \mathfrak{c}_j^{AS} = \langle \hat{\mathcal{M}_z}^{(i)} \hat{\mathcal{M}_z}^{(i+j)} \rangle_{\beta} =\tanh^2\bigg[ \dfrac{\beta}{2}(2\mathfrak{r}+\mathfrak{p})\bigg].
    \label{eq65}
\end{equation}
From Eq.~(\ref{eq65}), we notice that $(2\mathfrak{r}+\mathfrak{p})$ is always greater than \textit{zero}, since $\mathfrak{r}>0~\text{and}~\mathfrak{p}>0$, and there is no $\mathfrak{t}$-dependency in $\mathfrak{c}_j^{AS}$, so $\mathfrak{c}_j^{AS}$ = \textit{constant}, $\forall~ \mathfrak{t}$. From Eq.~(\ref{eq65}), we also find that for $T\rightarrow 0$ (or, $\beta \rightarrow \infty$), i.e., ZN limit, $\mathfrak{c}_j^{AS} \rightarrow 1,~\forall~\mathfrak{r}$, indicating \textit{positive} correlation, whereas, for $T\rightarrow \infty$ (or, $\beta \rightarrow 0$), i.e., IN limit, $\mathfrak{c}_j^{AS} \rightarrow 0$, indicating absence of correlation as players choose their strategies randomly. However, the results obtained via AS are incorrect, and we discuss this in Sec.~\ref{corr-pgg-analysis}.

\subsubsection{\underline{DS}}
Using \textit{Darwinian selection} method, we have the correlation (from Eq.~(\ref{eq41})) between strategies of the players at $1^{st}$ site and $j^{th}$ site, in terms of game parameters $\mathfrak{r},~\mathfrak{t},~\text{and}~\mathfrak{p}$, as,
\begin{equation}
    \mathfrak{c}_j^{DS} = \langle \hat{\mathcal{M}_z}^{(1)} \hat{\mathcal{M}_z}^{(j)} \rangle_{\beta} =\tanh^2\bigg[ \dfrac{\beta}{2}(\mathfrak{r}-\mathfrak{t}/2+\mathfrak{p})\bigg],
    \label{eq66}
\end{equation}
where, $j$ is the \textit{distance} from $1^{st}$ site. The correlation $\mathfrak{c}_j^{DS}$ is independent of the distance $j$ since, in DS, the main objective is to maximize the payoffs of the players of interest (fixed at the $1^{st}$ and $j^{th}$ site). When $T\rightarrow 0$ (or, $\beta \rightarrow \infty$), i.e., ZN limit, $\mathfrak{c}_j^{DS}\rightarrow +1,$ for $2\mathfrak{r}\neq (\mathfrak{t} - 2\mathfrak{p})$, indicating \textit{positive} correlation. This indicates that the strategies of all the players are the same (either all \textit{Cooperate}, or all \textit{Defect}), and they are perfectly correlated. This can be verified by Taylor expanding, up to \textit{first}-order, the expression of $\mathfrak{c}_j^{DS}$ in Eq.~(\ref{eq66}) about $\frac{1}{\beta}$. When $T\rightarrow \infty$ (or, $\beta \rightarrow 0$), i.e., IN limit, $\mathfrak{c}_j^{DS}\rightarrow 0$ since the players opt for random strategy selection, resulting in absence of any correlation. 

\subsubsection{\underline{ABM}}
We have explained the algorithm used to determine the correlation using \textit{Agent-based Method} in Sec.~\ref{sub-abm}. We have checked for both the cases of ZN as well as IN, and we have discussed them in Sec.~\ref{corr-pgg-analysis}. Here, we take $\Delta = -\Lambda_1$ (see, Eq.~(\ref{eq54})). Thus,
\begin{equation}
    \Delta = \begin{bmatrix}
            -2\mathfrak{r} & -\mathfrak{r}+\frac{\mathfrak{t}}{2}\\
            -\mathfrak{r} - \frac{\mathfrak{t}}{2} +\mathfrak{p} & \mathfrak{p}
        \end{bmatrix}.
        \label{eq67}
\end{equation}
When $T\rightarrow 0$ (or, $\beta \rightarrow \infty$), i.e., ZN limit, $\mathfrak{c}_j^{ABM} \rightarrow +1$, for $2\mathfrak{r} \neq (\mathfrak{t} - 2\mathfrak{p})$, indicating \textit{positive} correlation. This indicates that the strategies of all the players are the same (either all \textit{Cooperate}, or all \textit{Defect}), and they are perfectly correlated. When $T\rightarrow \infty$ (or, $\beta \rightarrow 0$), i.e., IN limit, $\mathfrak{c}_j^{ABM} \rightarrow 0$ since the players opt for random strategy selection, resulting in absence of any correlation.
\begin{figure*}[!ht]
    \centering
    \begin{subfigure}[b]{0.9\columnwidth}
        \centering
        \includegraphics[width = \textwidth]{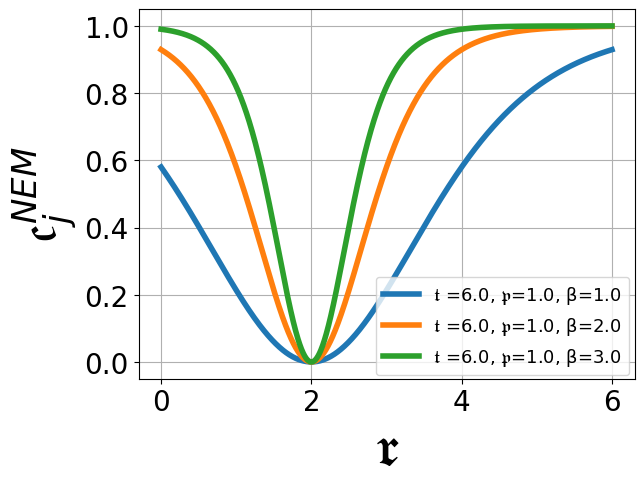}
        \caption{$\mathfrak{c}_j^{NEM}$ vs $\mathfrak{r}$}
        \label{fig4a}
    \end{subfigure}
    \begin{subfigure}[b]{0.9\columnwidth}
        \centering
        \includegraphics[width = \textwidth]{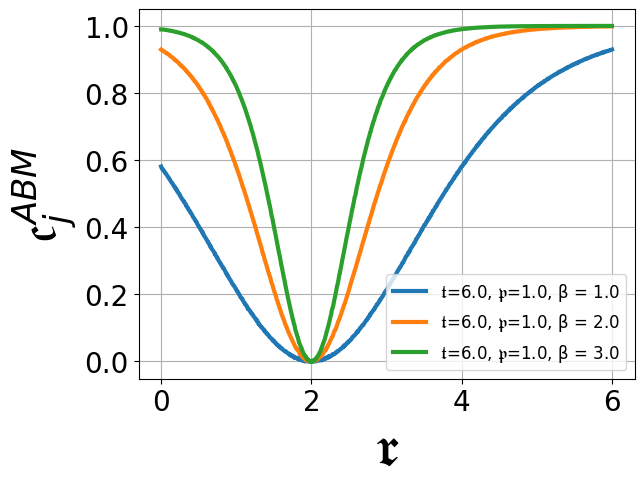}
        \caption{$\mathfrak{c}_j^{ABM}$ vs $\mathfrak{r}$}
        \label{fig4b}
    \end{subfigure}
    \begin{subfigure}[b]{0.9\columnwidth}
        \centering
        \includegraphics[width = \textwidth]{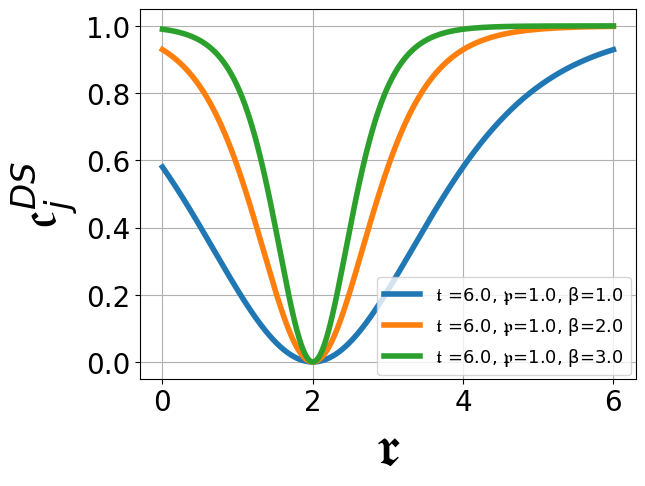}
        \caption{$\mathfrak{c}_j^{DS}$ vs $\mathfrak{r}$}
        \label{fig4c}
    \end{subfigure}
    \begin{subfigure}[b]{0.9\columnwidth}
        \centering
        \includegraphics[width = \textwidth]{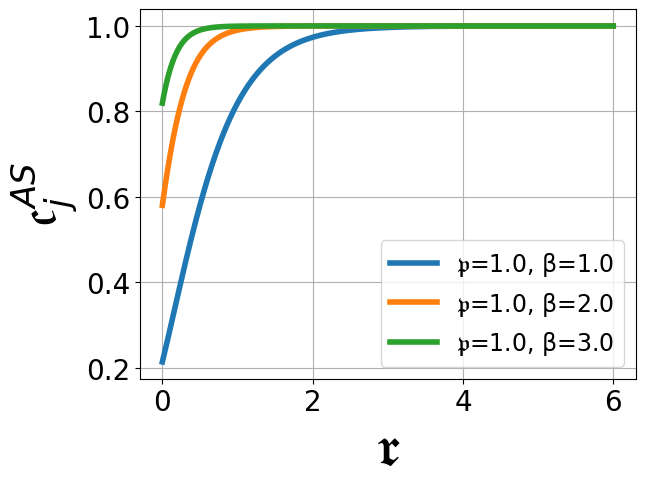}
        \caption{$\mathfrak{c}_j^{AS}$ vs $\mathfrak{r}$}
        \label{fig4d}
    \end{subfigure}
    \caption{\centering{\textbf{Correlation} $\mathfrak{c}_j$ vs \textbf{reward} $r$ for \textbf{cost} $t=6.0$ and \textbf{punishment} $p=1.0$ via NEM, AS, DS and ABM in PGG. {For $2r > t - 2p$, the Nash equilibria is (C, C), whereas, when $2r < t - 2p$, the Nash equilibria is (D, D), and again AS is the outlier.}}}
    \label{fig:4}
\end{figure*}

\begin{figure*}
    \centering
    \begin{subfigure}[b]{0.67\columnwidth}
        \centering
        \includegraphics[width = \textwidth]{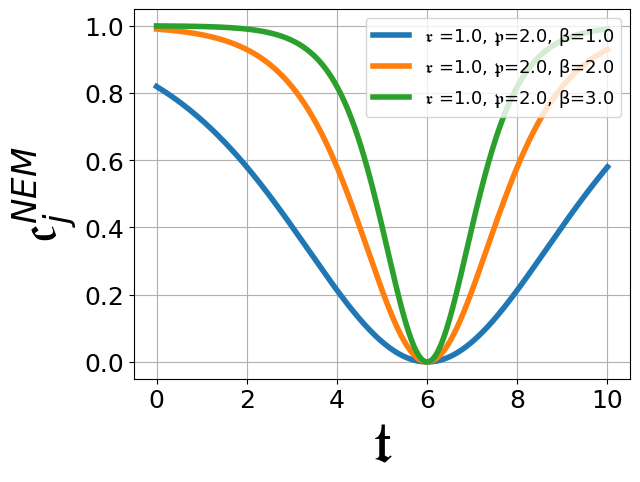}
        \caption{$\mathfrak{c}_j^{NEM}$ vs $t$}
        \label{fig5a}
    \end{subfigure}
    \begin{subfigure}[b]{0.67\columnwidth}
        \centering
        \includegraphics[width = \textwidth]{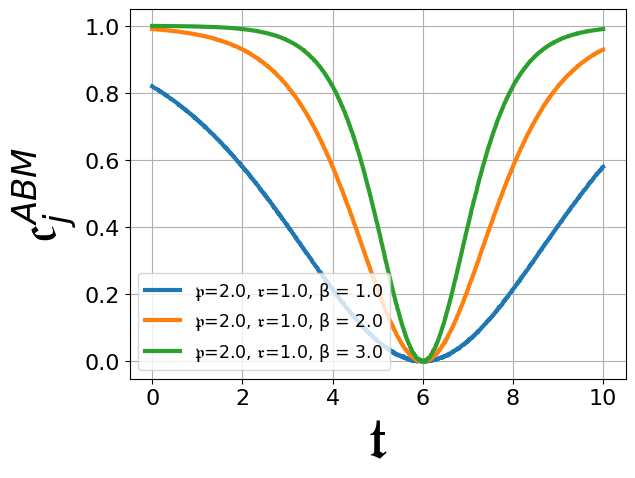}
        \caption{$\mathfrak{c}_j^{ABM}$ vs $t$}
        \label{fig5b}
    \end{subfigure}
    \begin{subfigure}[b]{0.67\columnwidth}
        \centering
        \includegraphics[width = \textwidth]{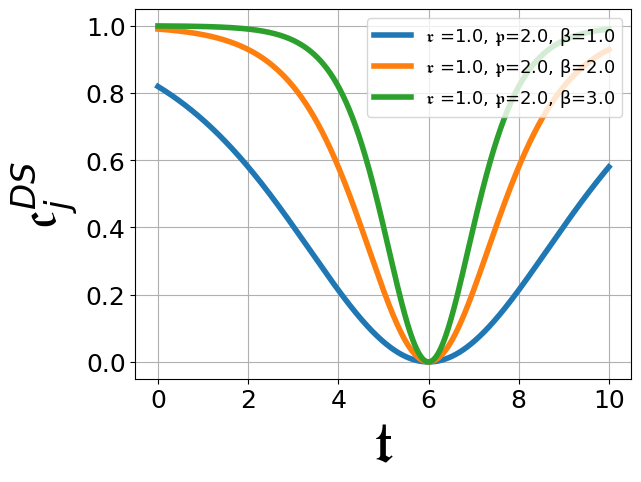}
        \caption{$\mathfrak{c}_j^{DS}$ vs $t$}
        \label{fig5c}
    \end{subfigure}
    \caption{\centering{\textbf{Correlation} $\mathfrak{c}_j$ vs \textbf{cost} $t$ for \textbf{reward} $r=1.0$ and \textbf{punishment} $p=2.0$ via NEM, DS and ABM in the PGG. {The results from NEM, ABM and DS match.}}}
    \label{fig:5}
\end{figure*} 

\subsubsection{\label{corr-pgg-analysis}Analysis of correlation for PGG}
\begin{figure*}[!ht]
    \centering
    \begin{subfigure}[b]{0.75\columnwidth}
        \centering
        \includegraphics[width = \textwidth]{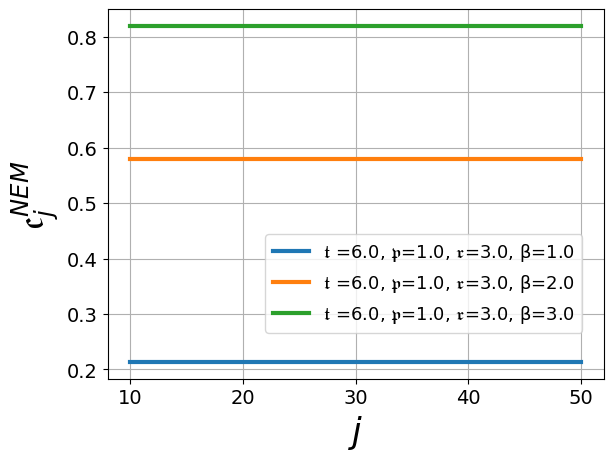}
        \caption{$\mathfrak{c}_j^{NEM}$ vs distance $j$}
        \label{fig2b-a}
    \end{subfigure}
    \begin{subfigure}[b]{0.75\columnwidth}
        \centering
        \includegraphics[width = \textwidth]{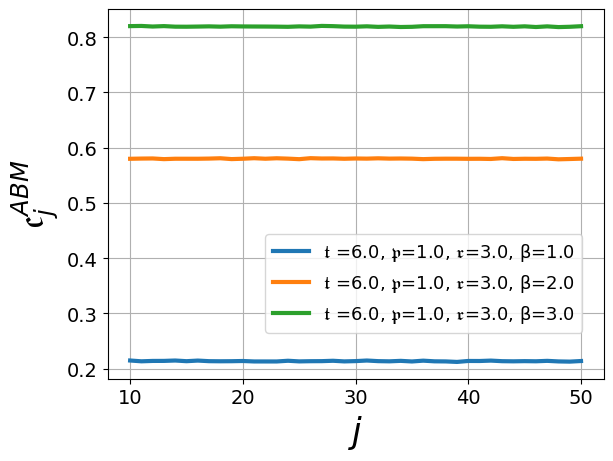}
        \caption{$\mathfrak{c}_j^{ABM}$ vs distance $j$}
        \label{fig2b-b}
    \end{subfigure}
    \caption{\centering{\textbf{Correlation} $\mathfrak{c}_j$ vs \textbf{distance} $j$ for \textbf{cost} $\mathfrak{t}=6.0$, \textbf{reward} $\mathfrak{r}=3.0$ and \textbf{punishment} $\mathfrak{p}=1.0$ via NEM and ABM in PGG. {NEM perfectly matches with ABM across distances.}}}
    \label{fig:2b-}
\end{figure*}

In this section, we will discuss the results obtained for the correlation via NEM, AS, DS and ABM, respectively. In Fig.~\ref{fig:4}, we have plotted the correlation against the reward $\mathfrak{r}$, and we observe that for NEM, DS and ABM, the correlation results agree perfectly, whereas, the results of AS are incorrect since it gives the incorrect Nash equilibria. In NEM, DS and AS, we observe that the Nash equilibrium is at $2\mathfrak{r}=\mathfrak{t} - 2\mathfrak{p}$. In PGG, the payoffs $(\mathcal{A, B, C, D})$ fulfil: $\mathcal{A+D=B+C}$ criterion, and upon fulfilment of this criterion, the \textit{distance}-dependent term (see, Eq.~(\ref{eq63})) for NEM has a negligible contribution to the correlation and the $(\cos^2 \varphi)$-term in Eq.~(\ref{eq63}) dominates. If we recall, the $(\cos^2 \varphi)$-term in Eq.~(\ref{eq63}) is nothing but the square of the average magnetization per player (see, Eq.~(\ref{eq17})), and this term is constant across all site distances. So, in the case of PGG, the results from NEM, ABM and DS match. For increasing values of $\mathfrak{r}$, in NEM, DS and ABM, we observe an initial decrease in the correlation when $2\mathfrak{r}$ approaches $(\mathfrak{t} - 2\mathfrak{p})$, reaching a minimum at $2\mathfrak{r}=\mathfrak{t} - 2\mathfrak{p}$, and then the correlation again increases as $2\mathfrak{r}$ becomes greater than $(\mathfrak{t}-2\mathfrak{p})$. This indicates that when $2\mathfrak{r}>\mathfrak{t}-2\mathfrak{p}$, the Nash equilibria is ($\mathbb{C, C}$), whereas, when $2\mathfrak{r}<\mathfrak{t}-2\mathfrak{p}$, the Nash equilibria is ($\mathbb{D, D}$). Similarly, in Fig.~\ref{fig:5}, we have plotted the correlation against the cost $\mathfrak{t}$, and we observe that for NEM, DS and ABM, the correlation results agree perfectly. However, the results of DS match with the results of NEM and ABM because of the same reasons discussed previously. Overall, the Nash equilibrium is the same for the results obtained via NEM and ABM. Now, if we try to look at the variation of correlation with the distance $j$ (for a fixed value of $\mathfrak{r}=3.0$, $\mathfrak{t}=6.0$ and $\mathfrak{p}=1.0$) as shown in Fig.~\ref{fig:2b-}, we find that the results of NEM perfectly match with the results of ABM and it also shows that across various distances, the major contributing term to the correlation for PGG is the square of average magnetization.  

\subsection{\label{paycap-pgg}Game Payoff Capacity}
In Ref.~\cite{ref4}, the authors have compared the three analytical methods with the numerical ABM by considering payoff per player as an indicator. It was shown that in the limiting cases of $\beta \rightarrow 0$ and $\beta \rightarrow\infty$, the payoff per player determined by NEM and ABM matched, whereas AS and DS gave incorrect results. We are looking for the variation of payoff capacity, i.e., change in the payoff per player due to a unit change in noise, for finite values of $\beta$. Here, we consider the modified energy matrix (see, Eq.~(\ref{e18})) for both NEM and ABM.
\subsubsection{\underline{NEM}}
For calculating the game payoff capacity, using \textit{Nash equilibrium mapping}, we determine the partition function $\zeta^{NEM}$ in terms of $\mathfrak{r}, \mathfrak{t}$ and $\mathfrak{p}$. For PGG, $\mathcal{J} = [\frac{\mathcal{(A-C)-(B-D)}}{4}] = 0,~\text{and}~\mathfrak{h} = [\frac{\mathcal{(A-C)+(B-D)}}{4}] = \frac{2\mathfrak{r}-\mathfrak{t} + 2\mathfrak{p}}{4}$ and hence, from Eq.~(\ref{e4}) and Table~\ref{table-analytical}, the NEM payoff capacity in terms of the game parameters $(\mathfrak{r},\mathfrak{t},\mathfrak{p})$ is,
\begin{gather}
    \mathbb{S}_{g}^{NEM} = \dfrac{[\frac{2\mathfrak{r}-\mathfrak{t}+2\mathfrak{p}}{4}]^2}{\cosh^2{[\frac{\beta}{4} (2\mathfrak{r}-\mathfrak{t}+2\mathfrak{p})]}}  .
    \label{e13}
\end{gather}
In Eq.~(\ref{e13}), when $T\rightarrow 0$ (or, $\beta \rightarrow \infty$), i.e., ZN limit, $\mathbb{S}_{g}^{NEM} \rightarrow 0,~\forall~\mathfrak{r},$ since in this case, the system is in equilibrium, i.e., \textit{Cooperation} (for transformed payoff per player $\langle \Lambda_1 \rangle = \frac{[2\mathfrak{r}-\mathfrak{t}+2\mathfrak{p}]}{4}$) becomes the Nash equilibrium strategy when $2\mathfrak{r}>(\mathfrak{t}-2\mathfrak{p})$, and \textit{Defection} (for transformed payoff per player $\langle \Lambda_1 \rangle = -\frac{[2\mathfrak{r}-\mathfrak{t}+2\mathfrak{p}]}{4}$) becomes the Nash equilibrium strategy when $2\mathfrak{r}<(\mathfrak{t}-2\mathfrak{p})$. Hence, in the ZN limit, we observe no change in the payoff per player (as the player's strategy does not change), which leads to vanishing payoff capacity. When $T\rightarrow \infty$ (or, $\beta \rightarrow 0$), i.e., IN limit, $\mathbb{S}_{g}^{NEM} \rightarrow \frac{(2\mathfrak{r}-\mathfrak{t}+2\mathfrak{p})^2}{16},~\forall~\mathfrak{r},$ because of the randomness in strategy selection by the players. The factor $\frac{(2\mathfrak{r}-\mathfrak{t}+2\mathfrak{p})^2}{16}$ indicates that in the IN limit, the rate of change of strategies critically depends on the \textit{reward} $\mathfrak{r}$ (for fixed $\mathfrak{t}, \mathfrak{p}$). For finite non-zero values of $\beta$, the payoff capacity $\mathbb{S}_{g}^{NEM}$ is always positive, which indicates that with increasing noise, the rate of change of payoff per player also increases.

\subsubsection{\underline{AS}}
For calculating the payoff capacity, using \textit{Aggregate selection} method, we determine the partition function $\zeta^{AS}$, from Eq.~(\ref{hdm-partition}), in terms of $\mathfrak{r}, \mathfrak{t}$ and $\mathfrak{p}$ as,
\begin{equation}
    \zeta^{AS} = (e^{2\beta\mathfrak{r}} + e^{-\beta \mathfrak{p}})^N,
    \label{e14a}
\end{equation}
where, the payoffs $\mathcal{A} = 2\mathfrak{r},~\mathcal{D} = -\mathfrak{p}$ respectively. From Eq.~(\ref{e14a}) and Table~\ref{table-analytical}, the AS payoff capacity in terms of the game parameters $(\mathfrak{r},\mathfrak{t},\mathfrak{p})$ is,
\begin{gather}
    \mathbb{S}_g^{AS} = \dfrac{(2\mathfrak{r}+\mathfrak{p})^2 e^{\beta(2\mathfrak{r} + \mathfrak{p})}}{(e^{\beta(2\mathfrak{r}+\mathfrak{p})}+1)^2}.
    \label{e14b}
\end{gather}
In Eq.~(\ref{e14b}), we find that the expression of $\mathbb{S}_g^{AS}$ is independent of the parameter $\mathfrak{t}$. When $T\rightarrow 0$ (or, $\beta \rightarrow \infty$), i.e., ZN limit, $\mathbb{S}_{g}^{AS} \rightarrow 0,~\text{for}~\mathfrak{r}>0,$ since in this case, the system is in equilibrium and when $\beta\rightarrow\infty$, the payoff per player $\langle \Lambda_1 \rangle = 2\mathfrak{r}$, i.e., \textit{Cooperation} becomes the Nash equilibrium strategy, $\forall~\mathfrak{r}$. 

\begin{figure*}[!ht]
    \centering
    \begin{subfigure}[b]{0.835\columnwidth}
        \centering
        \includegraphics[width = \textwidth]{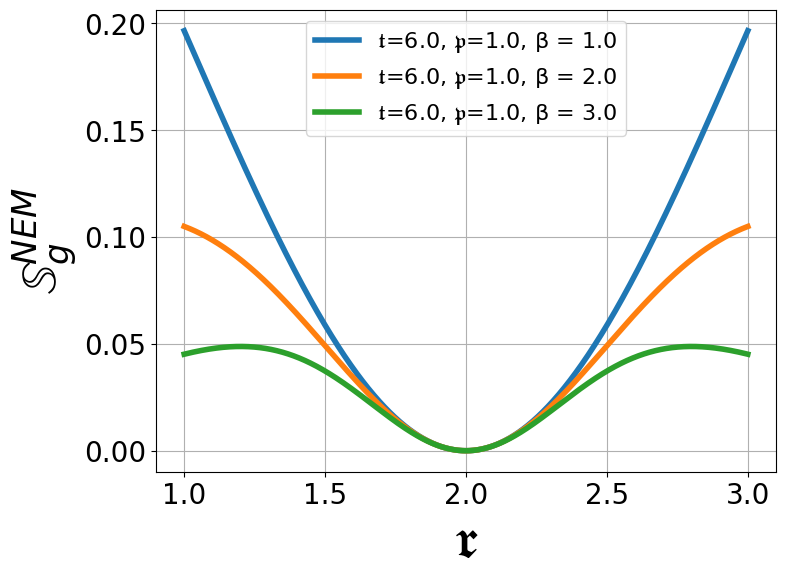}
        \caption{$\mathbb{S}_g^{NEM}$ vs $\mathfrak{r}$}
        \label{fig-pgg-spec-a}
    \end{subfigure}
    \begin{subfigure}[b]{0.835\columnwidth}
        \centering
        \includegraphics[width = \textwidth]{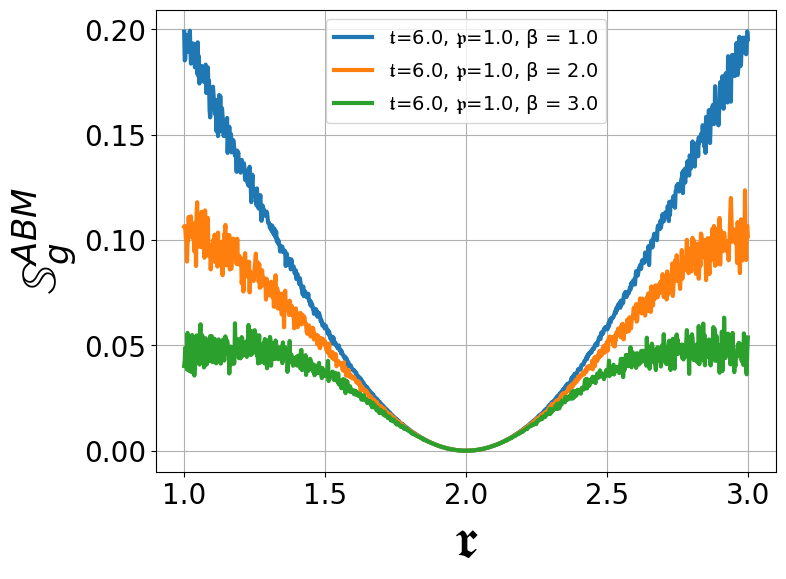}
        \caption{$\mathbb{S}_g^{ABM}$ vs $\mathfrak{r}$}
        \label{fig-pgg-spec-b}
    \end{subfigure}
    \begin{subfigure}[b]{0.835\columnwidth}
        \centering
        \includegraphics[width = \textwidth]{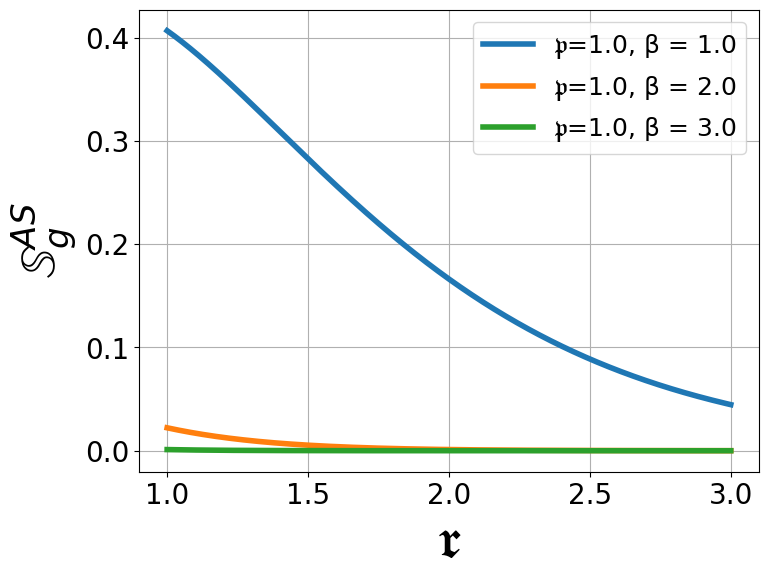}
        \caption{$\mathbb{S}_g^{AS}$ vs $\mathfrak{r}$}
        \label{fig-pgg-spec-c}
    \end{subfigure}
    \begin{subfigure}[b]{0.835\columnwidth}
        \centering
        \includegraphics[width = \textwidth]{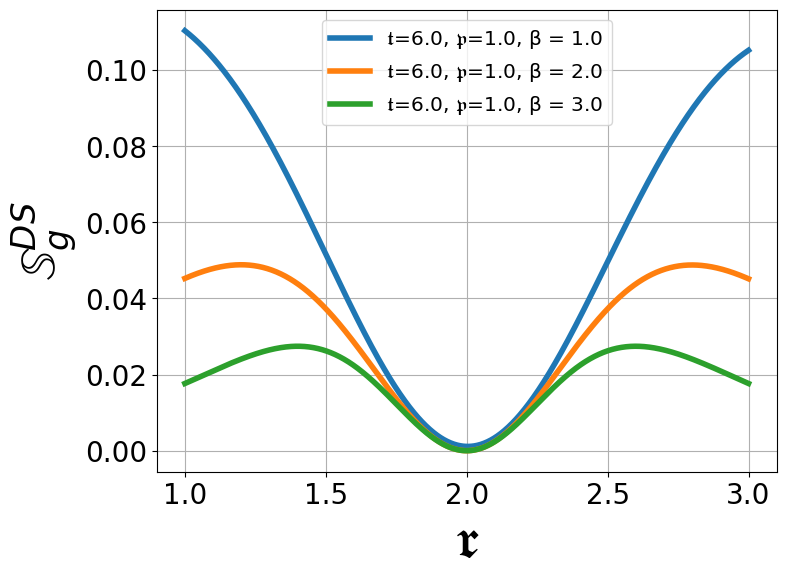}
        \caption{$\mathbb{S}_g^{DS}$ vs $\mathfrak{r}$}
        \label{fig-pgg-spec-d}
    \end{subfigure}
    \caption{\centering{\textbf{Payoff capacity} $\mathbb{S}_g$ vs \textbf{reward} $\mathfrak{r}$ for \textbf{cost} $\mathfrak{t}=6.0$ and \textbf{punishment} $\mathfrak{p}=1.0$ via NEM, AS, DS and ABM in PGG. {The inflexion point matches with the Nash equilibrium for NEM, DS and ABM while that from AS is an outlier.}}}
    \label{fig:pgg-spec}
\end{figure*} 

In NEM, we find that the Nash equilibrium strategy, when $2\mathfrak{r}<(\mathfrak{t}-2\mathfrak{p})$ in the ZN limit, is \textit{Defection}, which is clearly not seen in the case of AS. In AS, the system tries to minimize its total internal energy in order to maximize the cumulative payoff for all players, and this is the reason why all the players choose \textit{Cooperation} strategy. In the ZN limit, this leads to vanishing payoff capacity. When $T\rightarrow \infty$ (or, $\beta \rightarrow 0$), i.e., IN limit, $\mathbb{S}_{g}^{AS} \rightarrow \frac{(2\mathfrak{r}+\mathfrak{p})^2}{4}$ indicating the randomness in strategy selection by the players. For finite non-zero values of $\beta$, the payoff capacity $\mathbb{S}_{g}^{AS}$ is always positive, which indicates that with increasing noise, the rate of change of payoff per player also increases.

\subsubsection{\underline{DS}}
To calculate the payoff capacity, using \textit{Darwinian selection} method, we determine the partition function $\zeta^{DS}$, from Eq.~(\ref{dem-partition}), in terms of $\mathfrak{r}, \mathfrak{t}$ and $\mathfrak{p}$ as,
\begin{gather}
    \zeta^{DS} = e^{2\beta\mathfrak{r}} + e^{\beta(\mathfrak{r}-\frac{\mathfrak{t}}{2})} + e^{\beta(\mathfrak{r}-\mathfrak{p}+\frac{\mathfrak{t}}{2})} + e^{-\beta\mathfrak{p}},
    \label{e15}
\end{gather}
From $\zeta^{DS}$ in Eq.~(\ref{e15}) and Table~\ref{table-analytical}, the payoff capacity $\mathbb{S}_{g}^{DS}$, in terms of $\mathcal{A}=2\mathfrak{r},~\mathcal{B}=(\mathfrak{r}-\frac{\mathfrak{t}}{2}),~\mathcal{C}=(\mathfrak{r}+\frac{\mathfrak{t}}{2}-\mathfrak{p}),~\mathcal{D}=-\mathfrak{p}$, is,
\begin{gather}
    \mathbb{S}_{g}^{DS} = \dfrac{\mathcal{A}^2 e^{\beta \mathcal{A}} + \mathcal{B}^2 e^{\beta \mathcal{B}} + \mathcal{C}^2 e^{\beta \mathcal{C}} + \mathcal{D}^2 e^{\beta \mathcal{D}}}{\zeta^{DS}}\nonumber\\
    - \bigg(\dfrac{\mathcal{A}e^{\beta\mathcal{A}} + \mathcal{B}e^{\beta\mathcal{B}} + \mathcal{C}e^{\beta\mathcal{C}} + \mathcal{D}e^{\beta\mathcal{D}}}{\zeta^{DS}}\bigg)^2.
    \label{e16}
\end{gather}
In Eq.~(\ref{e16}), when $T\rightarrow 0$ (or, $\beta \rightarrow \infty$), i.e., ZN limit, $\mathbb{S}_{g}^{DS} \rightarrow 0,~\forall~\mathfrak{r},$ since in this case, the system is in equilibrium, i.e., when $2\mathfrak{r}>(\mathfrak{t}-2\mathfrak{p})$, \textit{Cooperation} becomes the Nash equilibrium strategy, and we have the payoff per player $\langle\Lambda_1\rangle = 2\mathfrak{r}$. However, when $2\mathfrak{r}<(\mathfrak{t}-2\mathfrak{p})$, we see $\langle\Lambda_1\rangle \neq -\mathfrak{p}$ (see, Eq.~\ref{e16case}), for both $\mathfrak{t}>\mathfrak{p}$ and $\mathfrak{t}<\mathfrak{p}$, which signifies that the Nash equilibrium strategy is not \textit{Defection}, but rather a mixed strategy of \textit{Cooperation} and \textit{Defection}, and this is shown below in Eq.~(\ref{e16case}). In ZN limit, we have,
\begin{gather}
    \text{When,}~2\mathfrak{r}<\mathfrak{t}-2\mathfrak{p},~\lim_{T\rightarrow 0}\langle\Lambda_1\rangle = 
    \begin{cases}
        \frac{(2\mathfrak{r}-\mathfrak{t})}{2},~\text{for}~\mathfrak{t}<\mathfrak{p}\\
        \frac{(2\mathfrak{r}-2\mathfrak{p}+\mathfrak{t})}{2},~\text{for}~\mathfrak{t}>\mathfrak{p}
    \end{cases}
    \label{e16case}
\end{gather}
From Eq.~(\ref{e16case}), we see that none of the payoffs corresponds to the \textit{Defection} strategy (i.e., payoff $-\mathfrak{p}$) associated with the players when $2\mathfrak{r}<(\mathfrak{t}-2\mathfrak{p})$, and this does not agree with the results of NEM, where \textit{defection} was predicted for $2\mathfrak{r}<(\mathfrak{t}-2\mathfrak{p})$. This leads to vanishing payoff capacity when $T\rightarrow 0$. 

\begin{figure*}[!ht]
    \centering
    \begin{subfigure}[b]{0.9\columnwidth}
        \centering
        \includegraphics[width = \textwidth]{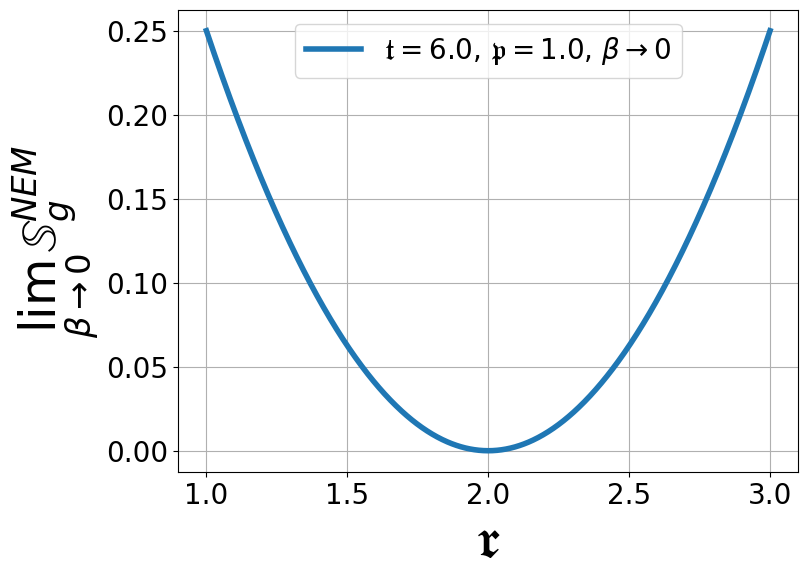}
        \caption{$\mathbb{S}_g^{NEM}$ vs $\mathfrak{r}$}
        \label{fig-pgg-spec-limit-a}
    \end{subfigure}
    \begin{subfigure}[b]{0.9\columnwidth}
        \centering
        \includegraphics[width = \textwidth]{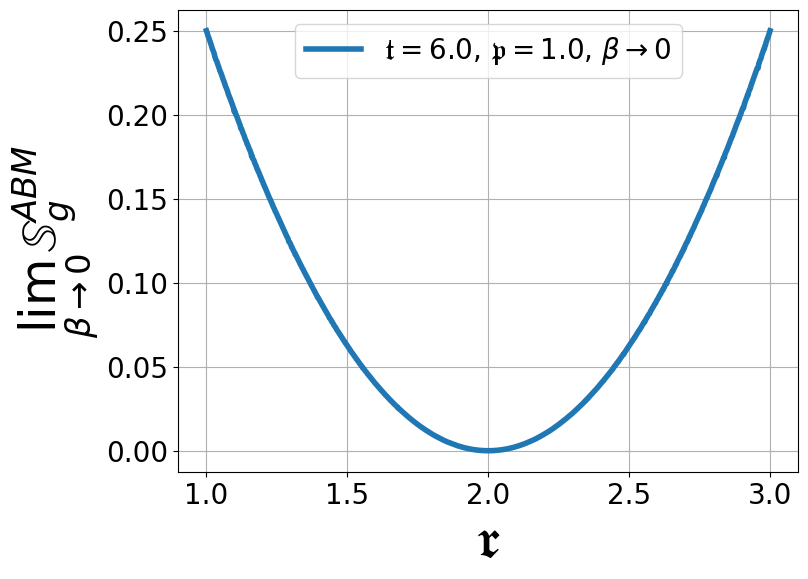}
        \caption{$\mathbb{S}_g^{ABM}$ vs $\mathfrak{r}$}
        \label{fig-pgg-spec-limit-b}
    \end{subfigure}
    \begin{subfigure}[b]{0.9\columnwidth}
        \centering
        \includegraphics[width = \textwidth]{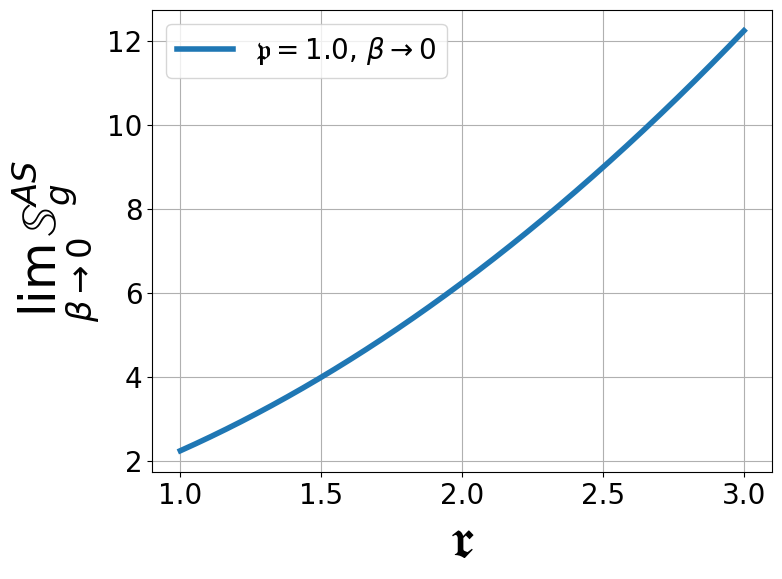}
        \caption{$\mathbb{S}_g^{AS}$ vs $\mathfrak{r}$}
        \label{fig-pgg-spec-limit-c}
    \end{subfigure}
    \begin{subfigure}[b]{0.9\columnwidth}
        \centering
        \includegraphics[width = \textwidth]{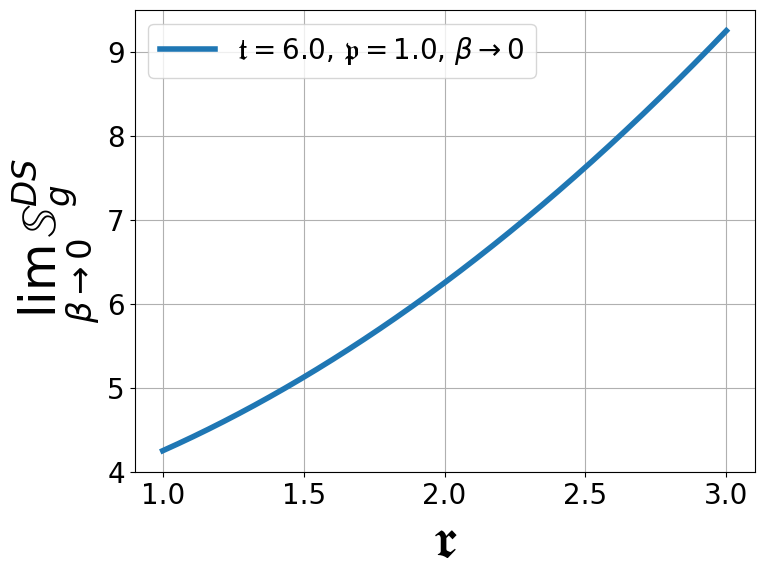}
        \caption{$\mathbb{S}_g^{DS}$ vs $\mathfrak{r}$}
        \label{fig-pgg-spec-limit-d}
    \end{subfigure}
    \caption{\centering{\textbf{Payoff capacity} $\mathbb{S}_g$ vs \textbf{reward} $\mathfrak{r}$, in IN limit, for \textbf{cost} $\mathfrak{t}=6.0$ and \textbf{punishment} $\mathfrak{p}=1.0$ via NEM, AS, DS and ABM in PGG. {In the IN or $\beta\rightarrow 0$ limit, AS and DS predict the inflexion point at $\mathfrak{r}\rightarrow 0$ which is not the Nash equilibria, while  NEM and ABM predict inflexion point to be $2\mathfrak{r}=(\mathfrak{t} - 2\mathfrak{p})$, which is the \textit{Nash equilibria}.}}}
    \label{fig:pgg-spec-limit}
\end{figure*} 

When $T\rightarrow \infty$ (or, $\beta \rightarrow 0$), i.e., IN limit, $\mathbb{S}_{g}^{DS} \rightarrow \frac{(4\mathfrak{r}^2 + 2\mathfrak{p}^2 +\mathfrak{t}^2 + 4\mathfrak{pr}-2\mathfrak{tp})}{8},$ indicating randomness in strategy selection by the players, leading to randomization of the payoffs. For finite non-zero values of $\beta$, the payoff capacity $\mathbb{S}_{g}^{DS}$ is always positive, i.e., with increasing noise, the rate of change of payoff per player also increases.  

\subsubsection{\underline{ABM}}
For determining the payoff capacity using the \textit{Agent-based method}, we designed our algorithm such that it finds the variance of the total energy of the system. In Eq.~(\ref{e3a}), we have shown how the specific heat $\mathbb{S}_V$, for the $1D$-Ising chain, is related to the variance of the total energy of the Ising chain. Similarly, for games, we have the numerical payoff capacity $\mathbb{S}_g^{ABM}$ as,
\begin{equation}
    \mathbb{S}_g^{ABM} = \dfrac{1}{N}[\langle \mathbb{E}^2\rangle - \langle \mathbb{E}\rangle^2],
    \label{e17}
\end{equation}
where $\mathbb{E}$ is the total energy of all the players in the game. The average energy per player, i.e., $\langle \mathbb{E}\rangle/N$, is equivalent to the negative of the individual player's average payoff, i.e., $\langle \Lambda_1\rangle$. The algorithm used in ABM is discussed in Sec.~\ref{sub-abm}, where we consider the Energy matrix $\Delta'$ (see, Ref.~\cite{ref4}) as,
\begin{equation}
    \Delta' = \dfrac{1}{4}\begin{bmatrix}
            (\mathfrak{t}-2\mathfrak{r} - 2\mathfrak{p}) & (\mathfrak{t}-2\mathfrak{r} - 2\mathfrak{p})\\
            (2\mathfrak{r}+2\mathfrak{p} - \mathfrak{t}) &  (2\mathfrak{r}+2\mathfrak{p} - \mathfrak{t})
        \end{bmatrix} = \begin{bmatrix}
            -\mathcal{A'} & -\mathcal{B'}\\
            -\mathcal{C'} & -\mathcal{D'}
        \end{bmatrix}.
        \label{e18}
\end{equation}
We have changed our energy matrix from $\Delta$ to $\Delta'$ (see, Eqs.~(\ref{eq62}, \ref{e18})) since while calculating the payoff capacity from the partition function $\zeta$, our original payoffs $(\mathcal{A} = 2\mathfrak{r},~ \mathcal{B} = \mathfrak{r}-\frac{\mathfrak{t}}{2},~ \mathcal{C} = \mathfrak{r} - \mathfrak{p} + \frac{\mathfrak{t}}{2},~ \mathcal{D} = -\mathfrak{p})$ gets transformed to new payoffs $(\mathcal{A'},\mathcal{B'},\mathcal{C'},\mathcal{D'})$ (see, Eq.~(\ref{e18})), and this gives us the new energy matrix $\Delta'$. When $T\rightarrow 0$ (or, $\beta \rightarrow \infty$), i.e., ZN limit, $\mathbb{S}_{g}^{ABM} \rightarrow 0,~\forall~\mathfrak{r},$ since in this case, the system is in equilibrium, i.e., \textit{Cooperation} (or, $\langle \Lambda_1 \rangle = 2\mathfrak{r}$) is the Nash equilibrium strategy when $2\mathfrak{r}>(\mathfrak{t}-2\mathfrak{p})$, while \textit{Defection} (or, $\langle \Lambda_1 \rangle = -\mathfrak{p}$) is the Nash equilibrium strategy when $2\mathfrak{r}<(\mathfrak{t}-2\mathfrak{p})$. In the ZN limit, we observe no change in the payoff per player (choosing either $\mathbb{C}$ or $\mathbb{D}$-strategy), which leads to zero payoff capacity.

\begin{table*}
\centering
\renewcommand{\arraystretch}{1.8}
\resizebox{\textwidth}{!}{%
\begin{tabular}{|cc|c|c|c|c|}
\hline
\multicolumn{2}{|c|}{\textit{\textbf{For PGG}}} & \textbf{NEM} & \textbf{ABM} & \textbf{AS} & \textbf{DS} \\ \hline
\multicolumn{1}{|c|}{} & $\beta \rightarrow 0$ & $\frac{1}{2}$, $\forall~\mathfrak{r}$ & $\frac{1}{2}$, $\forall~\mathfrak{r}$ & {\color[HTML]{FF0000} $1$, $\forall~\mathfrak{r}$} & $\frac{1}{2}$, $\forall~\mathfrak{r}$ \\ \cline{2-6} 
\multicolumn{1}{|c|}{\multirow{-2}{*}{\huge$\mathbf{\chi_\mathfrak{r}}$}} & $\beta \rightarrow \infty$ & 0, $\forall~\mathfrak{r} \neq \frac{(\mathfrak{t}-2\mathfrak{p})}{2}$ & 0, $\forall~\mathfrak{r} \neq \frac{(\mathfrak{t}-2\mathfrak{p})}{2}$ & {\color[HTML]{FF0000} 0, $\forall~\mathfrak{r}$} & 0, $\forall~\mathfrak{r} \neq \frac{(\mathfrak{t}-2\mathfrak{p})}{2}$ \\ \hline
\multicolumn{1}{|c|}{} & $\beta \rightarrow 0$ & $-\frac{1}{4}$, $\forall~\mathfrak{t}$ & $-\frac{1}{4}$, $\forall~\mathfrak{t}$ & {\color[HTML]{FF0000} 0, $\forall~\mathfrak{t}$} & $-\frac{1}{4}$, $\forall~\mathfrak{t}$ \\ \cline{2-6} 
\multicolumn{1}{|c|}{\multirow{-2}{*}{\huge$\mathbf{\chi_\mathfrak{t}}$}} & $\beta \rightarrow \infty$ & 0, $\forall~\mathfrak{t} \neq (2\mathfrak{r}+2\mathfrak{p})$ & 0, $\forall~\mathfrak{t} \neq (2\mathfrak{r}+2\mathfrak{p})$ & {\color[HTML]{FF0000} 0, $\forall~\mathfrak{t}$} & 0, $\forall~\mathfrak{t} \neq (2\mathfrak{r}+2\mathfrak{p})$ \\ \hline
\multicolumn{1}{|c|}{} & $\beta \rightarrow 0$ & $\frac{1}{2}$, $\forall~\mathfrak{p}$ & $\frac{1}{2}$, $\forall~\mathfrak{p}$ & $\frac{1}{2}$, $\forall~\mathfrak{p}$ & $\frac{1}{2}$, $\forall~\mathfrak{p}$ \\ \cline{2-6} 
\multicolumn{1}{|c|}{\multirow{-2}{*}{\huge$\mathbf{\chi_\mathfrak{p}}$}} & $\beta \rightarrow \infty$ & 0, $\forall~2\mathfrak{p} \neq (\mathfrak{t}-2\mathfrak{r})$ & 0, $\forall~2\mathfrak{p} \neq (\mathfrak{t}-2\mathfrak{r})$ & {\color[HTML]{FF0000} 0, $\forall~\mathfrak{p}$} & 0, $\forall~2\mathfrak{p} \neq (\mathfrak{t}-2\mathfrak{r})$ \\ \hline
\multicolumn{1}{|c|}{} & $\beta \rightarrow 0$ & 0, $\forall~\mathfrak{\{r,t\}}$ & 0, $\forall~\mathfrak{\{r,t\}}$ & 0, $\forall~\mathfrak{\{r,t\}}$ & 0, $\forall~\mathfrak{\{r,t\}}$ \\ \cline{2-6} 
\multicolumn{1}{|c|}{\multirow{-2}{*}{\huge$\mathbf{\mathfrak{c}_j}$}} & $\beta \rightarrow \infty$ & +1, $\forall~\mathfrak{t} \neq (2\mathfrak{r}+2\mathfrak{p})$ & +1, $\forall~\mathfrak{t} \neq (2\mathfrak{r}+2\mathfrak{p})$ & {\color[HTML]{FF0000} +1, $\forall~\mathfrak{\{r,t\}}$} & +1, $\forall~\mathfrak{t} \neq (2\mathfrak{r}+2\mathfrak{p})$ \\ \hline
\multicolumn{1}{|c|}{} & $\beta \rightarrow 0$ & $\frac{(2\mathfrak{r}-\mathfrak{t}+2\mathfrak{p})^2}{16}$, $\forall~\mathfrak{\{r,t,p\}}$ & $\frac{(2\mathfrak{r}-\mathfrak{t}+2\mathfrak{p})^2}{16}$, $\forall~\mathfrak{\{r,t,p\}}$ & {\color[HTML]{FF0000}$\frac{(2\mathfrak{r}+\mathfrak{p})^2}{4}$, $\forall~\mathfrak{\{r,t,p\}}$} & {\color[HTML]{FF0000}$\frac{(4\mathfrak{r}^2 +\mathfrak{t}^2 +2\mathfrak{p}^2 +4\mathfrak{rp}-2\mathfrak{tp})}{8}$, $\forall~\mathfrak{\{r,t,p\}}$} \\ \cline{2-6} 
\multicolumn{1}{|c|}{\multirow{-2}{*}{\large$\mathbf{\mathbb{S}_g}$}} & $\beta \rightarrow \infty$ & 0, $\forall~\mathfrak{\{r,t,p\}}$ & 0, $\forall~\mathfrak{\{r,t,p\}}$ & 0, $\forall~\mathfrak{\{r,t,p\}}$ & 0, $\forall~\mathfrak{\{r,t,p\}}$ \\ \hline
\end{tabular}%
}
\caption{PGG with reward $\mathfrak{r}$, cost $\mathfrak{t}$, punishment $\mathfrak{p}$, inter-site distance $j$, and measure of noise $\beta$.}
\label{tab:my-table2}
\end{table*}

When $T\rightarrow \infty$ (or, $\beta \rightarrow 0$), i.e., IN limit, the result of $\mathbb{S}_{g}^{ABM}$ exactly matches with the result of $\mathbb{S}_{g}^{NEM}$ and this indicates that $\lim_{\beta\rightarrow 0} \mathbb{S}_g^{ABM} = \lim_{\beta\rightarrow 0} \mathbb{S}_g^{ABM} = \frac{(2\mathfrak{r}-\mathfrak{t}+2\mathfrak{p})^2}{16},~\forall~\mathfrak{r},$ indicating the randomness in strategy selection by the players. For finite non-zero values of $\beta$, the payoff capacity $\mathbb{S}_{g}^{ABM}$ is always positive, which indicates that with increasing noise, the rate of change of payoff per player also increases.

\subsubsection{Analysis of payoff capacity for PGG}
Here, we will discuss the results obtained for the \textit{payoff capacity} $\mathbb{S}_g$ via NEM, AS, DS and ABM, respectively. As shown in Figs.~\ref{fig:pgg-spec} and \ref{fig:pgg-spec-limit}, we observe that for both the limiting cases as well as at finite non-zero values of $\beta$, the results of NEM and ABM match exactly and they have the same inflexion point at $2\mathfrak{r}=(\mathfrak{t} - 2\mathfrak{p})$, which indicates the \textit{Nash equilibria}. Meanwhile, for both the limiting cases as well as at finite non-zero values of $\beta$, AS has a  different inflexion point, at $\mathfrak{r}\rightarrow 0$, for arbitrary values of $\beta$. The results obtained via DS are quite interesting, where, for finite non-zero values of $\beta$, we observe the inflexion point to be $2\mathfrak{r}=(\mathfrak{t} - 2\mathfrak{p})$, which matches with results from NEM and ABM. However, when $\beta\rightarrow 0$, DS predicts the inflexion point at $\mathfrak{r}\rightarrow 0$, which does not agree with the results of NEM and ABM. In the $\beta\rightarrow 0$ limit, we observe that DS is an outlier when we consider payoff capacity as an indicator. AS gives $\mathbb{S}_{g}^{AS} = 0,~\forall~\mathfrak{r}$, without any regard to whether or not the condition $2\mathfrak{r} \neq (\mathfrak{t} - 2\mathfrak{p})$ is satisfied, since by AS, $\mathbb{S}_g^{AS}$ is independent of the parameter $\mathfrak{t}$. In Fig.~\ref{fig:pgg-spec}, for increasing values of $\beta$, i.e., decreasing noise in the system, we observe a decrease in the payoff capacity near the Nash equilibria, for the methods NEM, DS and ABM, which suggests that the rate of change in payoff per player is minimum at the Nash equilibrium.  

\subsection{Summary of PGG}
When we look at the results of \textit{reward susceptibility}, \textit{cost susceptibility}, \textit{punishment susceptibility}, \textit{correlation} and \textit{payoff capacity} for PGG, the results for NEM and ABM agree perfectly for all five indicators, whereas the results of AS are incorrect, in both limiting cases and for finite values of $\beta$, since it predicts an incorrect Nash equilibrium for all five indicators in question. In both the limiting cases as well as at finite non-zero values of $\beta$, DS gave us the same result as NEM and ABM for reward susceptibility $\chi_{\mathfrak{r}}$, cost susceptibility $\chi_{\mathfrak{t}}$, punishment susceptibility $\chi_{\mathfrak{p}}$, and correlation $\mathfrak{c}_j$. However, for payoff capacity $\mathbb{S}_g$, we observe that results from DS in the IN limit, i.e., $\beta\rightarrow 0$, predict a different Nash equilibrium, i.e., at $\mathfrak{r}= 0$, unlike NEM and ABM, both of which predict the Nash equilibrium at $\mathfrak{r} = \frac{(\mathfrak{t}-2\mathfrak{p})}{2}$. When $\beta$ is a finite non-zero value, DS predicts the correct Nash equilibrium point for the payoff capacity. In NEM and ABM, for all the indicators, we observe the Nash equilibrium point at $\mathfrak{r}=\frac{(\mathfrak{t} - 2\mathfrak{p})}{2}$. However, since the payoffs $(\mathcal{A, B, C, D})$ associated with this game satisfy: $\mathcal{A+D=B+C}$ criterion, the results of DS, for both the game susceptibility and correlation, also match the results of NEM and ABM. We have discussed the reasons behind this result in Sec.~\ref{corr-pgg-analysis}. The results in the ZN and IN limits for all five indicators are summarized in Table~\ref{tab:my-table2}. The AS results are incorrect because AS attempts to minimize the system's total energy in order to maximize the cumulative payoff for all the players. From Fig.~\ref{fig:2b-}, we find that the results of both NEM and ABM perfectly match with one another, and the correlation is constant across all distances $j$. This is mainly due to the fact that the major contributing term to the correlation, for PGG, is the square of average magnetization, and this remains constant across all distances for a particular $\beta$. With increasing $\beta$, the correlation approaches $+1,~\forall~j$'s, and this signifies that with decreasing noise in the system, the strategies of the players become more correlated.  

\section{\label{conclusion}Conclusion}
Depending on the values of game parameters, all three indicators, i.e., \textit{game susceptibility, correlation} and \textit{payoff capacity}, show that in the \textit{infinite} player limit, a significant fraction of players cooperate, even if \textit{Defection} is the Nash equilibrium in the \textit{two-player} limit. In the appendix, we show that the even if the pure Nash equilibrium for the HDG in the \textit{two-player} limit is one player plays \textit{Hawk} (i.e., \textit{Defect}) while other player plays \textit{Dove} (i.e., \textit{Cooperate}), in the thermodynamic limit all three indicators show that for increasing resource value $\mathfrak{r}$, and with fixed injury cost $\mathfrak{d}$, a larger than equal fraction of players tends to cooperate and switch to \textit{Dove} strategy. However, a slightly different result is obtained for the PGG, where a larger than equal fraction of players starts to cooperate, i.e., \textit{provide}, for the condition $\mathfrak{r}>\frac{(\mathfrak{t}-2\mathfrak{p})}{2}$, while this behaviour is reversed, i.e., a larger than equal fraction of players starts to defect (or, \textit{free-ride}) when the above condition is not met, i.e., $\mathfrak{r}<\frac{(\mathfrak{t}-2\mathfrak{p})}{2}$. 

If we consider and compare the results obtained in Ref.~\cite{ref4} and this paper, then individual player's average payoff $\langle \Lambda_1 \rangle$ and payoff capacity $\mathbb{S}_g$ serve as the best indicators to study cooperative behaviour among players in the thermodynamic limit. Using either indicator ($\langle \Lambda_1 \rangle$ or $\mathbb{S}_g$), we see the failure of both AS and DS methods, in both games, for limiting as well as finite values of $\beta$. The other indicators, like game susceptibility and correlation, can also be used to study cooperative behaviour among players in the thermodynamic limit, as we see in our work. However, the results obtained for these indicators (i.e., game magnetization, game susceptibility, and correlation) via DS method either agree or disagree with the results of NEM and ABM, depending on the type of game. Finally, for both games, we see that AS is a flawed method to study cooperative behaviour among players, in the thermodynamic limit.

The mathematical framework for all analytical methods, discussed in this paper, is based on the $1D$-Ising chain, and this can be extended to the $2D$-Ising model and other exactly solvable models in statistical mechanics. One can also extend our work to study the Prisoner's dilemma and check the results obtained via NEM, DS as well as AS, and compare them with the results of the numerical ABM. One can apply the same game indicators to study cooperative behaviour in quantum social dilemmas too. To conclude, this work highlights the flaws in Adami and Hintze's methodology (see, Ref.~\cite{ref5}) by emphasising the discrepancies between AS and DS methods when compared to the correct NEM and ABM approach. In Ref.~\cite{ref5}, the authors constructed the framework of AS method with the sole purpose of maximizing the payoffs of all players. For DS method, they adopted a similar approach where the \textit{player of interest} looks for the maximum payoff, without taking into consideration the decisions made by the neighbouring players. Although DS method gave better results than AS method for game susceptibility and correlation in PGG, the failure of these methods can be seen in the case of average payoff $\langle\Lambda_1 \rangle$ and payoff capacity $\mathbb{S}_g$, respectively. Our conclusion is that we should consider individual player's average payoff $\langle \Lambda_1\rangle$ and/or the payoff capacity $\mathbb{S}_g$ as the best available indicators to study cooperative behaviour among players in the thermodynamic limit, and NEM is the sole reliable analytical model that can be used to examine the emergence of cooperation among the players for \textit{one-shot} social dilemmas.
%\section*{Acknowledgements}

%\section*{Author Declarations}

%\subsection{Conflict of Interest}
%The authors have no conflicts to disclose.

%\subsection{Data Availability Statement}
%The data that supports the findings of this study are available in the article and Appendix.

%\subsection{Author Contributions}

\appendix

\section{\label{hdg}The Hawk-Dove Game (HDG)}
HDG\cite{ref9} is a \textit{bi-player, bi-strategy} social dilemma where both the players conflict over a shared resource `$\mathfrak{r}$', and this can lead to either gaining the resource or some damage to either depending on the strategy opted out by the players. If both the players choose the \textit{Hawk} (denoted by $\mathbb{H}$) strategy, then both of them suffer some damage which is given by $-\mathfrak{d}$, where $\mathfrak{d}>0$. If one of the players chooses the \textit{Hawk} strategy, and the other one chooses the \textit{Dove} (denoted by $\mathbb{D}$) strategy, then the player choosing $\mathbb{H}$ wins the resource denoted by the payoff $\mathfrak{r}$, and the player choosing $\mathbb{D}$ loses the resource, denoted by $-\mathfrak{r}$, where $\mathfrak{r}>0$. If both the players choose the $\mathbb{D}$-strategy, then both of them gain no resources, and they are denoted by \textit{zero} payoffs. The payoffs must satisfy the condition: $\mathfrak{d>r>}0$ since the resource benefits derived from such conflicts are overshadowed by the cost of injury to the players involved. For this game, the \textit{payoff matrix} $\Lambda$ is given as,
\begin{table}[H]
\centering
$\Lambda$ = 
\setlength{\extrarowheight}{2pt}
\begin{tabular}{*{4}{c|}}
    \multicolumn{2}{c}{} & \multicolumn{2}{c}{}\\\cline{3-4}
    \multicolumn{1}{c}{} &  & $\mathcal{S}_1 \equiv \mathbb{H}$  & $\mathcal{S}_2 \equiv \mathbb{D}$ \\\cline{2-4}
    \multirow{2}*{}  & $\mathcal{S}_1 \equiv \mathbb{H}$ & $(\mathfrak{-d, -d})$ & $(\mathfrak{r, -r})$ \\\cline{2-4}
    & $\mathcal{S}_2 \equiv \mathbb{D}$ & $(\mathfrak{-r, r})$ & $(0,0)$ \\\cline{2-4}
\end{tabular}
\end{table}
In terms of payoffs $(\mathcal{A,B,C,D})$, we have: $\mathcal{A} = -\mathfrak{d}$, $\mathcal{B} = -\mathcal{C} = \mathfrak{r}$ and $\mathcal{D} = 0$, respectively. The Nash equilibria for this game involving two pure strategies are $(\mathbb{H}, \mathbb{D})$ and $(\mathbb{D}, \mathbb{H})$ respectively. We can find another Nash equilibrium when both players opt for a mixed strategy, say opting for $\mathbb{H}$-strategy with a \textit{finite probability} of $\kappa$ and/or opting for $\mathbb{D}$-strategy with a \textit{finite probability} of $(1-\kappa)$, where $\kappa = \mathfrak{r/d}$ (see, Ref.~\cite{ref10}). We clearly see that the HDG does not satisfy the payoff criterion: $\mathcal{A+D = B+C}$, so AS is not applicable. Therefore we focus on NEM, DS and ABM only.

For a single player (we consider the \textit{row} player), the payoff matrix can be written as,
\begin{equation}
    \Lambda_1 = \begin{bmatrix}
            \mathfrak{-d} & \mathfrak{r}\\
            \mathfrak{-r} & 0 
        \end{bmatrix}
        \equiv \begin{bmatrix}
            \mathcal{A} & \mathcal{B}\\
            \mathcal{C} & \mathcal{D} 
        \end{bmatrix}.
\label{eq42}
\end{equation}
We will be using the payoff matrix $\Lambda_1$ given in Eq.~(\ref{eq42}) for further calculations related to the Hawk-Dove game.

\subsection{\label{sus-hdg}Game susceptibility}
There are two parameters in HDG: cost of injury ($\mathfrak{d}$) and resource ($\mathfrak{r}$), with $\mathfrak{d>r>} 0$. In the Ising model, magnetic susceptibility is defined as the change in magnetization as a response to a unit change in the external magnetic field $\mathfrak{h}$. However, in social dilemmas,  we work with game payoffs, instead of Ising parameters, and in one of the analytical methods, i.e., \textit{NEM}, we linearly map the Ising parameters to the game payoffs. For HDG, in NEM, we have $\mathcal{J} = -\frac{\mathfrak{d}}{4}$ and $\mathfrak{h} = \frac{2\mathfrak{r}-\mathfrak{d}}{4}$. Here, we observe that $\mathcal{J}$ is dependent on $\mathfrak{d}$ and independent of $\mathfrak{r}$, while $\mathfrak{h}$ is dependent on both $\mathfrak{r}$ and $\mathfrak{d}$. Hence, in analogy with magnetic susceptibility, we calculate the game susceptibility only with respect to the \textit{field payoff} $\mathfrak{r}$, as $\mathfrak{d}$ is the \textit{mixed payoff}, i.e., it appears in both $\mathcal{J}$ and $\mathfrak{h}$, and game susceptibility is only defined for field payoffs and not for mixed payoffs.

\subsubsection{\underline{NEM}}
For calculating the \textit{resource susceptibility}, using \textit{Nash equilibrium mapping}, we write the game magnetization ($\mu_g^{NEM}$) in terms of $\mathfrak{r}$ and $\mathfrak{d}$. From Eqs.~(\ref{eq22}, \ref{eq42}), the NEM game magnetization in terms of the game parameters is given as,
\begin{equation}
    \mu_g^{NEM} = \dfrac{\sinh{\beta (\frac{2\mathfrak{r}-\mathfrak{d}}{4})}}{\sqrt{\sinh^2{\beta (\frac{2\mathfrak{r}-\mathfrak{d}}{4})} + e^{\beta \mathfrak{d}}}}.
    \label{eq43}
\end{equation}
When $T\rightarrow 0$ (or, $\beta \rightarrow \infty$), i.e., \textit{zero noise} (ZN) limit, we find $\mu_g^{NEM} \rightarrow 0$, which indicates that the fraction of players choosing $\mathbb{H}$-strategy is equal to the fraction of players choosing $\mathbb{D}$-strategy to minimize their losses. This can be verified by Taylor expanding, up to \textit{first}-order, the expression of $\mu_g^{NEM}$, in Eq.~(\ref{eq43}), about $\frac{1}{\beta}$ and then determining the value of $\mu_g^{NEM}$ in the $\frac{1}{\beta}\rightarrow 0$ (or, $\beta\rightarrow\infty$) limit. From Eq.~(\ref{eq43}), in the ZN limit, we have,
\begin{gather}
    \lim_{\frac{1}{\beta}\rightarrow 0}\mu_g^{NEM} \bigg(\frac{1}{\beta}\bigg) = \dfrac{\sinh{\beta (\frac{2\mathfrak{r}-\mathfrak{d}}{4})}}{\sqrt{\sinh^2{\beta (\frac{2\mathfrak{r}-\mathfrak{d}}{4})} + e^{\beta \mathfrak{d}}}}\bigg|_{\frac{1}{\beta}\rightarrow 0} \nonumber\\
    = \underbrace{\mu_g^{NEM} [0]\bigg|_{\frac{1}{\beta}\rightarrow 0}}_{= 0} + \underbrace{\bigg(\frac{1}{\beta} \dfrac{\partial}{\partial (\frac{1}{\beta})}\mu_g^{NEM}[0]\bigg)\bigg|_{\frac{1}{\beta}\rightarrow 0}}_{=0} + \cancelto{0}{\mathcal{O}\bigg(\frac{1}{\beta^2}\bigg)}\nonumber\\
    \text{Thus,}~\lim_{\frac{1}{\beta}\rightarrow 0}\mu_g^{NEM} \rightarrow 0,~\text{in the ZN limit}.
    \label{eq43a}
\end{gather}
Further, for $T\rightarrow \infty$ (or, $\beta \rightarrow 0$), i.e., \textit{infinite noise} (IN) limit, $\mu_g^{NEM} \rightarrow 0$ since the players choose their strategies randomly, leading to equiprobable numbers for either $\mathbb{H}$ or $\mathbb{D}$-strategies. By Taylor expanding, up to \textit{first}-order, the expression of $\mu_g^{NEM}$, in Eq.~(\ref{eq43}), about $\beta$, we determine the value of $\mu_g^{NEM}$ in the $\beta\rightarrow 0$ limit. From Eq.~(\ref{eq43}), in the IN limit, we have,
\begin{gather}
    \lim_{\beta\rightarrow 0}\mu_g^{NEM} (\beta) = \dfrac{\sinh{\beta (\frac{2\mathfrak{r}-\mathfrak{d}}{4})}}{\sqrt{\sinh^2{\beta (\frac{2\mathfrak{r}-\mathfrak{d}}{4})} + e^{\beta \mathfrak{d}}}}\bigg|_{\beta\rightarrow 0} \nonumber\\
    = \underbrace{\mu_g^{NEM} [0]\bigg|_{\beta\rightarrow 0}}_{= 0} + \underbrace{\bigg(\beta \dfrac{\partial}{\partial \beta}\mu_g^{NEM}[0]\bigg)\bigg|_{\beta\rightarrow 0}}_{= \lim_{\beta\rightarrow 0
    }\frac{\beta(2\mathfrak{r}-\mathfrak{d})}{4} = 0} + \cancelto{0}{\mathcal{O}(\beta^2)}\nonumber\\
    \text{Thus,}~\lim_{\beta\rightarrow 0}\mu_g^{NEM} \rightarrow 0,~\text{in the IN limit}.
    \label{eq43b}
\end{gather}

From Eq.~(\ref{eq43}), we have the resource susceptibility as,
\begin{gather}
    \chi_\mathfrak{r}^{NEM} =\dfrac{1}{\beta} \dfrac{\partial}{\partial \mathfrak{r}}\mu_g^{NEM} =  \dfrac{e^{\beta \mathfrak{d}} \cosh\bigg[ \frac{\beta (2\mathfrak{r}-\mathfrak{d})}{4} \bigg]}{2\bigg(e^{\beta \mathfrak{d}} + \sinh^2\bigg[\frac{\beta (2\mathfrak{r}-\mathfrak{d})}{4} \bigg]\bigg)^{\frac{3}{2}}}.
    \label{eq44}
\end{gather} 
In Eq.~(\ref{eq44}), when $T\rightarrow 0$ (or, $\beta \rightarrow \infty$), i.e., ZN limit, $\chi_\mathfrak{r}^{NEM} \rightarrow 0$. This is because game magnetization $\mu_g^{NEM}$ in this limit is constant, i.e., \textit{zero} (from Eq.~(\ref{eq43}), we see that at $T\rightarrow 0$, $\mu_g^{NEM} \rightarrow 0$). 

When $T\rightarrow \infty$ (or, $\beta \rightarrow 0$), i.e., IN limit, $\chi_\mathfrak{r}^{NEM} \rightarrow \frac{1}{2}$ indicating the randomness in strategy selection by the players, i.e., when the noise in the system is very high, the rate of change of strategies averages out to $\frac{1}{2}$. This can be verified from Eq.~(\ref{eq43b}), where the \textit{first}-order correction (denoted by $\mu_g^{{NEM}{(1)}}$) of game magnetization $\mu_g^{NEM}$, Taylor expanded around $\beta$, is given as, $\mu_g^{{NEM}{(1)}} = \frac{\beta(2\mathfrak{r}-\mathfrak{d})}{4}$, and this gives us the \textit{zeroth}-order game susceptibility correction, in the IN limit, as,
\begin{equation}
\small{
    \lim_{\beta\rightarrow 0}\chi_\mathfrak{r}^{NEM(0)} = \lim_{\beta\rightarrow 0}\frac{1}{\beta} \frac{\partial}{\partial \mathfrak{r}}\mu_g^{NEM(1)} = \lim_{\beta\rightarrow 0}\frac{1}{\cancel{\beta}} \frac{\partial}{\partial \mathfrak{r}}\frac{\cancel{\beta}(2\mathfrak{r}-\mathfrak{d})}{4} =\frac{1}{2}}.
\end{equation}
The \textit{first} and \textit{higher} order terms in the Taylor expansion of $\chi_\mathfrak{r}^{NEM}$ (see, Eq.~(\ref{eq44})) about $\beta$ , in the IN limit, vanishes.

One thing to note is that for finite non-zero values of $\beta$, the resource susceptibility $\chi_\mathfrak{r}^{NEM}$ is always positive, which indicates that the net turnover of strategies from $\mathbb{H}$ to $\mathbb{D}$ is greater than the net turnover of strategies from $\mathbb{D}$ to $\mathbb{H}$. This implies that for increasing resource value $\mathfrak{r}$, the rate at which fraction of players change to $\mathbb{D}$-strategy, i.e., \textit{Cooperation}, is always more in comparison to the rate at which fraction of players change to $\mathbb{H}$-strategy, i.e., \textit{Defection}.

For a $1D$-Ising chain, magnetic susceptibility, in contrast to magnetization, is a far more accurate tool for measuring minute variations in the system's magnetic moment in the presence of an external magnetic field \cite{ref6}. Analogously, in social dilemmas, game susceptibility, in contrast to game magnetization, is a much more reliable tool for measuring changes in the fraction of players opting for $\mathbb{D}$ or $\mathbb{H}$-strategy, in response to a changing payoff, and studying the emergence of cooperative behaviour among the players in the \textit{infinite player} limit. 

Since HDG does not satisfy the criterion: ($\mathcal{A+D=B+C}$), for $\mathcal{A}=-\mathfrak{d},~\mathcal{B}=-\mathcal{C}=\mathfrak{r},~\mathcal{D}=0$, we cannot apply \textit{Aggregate selection} method to determine the game susceptibility.

\subsubsection{\underline{DS}}
For calculating the \textit{resource susceptibility}, using \textit{Darwinian selection} method, we will write the game magnetization ($\mu_g^{DS}$) in terms of $\mathfrak{r}$ and $\mathfrak{d}$. From Eqs.~(\ref{eq37}, \ref{eq42}), the DS game magnetization in terms of the game parameters is,
\begin{equation}
    \mu_g^{DS} = \langle \hat{\mathcal{M}_z}^{(1)}\rangle_\beta = \bigg(\dfrac{e^{-\beta \mathfrak{d}}+e^{\beta \mathfrak{r}}- e^{-\beta \mathfrak{r}}-1}{e^{-\beta \mathfrak{d}}+e^{\beta \mathfrak{r}}+ e^{-\beta \mathfrak{r}}+1}\bigg).
    \label{eq45}
\end{equation}
When $T\rightarrow 0$ (or, $\beta \rightarrow \infty$), i.e., \textit{zero noise} (ZN) limit, we find $\mu_g^{DS} \rightarrow 1$, which suggests that the $\mathbb{D}$-strategy is chosen by each and every player. Whereas, when $T\rightarrow \infty$ (or, $\beta \rightarrow 0$), i.e. \textit{infinite noise} (IN) limit, $\mu_g^{DS} \rightarrow 0$ since the players choose their strategies randomly, leading to an almost equiprobable selection of $\mathbb{H}$ and $\mathbb{D}$-strategies. From Eq.~(\ref{eq45}), we have the resource susceptibility as,
\begin{gather}
    \chi_\mathfrak{r}^{DS} = \dfrac{1}{\beta} \dfrac{\partial\mu_g^{DS}}{\partial\mathfrak{r}} =\dfrac{2 e^{\beta(\mathfrak{r}+\mathfrak{d})}[e^{\beta (2\mathfrak{r}+\mathfrak{d})} + 2e^{\beta (\mathfrak{r}+\mathfrak{d})} +1]}{[e^{\beta (2\mathfrak{r}+\mathfrak{d})} + (e^{\beta \mathfrak{d}} +1)e^{\beta \mathfrak{r}} + e^{\beta \mathfrak{d}}]^2}.
    \label{eq46}
\end{gather}
In Eq.~(\ref{eq46}), when $T\rightarrow 0$ (or, $\beta \rightarrow \infty$), i.e., ZN limit, $\chi_\mathfrak{r}^{DS} \rightarrow 0$ since in this case, all players choose the $\mathbb{D}$-strategy (i.e., $\mu_g^{DS} = 1$). Therefore, the rate of change of strategies from $\mathbb{D}$ to $\mathbb{H}$ or vice-versa vanishes in this limit.

When $T\rightarrow \infty$ (or, $\beta \rightarrow 0$), i.e., IN limit, $\chi_\mathfrak{r}^{DS} \rightarrow \frac{1}{2}$ because of the randomness in strategy selection by the players. This can also be verified by Taylor expanding the expression of $\mu_g^{DS}$ in Eq.~(\ref{eq45}), up to \textit{first}-order, about $\beta$. We have the \textit{first}-order correction in $\mu_g^{DS}$ (denoted by $\mu_g^{DS(1)}$) as, $\mu_g^{DS(1)} = \frac{\beta}{8}(4\mathfrak{r} - 2\mathfrak{d})$. This gives us the \textit{zeroth}-order game susceptibility correction, in the IN limit, as,
\begin{equation}
\small{
    \lim_{\beta\rightarrow 0}\chi_\mathfrak{r}^{DS(0)} = \lim_{\beta\rightarrow 0}\frac{1}{\beta} \frac{\partial}{\partial \mathfrak{r}}\mu_g^{DS(1)} = \lim_{\beta\rightarrow 0}\frac{1}{\cancel{\beta}} \frac{\partial}{\partial \mathfrak{r}}\frac{\cancel{\beta}(4\mathfrak{r}-2\mathfrak{d})}{8} =\frac{1}{2}}.
\end{equation}
The \textit{first} and \textit{higher} order terms in the Taylor expansion of $\chi_\mathfrak{r}^{DS}$ (see, Eq.~(\ref{eq46})) about $\beta$ , in the IN limit, vanishes. 

Here also, the resource susceptibility $\chi_\mathfrak{r}^{DS}$ is always positive for finite non-zero values of $\beta$, which indicates that the net turnover of strategies from $\mathbb{H}$ to $\mathbb{D}$ is more in comparison to the net turnover of strategies from $\mathbb{D}$ to $\mathbb{H}$. This implies that for increasing resource value $\mathfrak{r}$, the rate at which fraction of players change to $\mathbb{D}$-strategy is always greater than the rate at which fraction of players change to $\mathbb{H}$-strategy. 

\begin{figure*}[!ht]
    \centering
    \begin{subfigure}[b]{0.67\columnwidth}
        \centering
        \includegraphics[width = \textwidth]{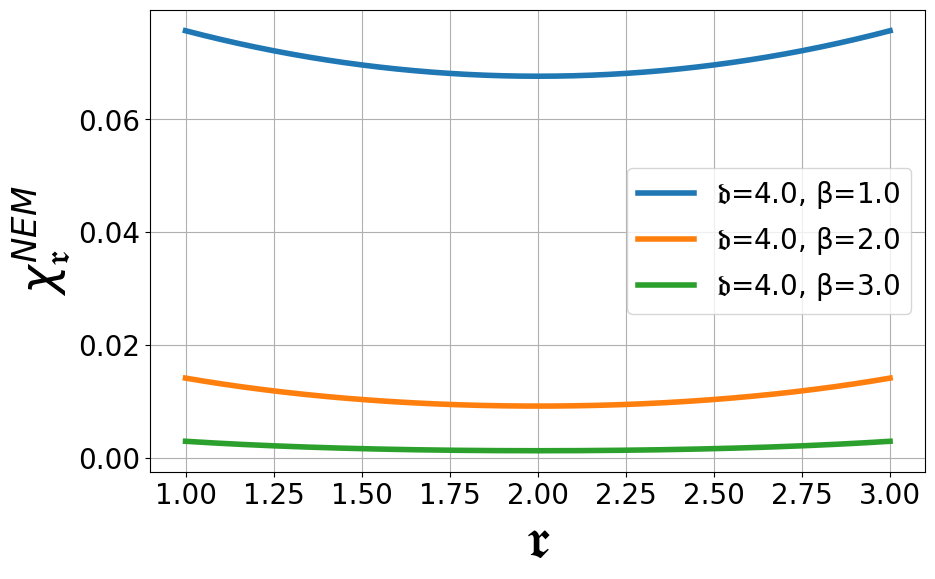}
        \caption{$\chi_\mathfrak{r}^{NEM}$ vs $\mathfrak{r}$}
        \label{fig1a}
    \end{subfigure}
    \begin{subfigure}[b]{0.67\columnwidth}
        \centering
        \includegraphics[width = \textwidth]{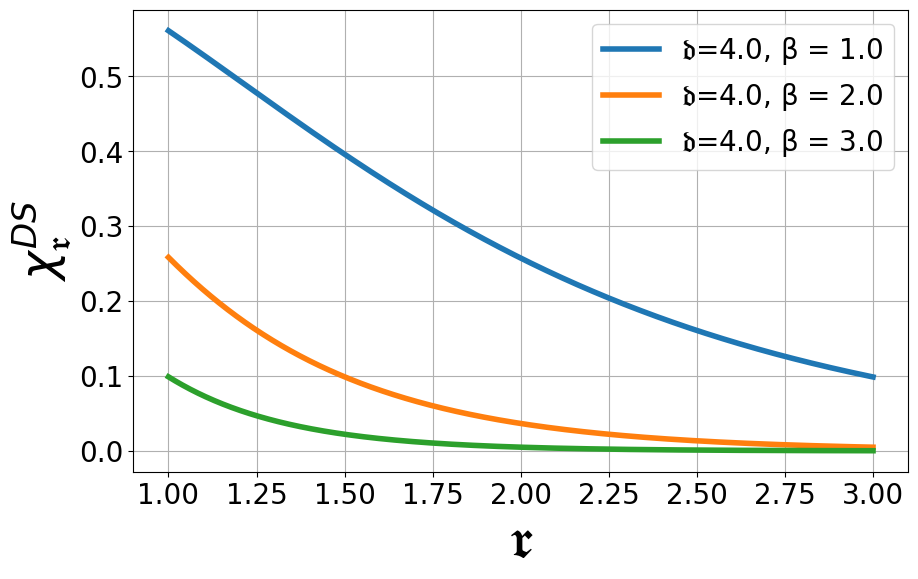}
        \caption{$\chi_\mathfrak{r}^{DS}$ vs $\mathfrak{r}$}
        \label{fig1b}
    \end{subfigure}
    \begin{subfigure}[b]{0.67\columnwidth}
        \centering
        \includegraphics[width = 5.65cm, height=3.65cm]{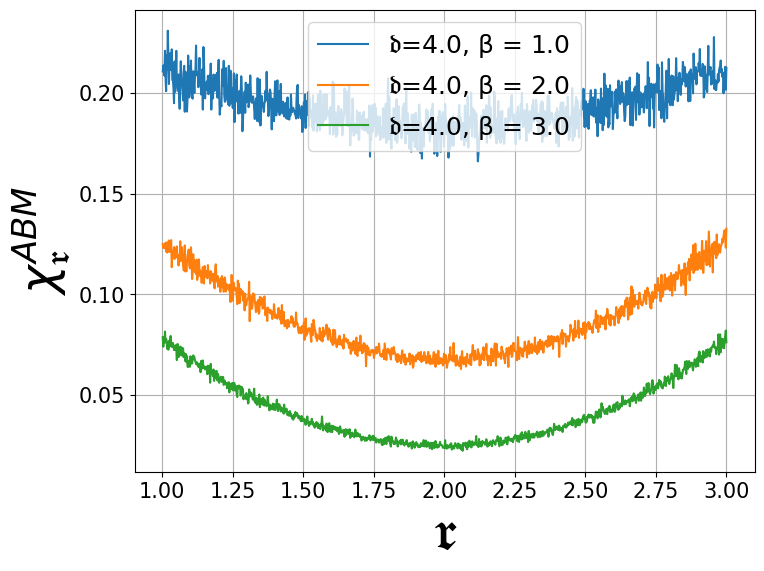}
        \caption{$\chi_\mathfrak{r}^{ABM}$ vs $\mathfrak{r}$}
        \label{fig1c}
    \end{subfigure}
    \caption{\centering{\textbf{Resource susceptibility} $\chi_\mathfrak{r}$ vs \textbf{resource value} $\mathfrak{r}$ for \textbf{cost of injury} $\mathfrak{d}=4.0$ via NEM, DS and ABM in Hawk-Dove game. {NEM and ABM follow a very similar trend with the same inflexion point at $\mathfrak{r}=\mathfrak{d}/2$, which also indicates the \textit{Nash equilibria}, whereas, DS is an outlier as it predicts the inflexion point at $\mathfrak{r}\rightarrow \mathfrak{d}$.}}}
    \label{fig:1}
\end{figure*}

Although in the limits $\beta\rightarrow 0$ and $\beta\rightarrow\infty$, results from DS and NEM are identical, when we look at Fig.~\ref{fig:1} for arbitrary $\beta$, we see a discrepancy in the Nash equilibrium predicted by NEM and DS. This discrepancy in the results of NEM and DS will be discussed in Sec.~\ref{sus-hdg-analysis}, where we will also compare the results obtained via ABM. 

\subsubsection{\underline{ABM}}
For determining the \textit{resource susceptibility} using  \textit{ABM}, we use Eq.~(\ref{eq13a}), which relates the magnetic susceptibility to the variance of magnetization. Since there exists a one-to-one mapping between the Ising parameters $(\mathcal{J}, \mathfrak{h})$ and the game parameters $(\mathcal{A, B, C, D})$, we can determine the game susceptibility from the fluctuations of the game magnetization $\mu_g$. For HDG, we have $(\mathfrak{d,r})$ as the possible payoffs and $\mathcal{A}=-\mathfrak{d},~\mathcal{B}=-\mathcal{C}=\mathfrak{r},~\mathcal{D}=0$. So, $(\mathcal{J}, \mathfrak{h})$ in terms of $(\mathfrak{r,d})$ is,
\begin{equation}
    \footnotesize{\mathcal{J} = \frac{\mathcal{(A-C)-(B-D)}}{4}= -\frac{\mathfrak{d}}{4},~ \mathfrak{h} = \frac{\mathcal{(A-C)+(B-D)}}{4} = \frac{2\mathfrak{r}-\mathfrak{d}}{4}.}
    \label{neweq1}
\end{equation}
From Eqs.~(\ref{eq1}, \ref{neweq1}), we can write the Hamiltonian $H$ in terms of $(\mathfrak{r,d})$ as,
\begin{gather}
    H = \dfrac{\mathfrak{d}}{4} \sum_{i=1}^{N} \mathfrak{s}_i \mathfrak{s}_{i+1} - \dfrac{\mathfrak{(2r-d)}}{4}\sum_{i=1}^{N} \mathfrak{s}_i \nonumber\\
    \text{The partition function is given as,}~\zeta  = \sum_{\{\mathfrak{s}_i\}} e^{-\beta H(\{\mathfrak{s}_i\})}, \nonumber\\
    \text{or,}~~\zeta = \sum_{\{\mathfrak{s}_i\}} e^{-\beta (\frac{\mathfrak{d}}{4} \sum_{i=1}^{N} \mathfrak{s}_i \mathfrak{s}_{i+1} - \frac{(2\mathfrak{r} - \mathfrak{d})}{4}\sum_{i=1}^{N}\mathfrak{s}_i)}. \nonumber\\
    \text{Thus,}~~ \dfrac{\partial \zeta}{\partial \mathfrak{r}} = \sum_{\{\mathfrak{s}_i\}} e^{-\beta H(\{\mathfrak{s}_i\})} ~\dfrac{\beta}{2}~ \Tilde{\mu}_g .\nonumber\\
    \text{So,} ~\chi_{\mathfrak{r}} = \dfrac{\partial \mu_g}{\beta\partial \mathfrak{r}} = \dfrac{\partial \langle \Tilde{\mu}_g \rangle}{\beta\partial \mathfrak{r}} = \dfrac{\partial}{\beta\partial \mathfrak{r}} \bigg[ \dfrac{1}{\zeta}\sum_{\{\mathfrak{s}_i\}} \Tilde{\mu}_g~ e^{-\beta H(\{\mathfrak{s}_i\})}  \bigg], \nonumber\\
    = -\dfrac{1}{\beta\zeta^2} \dfrac{\partial \zeta}{\partial \mathfrak{r}}\sum_{\{\mathfrak{s}_i\}} \Tilde{\mu}_g~ e^{-\beta H(\{\mathfrak{s}_i\})} + \dfrac{1}{\zeta} \sum_{\{\mathfrak{s}_i\}} \dfrac{1}{2}~\Tilde{\mu}_g^2~e^{-\beta H(\{\mathfrak{s}_i\})},  \nonumber\\
    = \bigg[-\dfrac{1}{\zeta^2} \sum_{\{\mathfrak{s}_i\}} e^{-\beta H(\{\mathfrak{s}_i\})} ~\dfrac{1}{2}~ \Tilde{\mu}_g \bigg]\sum_{\{\mathfrak{s}_i\}} \Tilde{\mu}_g~ e^{-\beta H(\{\mathfrak{s}_i\})} \nonumber\\
    + \dfrac{1}{\zeta} \sum_{\{\mathfrak{s}_i\}} \dfrac{1}{2}~\Tilde{\mu}_g^2~e^{-\beta H(\{\mathfrak{s}_i\})},\nonumber\\
    = -\dfrac{1}{2}\underbrace{\bigg[\dfrac{1}{\zeta} \sum_{\{\mathfrak{s}_i\}} \Tilde{\mu}_g~ e^{-\beta H(\{\mathfrak{s}_i\})}\bigg]^2}_{\langle \Tilde{\mu}_g \rangle^2} + \dfrac{1}{2} \underbrace{\bigg[ \dfrac{1}{\zeta} \sum_{\{\mathfrak{s}_i\}} \Tilde{\mu}_g^2~e^{-\beta H(\{\mathfrak{s}_i\})}\bigg]}_{\langle \Tilde{\mu}_g^2 \rangle}\nonumber\\
    \text{or,}~~ \chi_{\mathfrak{r}} = \dfrac{1}{\beta}\dfrac{\partial \mu_g}{\partial \mathfrak{r}} = \dfrac{1}{2} [  \langle \Tilde{\mu}_g^2 \rangle - \langle \Tilde{\mu}_g \rangle^2].
    \label{neweq2}
\end{gather}
Since, in this case, we are determining the game magnetization via ABM, so we have $\Tilde{\mu}_g = \mu_g^{ABM}$, i.e.,
\begin{equation}
    \chi_{\mathfrak{r}}^{ABM} = \dfrac{1}{2}[\langle (\mu_g^{ABM})^2 \rangle - \langle \mu_g^{ABM} \rangle^2],
    \label{eq47}
\end{equation}
where, for a particular configuration, $\mu_g^{ABM}$ denotes the game magnetization obtained for the system, each time the conditional loop runs for a particular value of $(\mathfrak{r}, \beta)$. The algorithm used in ABM is described in Sec.~\ref{sub-abm}, where we consider the Energy matrix $\Delta = -\Lambda_1$ (see, Eq.~(\ref{eq42})). We thus have,
\begin{equation}
    \Delta = \begin{bmatrix}
            \mathfrak{d} & -\mathfrak{r}\\
            \mathfrak{r} & 0 
        \end{bmatrix}.
        \label{eq48}
\end{equation}
When $T\rightarrow 0$ (or, $\beta \rightarrow \infty$), i.e., ZN limit, $\chi_{\mathfrak{r}}^{ABM} \rightarrow 0$ since, in this limit, both $\langle \mu_g^{ABM} \rangle \rightarrow 0$ and $\langle (\mu_g^{ABM})^2 \rangle \rightarrow 0$, i.e., the rate of change of strategies among the players vanishes, whereas, when $T\rightarrow \infty$ (or, $\beta \rightarrow 0$), i.e., IN limit, $\chi_{\mathfrak{r}}^{ABM} \rightarrow \frac{1}{2}$, indicating the randomness in strategy selection by the players. In the IN limit, $\langle \mu_g^{ABM} \rangle \rightarrow 0$, since the players choose their strategies randomly, leading to an equiprobable selection of $\mathbb{D}$ and $\mathbb{H}$-strategies. However, in the same $\beta \rightarrow 0$ limit, $\langle (\mu_g^{ABM})^2 \rangle \rightarrow 1$, and this leads to a value of $\frac{1}{2}$ for the game susceptibility (from Eq.~(\ref{eq47})), in the IN limit.

The resource susceptibility $\chi_{\mathfrak{r}}^{ABM}$ is always positive for finite non-zero values of $\beta$, which indicates that the rate of change of strategies from \textit{Hawk} $(\mathbb{H})$ to \textit{Dove} $(\mathbb{D})$ is more in comparison to the rate of change of strategies from $\mathbb{D}$ to $\mathbb{H}$. This implies that for increasing resource value $\mathfrak{r}$, the rate at which fraction of players change to $\mathbb{D}$-strategy is always more in comparison to the rate at which fraction of players change to $\mathbb{H}$-strategy.

\subsubsection{\label{sus-hdg-analysis}Analysis of game susceptibility for HDG}
Here, we will discuss the results obtained for the \textit{resource susceptibility} $\chi_{\mathfrak{r}}$ via NEM, DS and ABM, respectively. As shown in Fig.~\ref{fig:1}, we observe that for all three models, the \textit{Dove} $(\mathbb{D})$-strategy, i.e., \textit{Cooperation}, remains the dominant strategy, i.e., the rate of change of strategies from \textit{Hawk} $(\mathbb{H})$ to \textit{Dove} $(\mathbb{D})$ is more in comparison to the rate of change of strategies from $\mathbb{D}$ to $\mathbb{H}$. However, in NEM and ABM, for increasing values of $\mathfrak{r}$, we observe an initial decrease in the rate of change of strategies from $\mathbb{H}$ to $\mathbb{D}$ when $\mathfrak{r}$ approaches $\mathfrak{d}/2$, reaching a minimum at $\mathfrak{r}=\mathfrak{d}/2$, and then the rate again increases as $\mathfrak{r}$ approaches $\mathfrak{d}$. In NEM and ABM, for increasing $\mathfrak{r}$, initially, the rate of change of strategies from $\mathbb{H}$ to $\mathbb{D}$ will be less, mainly due to the increasing resource value that will compel a small yet significant fraction of players to choose $\mathbb{H}$-strategy over the $\mathbb{D}$-strategy. The rate of change of strategies is minimal at the inflexion point $\mathfrak{r}=\frac{\mathfrak{d}}{2}$. However, as $\mathfrak{r}$ further approaches $\mathfrak{d}$, the rate of change of strategies from $\mathbb{H}$ to $\mathbb{D}$ increases since most of the players will now adopt $\mathbb{D}$-strategy, rather than $\mathbb{H}$-strategy, so as to avoid inflicting some damage cost which would have been caused due to a larger fraction of $\mathbb{H}$-players in the population. The results obtained via NEM and ABM follow a very similar trend where they both have the same inflexion point at $\mathfrak{r}=\mathfrak{d}/2$, which also indicates the \textit{Nash equilibria}, whereas, DS is an outlier as it predicts the inflexion point at $\mathfrak{r}\rightarrow \mathfrak{d}$.

\subsection{\label{corr-hdg}Correlation}
A spin-spin correlation is a measure of the spin-order in a system \cite{ref8}, and it shows how spins at different positions are co-related. In this section, we discuss the correlation between players in a social dilemma, and its variation with resource value $\mathfrak{r}$. We replace spins ($\uparrow, \downarrow$) with the two strategies ($\mathcal{S}_1, \mathcal{S}_2$) available to the players of the HDG. We have discussed the different methods to calculate correlations in Sec.~\ref{theory}.

\subsubsection{\underline{NEM}}
\begin{figure*}[!ht]
    \centering
    \includegraphics[width = 0.9\textwidth]{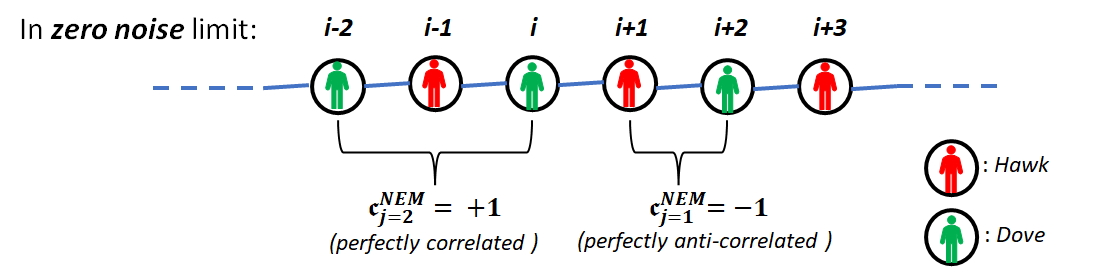}
    \caption{\centering{\textbf{NEM}: For HDG, in the ZN limit, the players choose the opposite strategy with regard to their nearest neighbours' strategies. For \textit{odd} $j$'s (say, $j=1$), the strategies are perfectly \textit{anti-correlated}, whereas, for \textit{even} $j$'s (say, $j=2$), the strategies are perfectly \textit{correlated}}.}
    \label{fig:hdg-il}
\end{figure*}

Using \textit{Nash equilibrium mapping}, the expression for the correlation between the strategies of the players at the $i^{th}$ and the $(i+j)^{th}$ sites is given as,
\begin{equation}
    \mathfrak{c}_j^{NEM} = \langle \mathfrak{s}_i \mathfrak{s}_{i+j} \rangle = \cos^2 \varphi + \bigg(\dfrac{\Omega_{-}}{\Omega_{+}} \bigg)^j \sin^2 \varphi,
    \label{eq49}
\end{equation}
where, we have $j$ as the \textit{distance} from the $i^{th}$ site and from Eqs.~(\ref{eq26}, \ref{eq27}), we have $(\cos^2\varphi,~ \sin^2\varphi,~\Omega_{\pm})$ in terms of the resource value $\mathfrak{r}$ and cost of injury $\mathfrak{d}$ as,
\begin{gather}
    \cos^2\varphi = \dfrac{e^{-\frac{\beta \mathfrak{d}}{2}} \sinh^2\bigg[\dfrac{\beta (2\mathfrak{r}-\mathfrak{d})}{4} \bigg]}{\mathfrak{Y}} = 1-\sin^2\varphi, \nonumber\\
    \text{and,}~\Omega_{\pm} = e^{-\frac{\beta \mathfrak{d}}{4}} \cosh\bigg[\frac{\beta (2\mathfrak{r}-\mathfrak{d})}{4} \bigg]\pm  \sqrt{\mathfrak{Y}} \nonumber\\
    \text{where,}~\mathfrak{Y} = e^{\frac{\beta \mathfrak{d}}{2}} + e^{-\frac{\beta \mathfrak{d}}{2}} \sinh^2\bigg[\dfrac{\beta (2\mathfrak{r}-\mathfrak{d})}{4} \bigg].
    \label{eq50}
\end{gather}
In HDG, for \textit{even} values of $j$ (say, $j=10, 12,...$), when $T\rightarrow 0$ (or, $\beta \rightarrow \infty$), i.e., ZN limit, the correlation in Eq.~(\ref{eq49}); $\mathfrak{c}_j^{NEM} \rightarrow +1$, indicating positive correlation, i.e., same strategy of both the players, whereas, for \textit{odd} values of $j$, in the same limit of $T\rightarrow 0$, we have the correlation; $\mathfrak{c}_j^{NEM} \rightarrow -1$, indicating negative correlation, i.e., opposite strategy of both the players. This can be verified by Taylor expanding the expression of $\mathfrak{c}_j^{NEM}$ in Eq.~(\ref{eq49}), up to \textit{first}-order, about $\frac{1}{\beta}$. An illustration of this case is shown in Fig.~\ref{fig:hdg-il}, where we see that, in the ZN limit, an equal number of players adopt an alternating fixed strategy (notice the site index) due to the absence of \textit{noise}, and hence, they show maximum correlation (both \textit{positive} and \textit{negative} depending on $j$). For the $1D$-Ising chain at \textit{zero} temperature, an alternating arrangement of spins, across all sites, corresponds to the minimum internal energy $\langle \mathbb{E} \rangle$ of the system. Analogously, in HDG, the alternating arrangement of $\mathbb{H}$ and $\mathbb{D}$-players (to ensure that the nearest neighbours do not fight over the resource to inflict some damage) corresponds to the best feasible payoff, i.e., Nash equilibrium, in the ZN limit. When $T\rightarrow 0$ (or, $\beta \rightarrow \infty$), the two-player pure Nash equilibrium, i.e. $(\mathbb{D}, \mathbb{H})$ or $(\mathbb{H}, \mathbb{D})$, becomes the Nash equilibrium for all the players in the thermodynamic limit, i.e., two nearest neighbours/players always opt for the opposite strategies, as illustrated in Fig.~\ref{fig:hdg-il}.

When $T\rightarrow \infty$, i.e. IN limit, the correlation; $\mathfrak{c}_j^{NEM} \rightarrow 0$ irrespective of the even/odd value of $j$ since the players choose their strategies randomly, leading to the absence of any correlation. Since Hawk-Dove game does not satisfy the criterion: ($\mathcal{A+D=B+C}$), for $\mathcal{A}=-\mathfrak{d},~\mathcal{B}=-\mathcal{C}=\mathfrak{r},~\mathcal{D}=0$, we cannot apply \textit{Aggregate selection} method to determine the correlation in this game.

\subsubsection{\underline{DS}}
Using \textit{DS} method, the expression for the correlation, from Eq.~(\ref{eq41}), between the strategies of the players at the $1^{st}$ site and the $j^{th}$ site, in terms of game parameters $\mathfrak{r}~\text{and}~\mathfrak{d}$, is given as,
\begin{equation}
    \mathfrak{c}_j^{DS} = \langle \hat{\mathcal{M}_z}^{(1)} \hat{\mathcal{M}_z}^{(j)} \rangle_{\beta} =\bigg[ \dfrac{e^{-\beta \mathfrak{d}} + e^{\beta \mathfrak{r}} - e^{-\beta \mathfrak{r}} - 1}{e^{-\beta \mathfrak{d}} + e^{\beta \mathfrak{r}} + e^{-\beta \mathfrak{r}} + 1}\bigg]^2,
    \label{eq51}
\end{equation}
where, we have $j$ as the \textit{distance} from the $1^{st}$ site. Here, we notice that the correlation is independent of the distance $j$ since, in DS, the main objective is to maximize the payoffs of the player(s) of interest, and for our case, the correlation will exist only in between the two players of interest (fixed at the $1^{st}$ and $j^{th}$ site), and they will be the same irrespective of the distance. When $T\rightarrow 0$ (or, $\beta \rightarrow \infty$), i.e., ZN limit, the correlation; $\mathfrak{c}_j^{DS} \rightarrow 1$, indicating positive correlation, whereas, when $T\rightarrow \infty$ (or, $\beta \rightarrow 0$), i.e., IN limit, $\mathfrak{c}_j^{DS} \rightarrow 0$, denoting the absence of correlation since the players choose their strategies randomly.

\subsubsection{\underline{ABM}}
We have explained the algorithm used to determine the Correlation using \textit{ABM} in Sec.~\ref{sub-abm}. We have checked for both the cases of ZN as well as IN, and we have discussed them in Sec.~\ref{corr-hdg-analysis}. For the payoff matrix given in Eq.~(\ref{eq42}), we have the energy matrix as,
\begin{equation}
    \Delta = 
    \begin{bmatrix}
        \mathfrak{d} & -\mathfrak{r}\\
        \mathfrak{r} & 0
    \end{bmatrix}.
    \label{eq52}
\end{equation}
For \textit{even} values of distance $j$ (say, $j=10, 12,...$), we find that for $T\rightarrow 0$ (or, $\beta \rightarrow \infty$), i.e., ZN limit, the correlation $\mathfrak{c}_j^{ABM} \rightarrow +1$, indicating \textit{positive} correlation, i.e., same strategy of both the players, whereas, for \textit{odd} values of $j$, in the same limit of $T\rightarrow 0$, we have the correlation $\mathfrak{c}_j^{ABM} \rightarrow -1$, indicating \textit{negative} correlation, i.e., opposite strategy of both the players. When $T\rightarrow 0$, the strategies of the players do not change due to the absence of \textit{noise} (similar to what we discussed for NEM), and hence they show maximum correlation (both \textit{positive} and \textit{negative} depending on \textit{odd/even} values of $j$). When $T\rightarrow \infty$ (or, $\beta \rightarrow 0$), i.e., IN limit, the correlation $\mathfrak{c}_j^{ABM} \rightarrow 0$ irrespective of the even/odd value of $j$ since the players choose their strategies randomly, leading to the absence of correlation. The results from ABM match with the NEM results.

\subsubsection{\label{corr-hdg-analysis}Analysis of correlation for HDG}
\begin{figure*}[!ht]
    \centering
    \begin{subfigure}[b]{0.68\columnwidth}
        \centering
        \includegraphics[width = \textwidth]{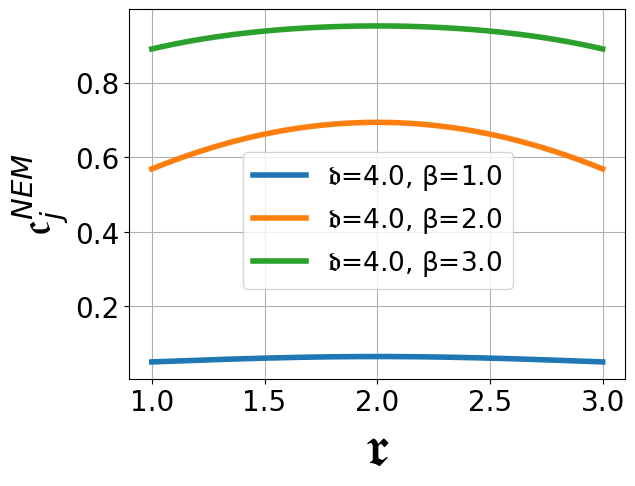}
        \caption{$\mathfrak{c}_j^{NEM}$ vs $\mathfrak{r}$ (even $j=10$)}
        \label{fig2a}
    \end{subfigure}
    \begin{subfigure}[b]{0.69\columnwidth}
        \centering
        \includegraphics[width = \textwidth]{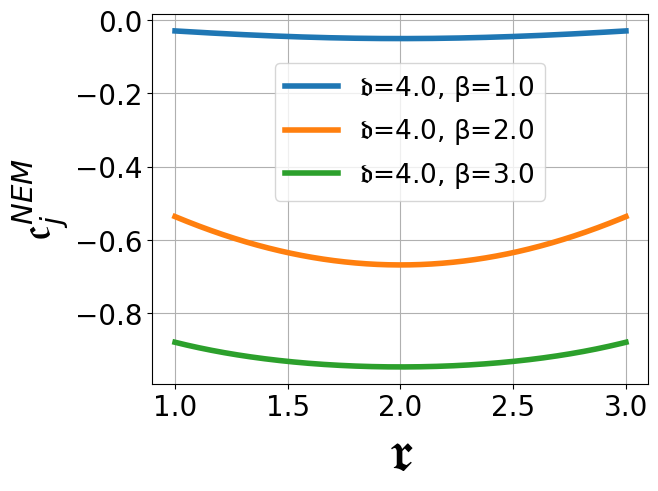}
        \caption{$\mathfrak{c}_j^{NEM}$ vs $\mathfrak{r}$ (odd $j=11$)}
        \label{fig2b}
    \end{subfigure}
        \begin{subfigure}[b]{0.68\columnwidth}        
        \centering
        \includegraphics[width = \textwidth]{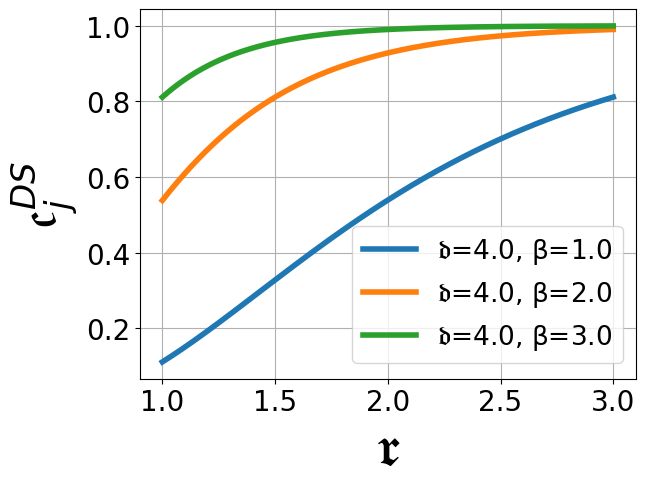}
        \caption{$\mathfrak{c}_j^{DS}$ vs $\mathfrak{r}$}
        \label{fig2c}
    \end{subfigure}
    \begin{subfigure}[b]{0.72\columnwidth}        
        \centering
        \includegraphics[width = \textwidth]{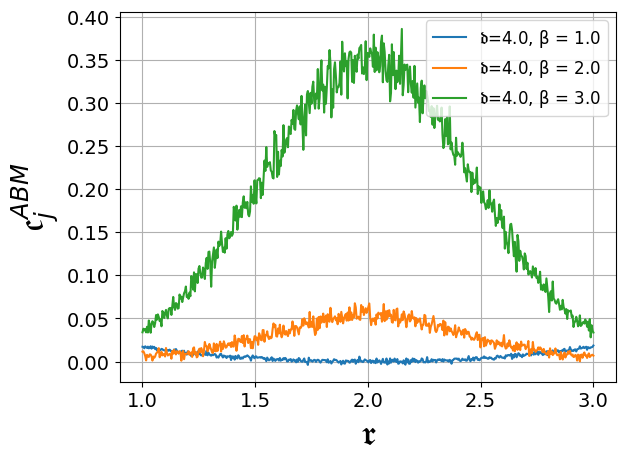}
        \caption{$\mathfrak{c}_j^{ABM}$ vs $\mathfrak{r}$ (even $j=10$)}
        \label{fig2d}
    \end{subfigure}
    \begin{subfigure}[b]{0.72\columnwidth}        
        \centering
        \includegraphics[width = \textwidth]{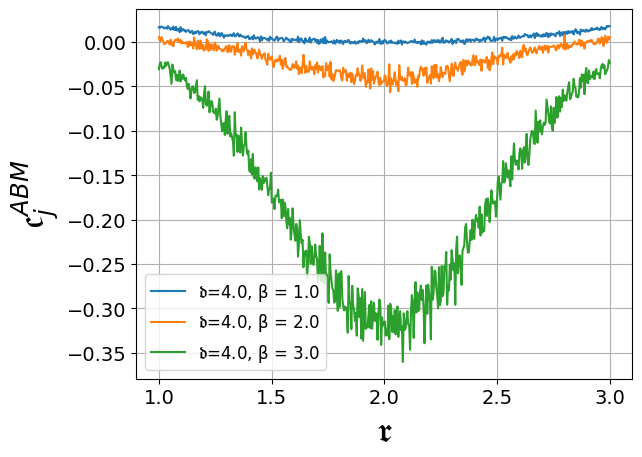}
        \caption{$\mathfrak{c}_j^{ABM}$ vs $\mathfrak{r}$ (odd $j=11$)}
        \label{fig2e}
    \end{subfigure}
    \caption{\centering{\textbf{Correlation} $\mathfrak{c}_j$ vs \textbf{resource value} $\mathfrak{r}$ for \textbf{cost of injury} $\mathfrak{d}=4.0$ via NEM, DS and ABM in HDG. { NEM and ABM follow a very similar pattern where they both have the same inflexion point at $\mathfrak{r}=\mathfrak{d}/2$ and this is the \textit{Nash equilibria}, whereas, DS is an outlier.}}}
    \label{fig:2}
\end{figure*}

Here, we will discuss the results obtained for the correlation via NEM, DS and ABM, respectively. As shown in Fig.~\ref{fig:2}, for both even and odd values of distance $j$, we observe that the correlation results obtained via NEM and ABM follow a very similar pattern where they both have the same inflexion point at $\mathfrak{r}=\mathfrak{d}/2$ and this is the \textit{Nash equilibria}, whereas, DS is an outlier. In NEM and ABM, for increasing values of $\mathfrak{r}$, we observe an initial increase in the correlation (both \textit{positive} and \textit{negative}) when $\mathfrak{r}$ approaches $\mathfrak{d}/2$, reaching the highest, positive or negative value (depending on \textit{odd} or \textit{even} distance) at $\mathfrak{r}=\mathfrak{d}/2$, and then the correlation again decreases as $\mathfrak{r}$ approaches $\mathfrak{d}$. This indicates that at \textit{Nash equilibria}, the strategies of the players are maximally correlated, and the system favours this condition. From NEM and ABM, we also found that at $\beta \rightarrow \infty$, the correlation $\mathfrak{c}_j \rightarrow \pm 1$ (+1 for even $j$'s and -1 for odd $j$'s), whereas, in DS, $\mathfrak{c}_j^{DS} \rightarrow +1,~\forall~j$, i.e., all players choose either the $\mathbb{D}$-strategy or the $\mathbb{H}$-strategy. If all players choose $\mathbb{H}$-strategy, then it would lead to some injury $\mathfrak{d}$ to all of them, and if all players choose $\mathbb{D}$-strategy, then it would result in no resource gain for the players. These are not ideal equilibrium conditions, and hence, the results of DS are incorrect. However, when $\beta \rightarrow 0$, the correlations from NEM, ABM and DS all vanish. 

\begin{figure*}[!ht]
    \centering
    \begin{subfigure}[b]{0.87\columnwidth}
        \centering
        \includegraphics[width = \textwidth]{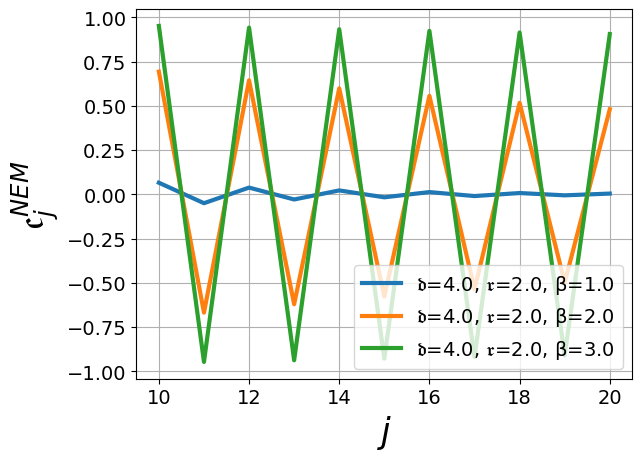}
        \caption{$\mathfrak{c}_j^{NEM}$ vs distance $j$}
        \label{fig2a-a}
    \end{subfigure}
    \begin{subfigure}[b]{0.87\columnwidth}
        \centering
        \includegraphics[width = \textwidth]{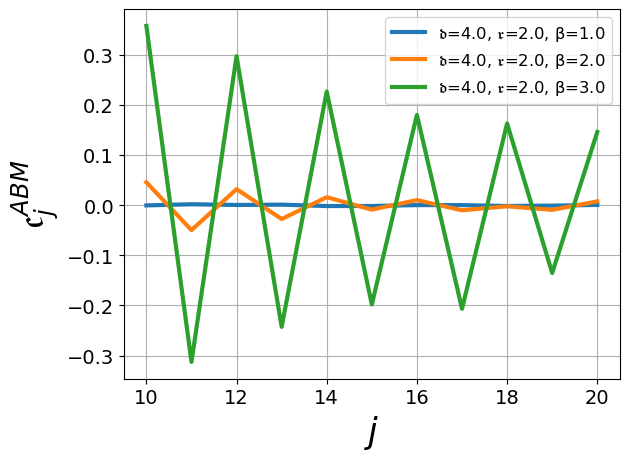}
        \caption{$\mathfrak{c}_j^{ABM}$ vs distance $j$}
        \label{fig2a-b}
    \end{subfigure}
    \caption{\centering{\textbf{Correlation} $\mathfrak{c}_j$ vs \textbf{distance} $j$ for \textbf{cost of injury} $\mathfrak{d}=4.0$, \textbf{resource value} $\mathfrak{r}=2.0$ via NEM and ABM in HDG. { Change of sign in correlation for even/odd distances.}}}
    \label{fig:2a-}
\end{figure*}

When we look at the variation of correlation with distance $j$ (for a fixed value of $\mathfrak{r}=2.0$ and $\mathfrak{d}=4.0$), we find that the results of NEM are in close agreement with the results of ABM. In Fig.~\ref{fig:2a-}, we also notice the change of sign in correlation for even/odd distances. 

\subsection{\label{paycap-hdg}Game Payoff Capacity}
In Ref.~\cite{ref4}, the authors have compared NEM and DS results with the numerical ABM results by considering the \textit{average payoff} per player as an indicator. It was shown that in the limiting cases of $T\rightarrow 0$ and $T\rightarrow\infty$, the \textit{average payoff} per player determined by NEM and ABM matched, whereas DS gave an incorrect result. The \textit{energy matrix} (or negative of the \textit{payoff matrix}) considered for NEM and ABM was different, although they were mapped to one another by a linear transformation, and the Nash equilibrium under such transformation remains \textit{invariant}. Hence, both of them predicted the same Nash equilibrium in the limiting cases. Since we are looking for the variation of the payoff capacity, i.e., change in the payoff per player due to a unit change in noise, we will consider the modified energy matrix (see, Eq.~(\ref{e12})) for both NEM and ABM. The modified energy (or, negative payoff) matrix is defined for NEM, and since we compare \textit{like with like}, we modify the energy matrix for ABM too (see, Eqs.~(\ref{eq52}, \ref{e12})). 
\subsubsection{\underline{NEM}}
\begin{figure*}[!ht]
    \centering
    \begin{subfigure}[b]{0.69\columnwidth}
        \centering
        \includegraphics[width = \textwidth]{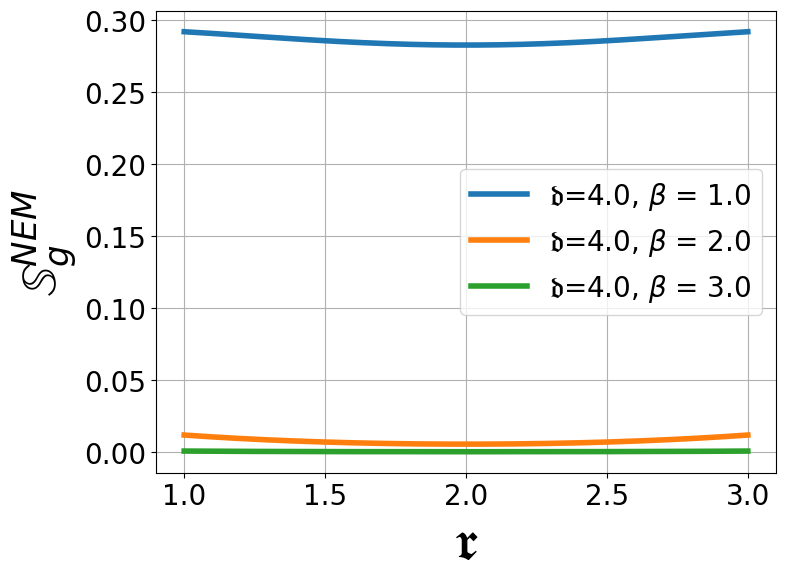}
        \caption{$\mathbb{S}_g^{NEM}$ vs $\mathfrak{r}$}
        \label{fig-hdg-spec-a}
    \end{subfigure}
    \begin{subfigure}[b]{0.67\columnwidth}
        \centering
        \includegraphics[width = \textwidth]{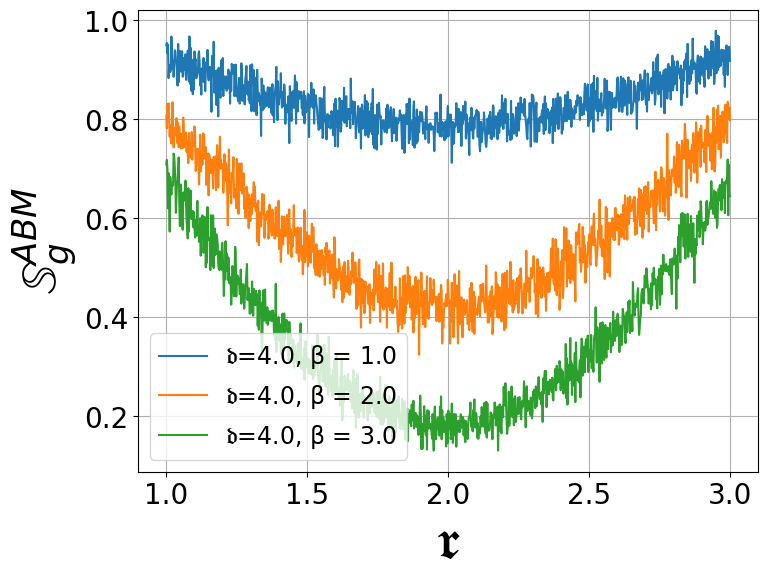}
        \caption{$\mathbb{S}_g^{ABM}$ vs $\mathfrak{r}$}
        \label{fig-hdg-spec-b}
    \end{subfigure}
    \begin{subfigure}[b]{0.69\columnwidth}
        \centering
        \includegraphics[width = \textwidth]{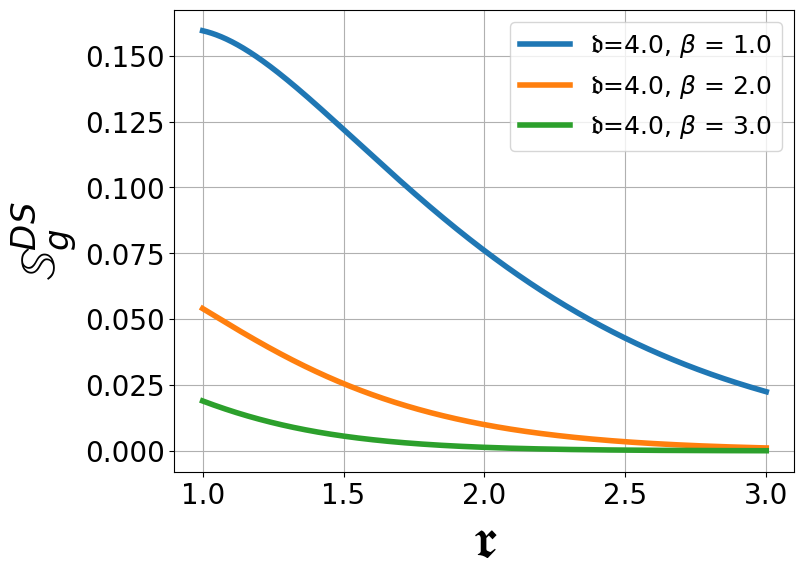}
        \caption{$\mathbb{S}_g^{DS}$ vs $\mathfrak{r}$}
        \label{fig-hdg-spec-c}
    \end{subfigure}
    \caption{\centering{\textbf{Payoff capacity} $\mathbb{S}_g$ vs \textbf{resource value} $\mathfrak{r}$ for \textbf{cost of injury} $\mathfrak{d}=4.0$ via NEM, DS and ABM in HDG. {NEM and ABM show similar nature with inflexion at $\mathfrak{r}=\mathfrak{d}/2$, while DS doesn\rq{}t have an inflexion point in the applicable range.}}}
    \label{fig:hdg-spec}
\end{figure*} 
\begin{figure*}[!ht]
    \centering
    \begin{subfigure}[b]{0.67\columnwidth}
        \centering
        \includegraphics[width = \textwidth]{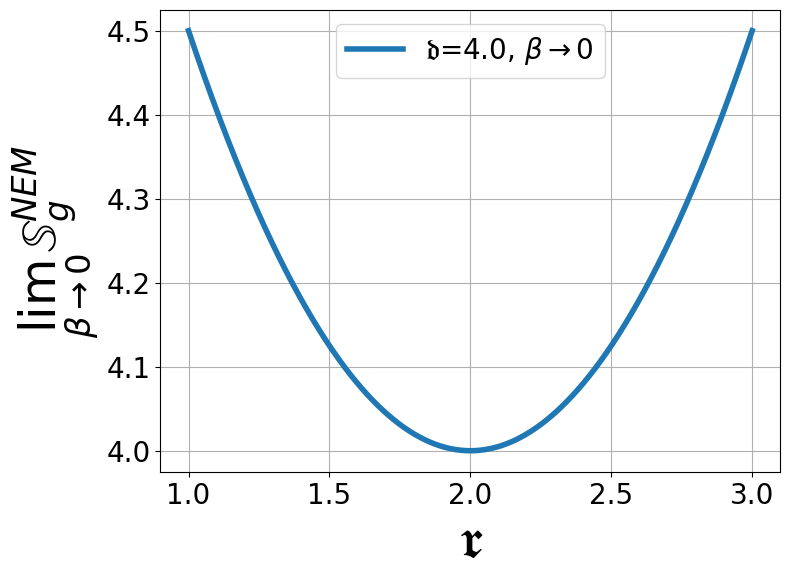}
        \caption{$\mathbb{S}_g^{NEM}$ vs $\mathfrak{r}$}
        \label{fig-hdg-spec-a-limit}
    \end{subfigure}
    \begin{subfigure}[b]{0.68\columnwidth}
        \centering
        \includegraphics[width = \textwidth]{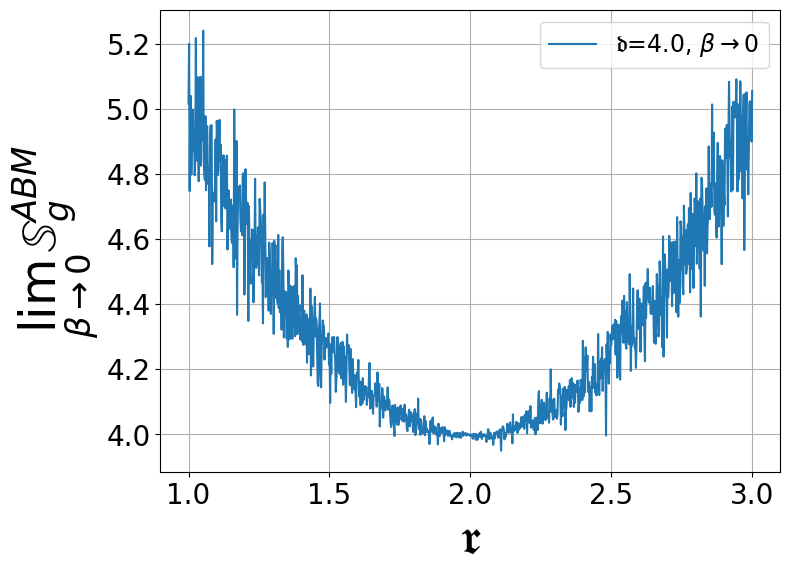}
        \caption{$\mathbb{S}_g^{ABM}$ vs $\mathfrak{r}$}
        \label{fig-hdg-spec-b-limit}
    \end{subfigure}
    \begin{subfigure}[b]{0.66\columnwidth}
        \centering
        \includegraphics[width = \textwidth]{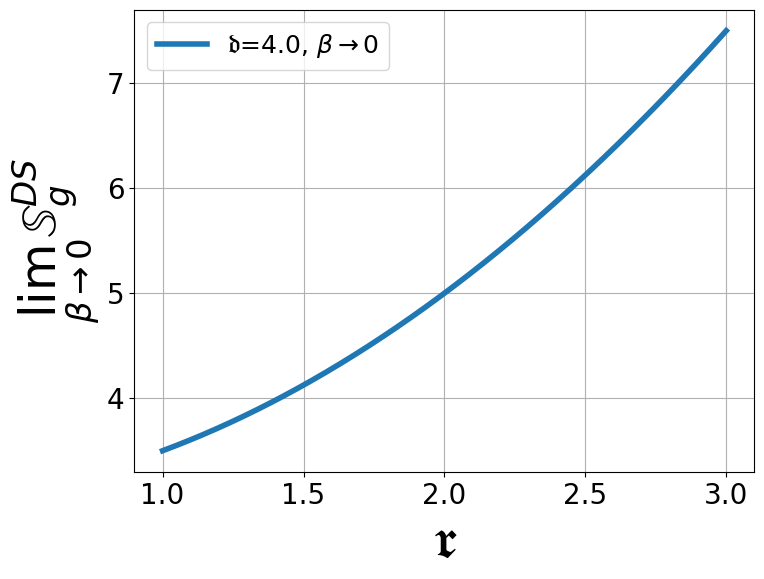}
        \caption{$\mathbb{S}_g^{DS}$ vs $\mathfrak{r}$}
        \label{fig-hdg-spec-c-limit}
    \end{subfigure}
    \caption{\centering{\textbf{Payoff capacity} $\mathbb{S}_g$ vs \textbf{resource value} $\mathfrak{r}$, in IN limit $(\beta\rightarrow 0)$, for \textbf{cost of injury} $\mathfrak{d}=4.0$ via NEM, DS and ABM in HDG.    
    {NEM and ABM agree while DS is an outlier.}}}
    \label{fig:hdg-spec-limit}
\end{figure*} 
For calculating the \textit{payoff capacity}, using \textit{Nash equilibrium mapping}, we write the partition function $\zeta^{NEM}$ in terms of $\mathfrak{r}$ and $\mathfrak{d}$. From Eq.~(\ref{e4}) and Table~\ref{table-analytical}, the NEM partition function in terms of the game parameters $(\mathfrak{r},\mathfrak{d})$ is given as,
\begin{gather}
    \zeta^{NEM} = e^{\beta (\mathfrak{r}-\mathfrak{d})} + e^{-\beta \mathfrak{r}} + 2e^{\beta\frac{\mathfrak{d}}{2}} ,
    \label{e7}
\end{gather}
where, $2(\mathcal{J}+\mathfrak{h}) = (\mathfrak{r}-\mathfrak{d})$, $2(\mathcal{J}-\mathfrak{h}) = -\mathfrak{r}$ and $2\mathcal{J} = -\frac{\mathfrak{d}}{2}$, respectively. Thus, from Eq.~(\ref{e4}), we have the payoff capacity $\mathbb{S}_{g}^{NEM}$ as,
\begin{gather}
    \mathbb{S}_{g}^{NEM} = \dfrac{\frac{\mathfrak{d}^2 e^\frac{\beta\mathfrak{d}}{2}}{2}+(\mathfrak{r}-\mathfrak{d})^2 e^{\beta(\mathfrak{r}-\mathfrak{d})}+\mathfrak{r}^2 e^{-\beta\mathfrak{r}}}{2e^\frac{\beta\mathfrak{d}}{2}+e^{\beta(\mathfrak{r}-\mathfrak{d})}+e^{-\beta\mathfrak{r}}}\nonumber\\
    -\bigg(\dfrac{\mathfrak{d} e^\frac{\beta\mathfrak{d}}{2}+(\mathfrak{r}-\mathfrak{d})e^{\beta(\mathfrak{r}-\mathfrak{d})}-\mathfrak{r}e^{-\beta \mathfrak{r}}}{2e^\frac{\beta\mathfrak{d}}{2}+e^{\beta(\mathfrak{r}-\mathfrak{d})}+e^{-\beta \mathfrak{r}}}\bigg)^2.
    \label{e8}
\end{gather}
In Eq.~(\ref{e8}), when $T\rightarrow 0$ (or, $\beta \rightarrow \infty$), i.e., ZN limit, $\mathbb{S}_{g}^{NEM} \rightarrow 0,~\forall~\mathfrak{r}$ since in this case, the social dilemma game players have attained the minimum internal energy, or, maximum payoff. In terms of the payoff per player $\langle \Lambda_1 \rangle = -\langle\mathbb{E}\rangle$, when $\beta\rightarrow\infty$, an equal number of players choose either $\mathbb{D}$-strategy or $\mathbb{H}$-strategy, leading to $\langle\Lambda_1\rangle = \frac{\mathfrak{d}}{4}$ via the transformed payoff matrix (see, Ref.~\cite{ref4}), and this further gives us the payoff capacity equals to \textit{zero}. When $T\rightarrow \infty$ (or, $\beta \rightarrow 0$), i.e., IN limit, $\mathbb{S}_{g}^{NEM} \rightarrow \frac{(4\mathfrak{r}^2 - 4\mathfrak{rd} + 3\mathfrak{d}^2)}{8},~\forall~0<\mathfrak{r}<\mathfrak{d}$ because of the randomness in strategy selection by the players, which leads to the randomization of payoffs. This can be verified by Taylor expanding $\ln{\zeta^{NEM}}$ (see, Eqs.~(\ref{e3}, \ref{e7})), up to the \textit{second}-order, about $\beta$. From the \textit{second}-order correction of $\ln{\zeta^{NEM}}$, i.e., $\ln{\zeta^{NEM(2)}}$, we have the \textit{zeroth}-order payoff capacity correction, in the IN limit, as,
\begin{equation}
\small{
    \lim_{\beta\rightarrow 0}\mathbb{S}_{g}^{NEM(0)} = \lim_{\beta\rightarrow 0}\frac{1}{\beta^2} \frac{\partial}{\partial \mathfrak{\beta^2}}\ln{\zeta^{NEM(2)}} =\frac{(4\mathfrak{r}^2 - 4\mathfrak{rd} + 3\mathfrak{d}^2)}{8}}.
\end{equation}
The \textit{first} and \textit{higher} order terms in the Taylor expansion of $\mathbb{S}_{g}^{NEM}$ (see, Eq.~(\ref{e8})) about $\beta$ , in the IN limit, vanishes. The non-zero value of $\mathbb{S}_{g}^{NEM}$ in the IN limit indicates that the payoff per player changes by a finite amount when the noise in the system changes by a unit in the $T\rightarrow\infty$ limit. For finite non-zero values of $\beta$, the payoff capacity $\mathbb{S}_{g}^{NEM}$ is always positive, which indicates that with increasing noise, the rate of change of payoff per player also increases.  

Since Hawk-Dove game does not fulfil the criterion: ($\mathcal{A+D=B+C}$), for $\mathcal{A}=-\mathfrak{d},~\mathcal{B}=-\mathcal{C}=\mathfrak{r},~\mathcal{D}=0$ (in original payoff matrix), we cannot apply \textit{Aggregate selection} method to determine the payoff capacity in this game.

\begin{table*}[!ht]
\centering
\renewcommand{\arraystretch}{1.8}
\resizebox{0.65\textwidth}{!}{%
\begin{tabular}{|cc|c|c|c|}
\hline
\multicolumn{2}{|c|}{\textit{\textbf{For HDG}}} & \textbf{NEM} & \textbf{ABM} & \textbf{DS} \\ \hline
\multicolumn{1}{|c|}{} & \textit{$\beta \rightarrow 0$} & \begin{tabular}[c]{@{}c@{}}$\frac{1}{2},~\forall~\mathfrak{r}$\end{tabular} & \begin{tabular}[c]{@{}c@{}}$\frac{1}{2},~\forall~\mathfrak{r}$\end{tabular} & \begin{tabular}[c]{@{}c@{}}$\frac{1}{2},~\forall~\mathfrak{r}$\end{tabular} \\ \cline{2-5} 
\multicolumn{1}{|c|}{\multirow{-2}{*}{\textbf{\huge${\mathbf{\chi_\mathfrak{r}}}$}}} & \textit{$\beta \rightarrow \infty$} & \begin{tabular}[c]{@{}c@{}}$0,~\forall~\mathfrak{r}$\end{tabular} & \begin{tabular}[c]{@{}c@{}}$0,~\forall~\mathfrak{r}$\end{tabular} & \begin{tabular}[c]{@{}c@{}}$0,~\forall~\mathfrak{r}$\end{tabular} \\ \hline
\multicolumn{1}{|c|}{} & \textit{$\beta \rightarrow 0$} & \begin{tabular}[c]{@{}c@{}}$0$\end{tabular} & \begin{tabular}[c]{@{}c@{}}$0$\end{tabular} & \begin{tabular}[c]{@{}c@{}}$0$\end{tabular} \\ \cline{2-5} 
\multicolumn{1}{|c|}{\multirow{-2}{*}{\textbf{\begin{tabular}[c]{@{}c@{}}\huge$\mathbf{\mathfrak{c}_j}$\end{tabular}}}} & \textit{$\beta \rightarrow \infty$} & \begin{tabular}[c]{@{}c@{}}$+ 1$, for even $j$\\ $-1$, for odd $j$\end{tabular} & \begin{tabular}[c]{@{}c@{}}$+1$, for even $j$\\ $-1$, for odd $j$\end{tabular} & \cellcolor[HTML]{FFFFFF}{\color[HTML]{FE0000} \begin{tabular}[c]{@{}c@{}}$+1$, $\forall~ j$\end{tabular}} \\ \hline
\multicolumn{1}{|c|}{} & \textit{$\beta \rightarrow 0$} & \begin{tabular}[c]{@{}c@{}}$\frac{(4\mathfrak{r}^2 - 4\mathfrak{rd} + 3\mathfrak{d}^2)}{8},~\forall~\mathfrak{r}$\end{tabular} & \begin{tabular}[c]{@{}c@{}}$\frac{(4\mathfrak{r}^2 - 4\mathfrak{rd} + 3\mathfrak{d}^2)}{8},~\forall~\mathfrak{r}$\end{tabular} & \cellcolor[HTML]{FFFFFF}{\color[HTML]{FE0000}\begin{tabular}[c]{@{}c@{}}$\frac{(8\mathfrak{r}^2 + 3\mathfrak{d}^2)}{16},~\forall~\mathfrak{r}$\end{tabular}} \\ \cline{2-5} 
\multicolumn{1}{|c|}{\multirow{-2}{*}{\textbf{\large${\mathbf{\mathbb{S}_g}}$}}} & \textit{$\beta \rightarrow \infty$} & \begin{tabular}[c]{@{}c@{}}$0,~\forall~\mathfrak{r}$\end{tabular} & \begin{tabular}[c]{@{}c@{}}$0,~\forall~\mathfrak{r}$\end{tabular} & \begin{tabular}[c]{@{}c@{}}$0,~\forall~\mathfrak{r}$\end{tabular} \\ \hline
\end{tabular}%
}
\caption{\centering{HDG with resource value $\mathfrak{r}$, cost of injury $\mathfrak{d}$, inter-site distance $j$, and measure of noise $\beta$.}}
\label{tab:my-table1}
\end{table*}

\subsubsection{\underline{DS}}
To ascertain the \textit{payoff capacity}, using \textit{DS} method, we write the partition function $\zeta^{DS}$, from Eq.~(\ref{dem-partition}), in terms of $\mathfrak{r}$ and $\mathfrak{d}$. We have,
\begin{gather}
    \zeta^{DS} = (e^{-\beta\mathfrak{d}} + e^{\beta\mathfrak{r}} + e^{-\beta\mathfrak{r}} + 1),
    \label{e9}
\end{gather}
where, $\mathcal{A}=-\mathfrak{d},~\mathcal{B}=-\mathcal{C}=\mathfrak{r},~\mathcal{D}=0$, respectively. From $\zeta^{DS}$ in Eq.~(\ref{e9}), we have the payoff capacity $\mathbb{S}_{g}^{DS}$ as,
\begin{equation}
\footnotesize{
    \mathbb{S}_{g}^{DS} = \dfrac{\mathfrak{d}^2 e^{-\beta\mathfrak{d}}+\mathfrak{r}^2 e^{\beta\mathfrak{r}}+\mathfrak{r}^2 e^{-\beta\mathfrak{r}}}{\zeta^{DS}}-\bigg(\dfrac{-\mathfrak{d}e^{-\beta\mathfrak{d}}+\mathfrak{r}e^{\beta\mathfrak{r}}-\mathfrak{r}e^{-\beta\mathfrak{r}}}{\zeta^{DS}}\bigg)^2.}
    \label{e10}
\end{equation}
In Eq.~(\ref{e10}), when $T\rightarrow 0$ (or, $\beta \rightarrow \infty$), i.e., ZN limit, $\mathbb{S}_{g}^{DS} \rightarrow 0,~\forall~\mathfrak{r}$ since the system is in equilibrium, i.e., it has attained the minimum internal energy of $\langle\mathbb{E}\rangle =  -\mathfrak{r}$. In terms of the game payoff per player, when $\beta\rightarrow\infty$, $\langle\Lambda_1\rangle = -\langle\mathbb{E}\rangle = \mathfrak{r},$ and this payoff does not change for any unit change in noise, in the $T\rightarrow 0$ limit. Even though the payoff capacity via NEM and DS matches in the ZN limit, the payoff per player varies for both NEM and DS, which clearly shows the failure of DS since all players cannot get the resource $\mathfrak{r}$ without inflicting some injury $\mathfrak{d}$. When $T\rightarrow \infty$ (or, $\beta \rightarrow 0$), i.e., IN limit, $\mathbb{S}_{g}^{DS} \rightarrow \frac{(8\mathfrak{r}^2 + 3\mathfrak{d}^2)}{16},~\forall~0<\mathfrak{r}<\mathfrak{d}$ because of the randomness in strategy selection by the players. This can also be verified by Taylor expanding $\ln{\zeta^{DS}}$ (see, Eqs.~(\ref{e3}, \ref{e9})), up to the \textit{second}-order, about $\beta$ (similar to what we did in the case of NEM). 

When $T\rightarrow \infty$, the players randomly change their strategies because of the noise, and the rate at which they change their strategy (equivalent to the rate of change of payoffs associated with the strategies) is given by $\lim_{\beta\rightarrow 0}\mathbb{S}_{g}^{DS}= \frac{(8\mathfrak{r}^2 + 3\mathfrak{d}^2)}{16}$. Here also, for finite non-zero values of $\beta$, the payoff capacity $\mathbb{S}_{g}^{DS}$ is always positive, which indicates that with increasing noise, the rate of change of payoff per player also increases.

\subsubsection{\underline{ABM}}
For determining the \textit{payoff capacity} using the \textit{ABM}, we designed our algorithm such that it finds the variance of the total energy of the system. In Eq.~(\ref{e3a}), we have shown how the specific heat $\mathbb{S}_V$, for the $1D$-Ising chain, is related to the variance of the total energy of the Ising chain. Similarly, for games, we have the numerical payoff capacity $\mathbb{S}_g^{ABM}$ as,
\begin{equation}
    \mathbb{S}_g^{ABM} = \dfrac{1}{N}[\langle \mathbb{E}^2\rangle - \langle \mathbb{E}\rangle^2],
    \label{e11}
\end{equation}
where $\mathbb{E}$ is the total energy (i.e., negative of total payoff) of all the players in the game. The average energy per player, i.e., $\langle \mathbb{E}\rangle/N$, is equivalent to the negative of the individual player's average payoff, i.e., $\langle \Lambda_1\rangle$. The algorithm used in ABM is described in Sec.~\ref{sub-abm}, where we consider the Energy matrix $\Delta'$ (see, Ref.~\cite{ref4}) as,
\begin{equation}
    \Delta' = \begin{bmatrix}
            \frac{\mathfrak{d}-\mathfrak{r}}{2} & -\frac{\mathfrak{r}}{2}\\
            \frac{\mathfrak{r}-\mathfrak{d}}{2} & \frac{\mathfrak{r}}{2}
        \end{bmatrix} = \begin{bmatrix}
            -\mathcal{A'} & -\mathcal{B'}\\
            -\mathcal{C'} & -\mathcal{D'}
        \end{bmatrix}.
        \label{e12}
\end{equation}
We have changed our energy matrix from $\Delta$ to $\Delta'$ (see, Eqs.~(\ref{eq48}, \ref{e12})) since while calculating the payoff capacity from the partition function $\zeta$, our original payoffs $(\mathcal{A} = -\mathfrak{d},~ \mathcal{B} = \mathfrak{r},~ \mathcal{C} = -\mathfrak{r},~ \mathcal{D} = 0)$ get transformed to new payoffs $(\mathcal{A'} = \frac{\mathfrak{r}-\mathfrak{d}}{2},~ \mathcal{B'} = \frac{\mathfrak{r}}{2},~ \mathcal{C'} = \frac{\mathfrak{d}-\mathfrak{r}}{2},~ \mathcal{D'} =-\frac{\mathfrak{r}}{2})$, and this gives us the new energy matrix $\Delta'$ in Eq.~(\ref{e12}). When $T\rightarrow 0$ (or, $\beta \rightarrow \infty$), i.e., ZN limit, $\mathbb{S}_{g}^{ABM} \rightarrow 0,~\forall~\mathfrak{r},$ since in this case, the system is in equilibrium, i.e., it has attained the minimum internal energy of $\langle\mathbb{E}\rangle^{ABM} =  -\frac{\mathfrak{d}}{4}$. In terms of payoff per player, $\langle\Lambda_1\rangle^{ABM} = -\langle\mathbb{E}\rangle^{ABM} = \frac{\mathfrak{d}}{4}$, when $\beta\rightarrow\infty$, an equal number of players choose either $\mathbb{H}$ or $\mathbb{D}$-strategy, which further leads to vanishing payoff capacity. 

When $T\rightarrow \infty$ (or, $\beta \rightarrow 0$), i.e., IN limit, the results of ABM are in good agreement with the results of NEM (see, Fig.~\ref{fig:hdg-spec-limit}), and this indicates that $\lim_{\beta\rightarrow 0}\mathbb{S}_{g}^{ABM} \approx \frac{(4\mathfrak{r}^2 - 4\mathfrak{rd} + 3\mathfrak{d}^2)}{8},~\forall~0<\mathfrak{r}<\mathfrak{d}$. For finite non-zero values of $\beta$, the payoff capacity $\mathbb{S}_{g}^{ABM}$ is always positive, which indicates that with increasing noise, the rate of change of payoff per player also increases.

\subsubsection{Analysis of payoff capacity for HDG}
Finally, we discuss the results obtained for the \textit{payoff capacity} $\mathbb{S}_g$ via NEM, DS and ABM, respectively. As shown in Figs.~\ref{fig:hdg-spec}, \ref{fig:hdg-spec-limit}, we observe that for both limiting cases and finite non-zero values of $\beta$, NEM and ABM have the same inflexion point at $\mathfrak{r}=\mathfrak{d}/2$, and this indicates the \textit{Nash equilibria}, whereas, DS gives an incorrect result where it predicts the inflexion point at $\mathfrak{r}\rightarrow 0$ for different values of $\beta$. Also, in the $T\rightarrow\infty$ limit, DS gives a completely wrong result where it predicts the Nash equilibrium to be at $\mathfrak{r}\rightarrow 0$ (see, Fig.~\ref{fig:hdg-spec-limit}). In Fig.~\ref{fig:hdg-spec}, for increasing values of $\beta$, i.e., decreasing noise in the system, we observe a decrease in the payoff capacity near the Nash equilibrium, for both NEM and ABM, which suggests that the rate of change in payoff per player, with respect to the system's noise, is minimum at the Nash equilibrium.

\subsection{Summary of HDG}
When we look at the results of \textit{resource susceptibility}, \textit{correlation}, and \textit{payoff capacity} for HDG, NEM and ABM results were in good agreement. At finite values of $\beta$, both of them predicted the same Nash equilibrium at $\mathfrak{r}=\mathfrak{d}/2$ for all the three indicators in question. On the other hand, DS gave a different result to that of NEM and ABM. In the $\beta\rightarrow 0$ (or, IN) limit, the payoff capacity $\mathbb{S}_g$ via DS was different from the $\mathbb{S}_g$ obtained via NEM and ABM. Moreover, in the $\beta\rightarrow \infty$ (or, ZN) limit, DS gave an incorrect result for the correlation. The results for the HDG  as discussed in Sec.~\ref{hdg}, in the ZN and IN limits, are summarized in Table~\ref{tab:my-table1}. As shown in Fig.~\ref{fig:2a-}, the results obtained for the variation of correlation with the inter-site distance $j$, via NEM and ABM, are in close agreement with each other. We observe the change of sign in correlation for odd/even distances $j$. For increasing values of $\beta$, the correlation value approaches $+1$ for even $j$'s and $-1$ for odd $j$'s, respectively in case of HDG

\end{document}